\documentclass[prd,amssymb,amsmath,amsfonts,nofootinbib,reprint,showpacs,longbibliography]{revtex4-1}

\usepackage{graphicx}
\usepackage{lmodern}
\usepackage{amsmath,amssymb}
\usepackage{mathrsfs}
\usepackage{amsfonts}
\usepackage[utf8]{inputenc}
\usepackage{url}
\usepackage[colorlinks]{hyperref}
\usepackage[table]{xcolor}
\usepackage{multirow}
\usepackage[normalem]{ulem}
\usepackage{lipsum}

\usepackage{marvosym}
\usepackage{enumerate}
\usepackage{color,soul}
\usepackage{acronym}

\usepackage{bm}
\usepackage{color}
\usepackage{commath}
\allowdisplaybreaks
\usepackage{multirow}

\newcommand{\stf}[1]{{\langle {#1} \rangle}}

\newcommand{\mbf}[1]{\mathbf{#1}}

\newcommand{\hS}{\hat{S}}

\newcommand{\mce}{\mathcal{E}}
\newcommand{\mcb}{\mathcal{B}}

\def\TEOBResumS{\texttt{TEOBResumS}}

\def\TEOBResumSDali{\texttt{TEOBResumS-Dalí}}

\newacro{adm}[ADM]{Arnowitt-Deser-Misner}
\newacro{bbh}[BBH]{binary black hole}
\newacro{bh}[BH]{black hole}
\newacroplural{bh}[BHs]{black holes}
\newacro{bhns}[BHNS]{black hole-neutron star}
\newacro{bhpt}[BHPT]{black hole perturbation theory}
\newacro{bns}[BNS]{binary neutron star}
\newacro{bf}[BF]{Bayes' factor}
\newacro{cbc}[CBC]{compact binary coalescence}
\newacro{ce}[CE]{Cosmic Explorer}
\newacro{da}[DA]{data analysis}
\newacro{et}[ET]{Einstein Telescope}
\newacro{eob}[EOB]{Effective-One-Body}
\newacro{eom}[EOM]{equations of motion}
\newacro{fd}[FD]{frequency domain}
\newacro{fft}[FFT]{Fast Fourier transform}
\newacro{gw}[GW]{gravitational-wave}
\newacroplural{gw}[GWs]{Gravitational-waves}
\newacro{gr}[GR]{general relativity}
\newacro{grb}[GRB]{gamma-ray burst}
\newacro{grhd}[GRHD]{general-relativistic hydrodynamics}
\newacro{gwosc}[GWOSC]{Gravitational Wave Open Science Center}
\newacro{gwtc1}[GWTC-1]{the first gravitational-wave transients catalog}
\newacro{gsf}[GSF]{Gravitational Self Force}
\newacro{hm}[HM]{Higher mode}
\newacroplural{hm}[HMs]{Higher modes}
\newacro{ifo}[IFO]{interferometer}
\newacro{imr}[IMR]{inspiral-merger-ringdown}
\newacro{im}[IMR]{inspiral-to-merger}
\newacro{kagra}[KAGRA]{Kamioka Gravitational Wave Detector}
\newacro{ligo}[LIGO]{Laser Interferometer Gravitational-Wave Observatory}
\newacro{lisa}[LISA]{Laser Interferometer Space Antenna}
\newacro{lr}[LR]{Light Ring}
\newacro{lso}[LSO]{Last Stable Orbit}
\newacro{lvc}[LVC]{LIGO-Virgo Collaboration}
\newacro{lvk}[LVK]{LIGO-Virgo-Kagra Collaboration}
\newacro{lo}[LO]{leading order}
\newacro{ns}[NS]{neutron star}
\newacroplural{ns}[NSs]{neutron stars}
\newacro{nr}[NR]{numerical relativity}
\newacro{nqc}[NQCs]{Next-to-quasicircular corrections}
\newacro{nlo}[NLO]{next-to-leading order}
\newacro{nnlo}[NNLO]{next-to-next-to-leading order}
\newacro{n3lo}[N3LO]{next-to-next-to-next-to-leading order}
\newacro{n4lo}[N3LO]{next-to-next-to-next-to-next-to-leading order}
\newacro{ode}[ODE]{Ordinary Differential Equation}
\newacroplural{ode}[ODEs]{Ordinary Differential Equations}
\newacro{pn}[PN]{post-Newtonian}
\newacro{pm}[PM]{post-Minkowskian}
\newacro{pe}[PE]{parameter estimation}
\newacro{psd}[PSD]{power spectral density}
\newacro{pa}[PA]{post-adiabatic}
\newacro{qnm}[QNM]{quasi-normal mode}
\newacro{qc}[QC]{quasi-circular}
\newacro{snr}[SNR]{signal-to-noise ratio}
\newacro{spa}[SPA]{stationary-phase approximation}
\newacro{sxs}[SXS]{Simulating eXtreme Spacetimes}
\newacro{td}[TD]{time domain}
\newacro{ng}[NG]{Nect Generation}

\definecolor{cyan}{rgb}{0,0.9,0.9}
\definecolor{orange}{rgb}{0.9,0.5,0}
\definecolor{magenta}{rgb}{1,0,1}
\definecolor{purple}{rgb}{0.8,0.4,0.8}
\definecolor{gray}{rgb}{0.8242,0.8242,0.8242}
\definecolor{dodgerblue}{rgb}{0.12, 0.56, 1.0}

\newcommand{\eo}[0]{\hat{E}_0}
\newcommand{\lo}[0]{\hat{L}_0}

\newcommand{\dbphi}[0]{\dot{\bar{\varphi}}}
\newcommand{\dphi}[0]{\dot{\varphi}}

\begin{document}

\title{Spin the black circle: horizon absorption on non-circular, planar binary black hole dynamics}
\thanks{``Spin the black circle'' is a song by Pearl Jam from the album \emph{Vitalogy} (1994). The ``black circle'' here, of course, refers to the black hole horizon
rather than a vinyl record. We invite the reader to listen to the song while reading this paper.}
\author{Danilo \surname{Chiaramello}${}^{1,2}$}
\author{Rossella \surname{Gamba}${}^{3,4}$}

\affiliation{${}^{1}$Dipartimento di Fisica, Universit\`a di Torino, Via P. Giuria 1, 10125 Torino, Italy}
\affiliation{${}^{2}$ INFN Sezione di Torino, Torino, 10125, Italy}
\affiliation{${}^{3}$ Institute for Gravitation \& the Cosmos, The Pennsylvania State University, University Park PA 16802, USA}
\affiliation{${}^{4}$ Department of Physics, University of California, Berkeley, CA 94720, USA}

\begin{abstract}
Binary systems of black holes emit gravitational waves as they move through their orbits.
While most of the emitted radiation escapes to future null infinity, a small fraction is 
absorbed by the black holes themselves. This is known as horizon absorption or tidal heating/torquing,
and causes the black holes' masses and spins to change as the system evolves. 
In this work, we quantify the effects of the horizon fluxes on binary black hole dynamics by computing 
them up to next-to-next-to-leading order on generic planar orbits, also exploring physically
motivated factorizations of the results.
We integrate these fluxes over unbound, hyperbolic-like trajectories obtained with the 
Effective-One-Body model \TEOBResumSDali{}. We discuss the resulting phenomenology across a sizable slice
of the relevant parameter space, finding a very small effect in most cases, except
on highly energetic orbits. However, the predicted mass and spin variations
are quantitatively and qualitatively very sensitive to the analytical representation chosen 
for the fluxes in that regime.
We then perform comparisons with numerical relativity data of induced spins from hyperbolic encounters
of initially nonrotating black holes,
finding that the next-to-next-to-leading order factorized expressions we derive are crucial to reproduce the data.
An optimization on the initial conditions (energy, angular momentum) is necessary for this, however,
with differences of up to $9\%$ between the numerical and optimal initial data.
Finally, we use our analytical expressions to model possible astrophysical implications for black holes
in globular clusters.
\end{abstract}

\date{\today}
\maketitle

\acresetall

\section{Introduction}

Coalescing systems of \acp{bh} are among the most promising
sources of \acp{gw} for ground-based detectors such as LIGO, Virgo and 
KAGRA~\cite{KAGRA:2013rdx, LIGOScientific:2014pky,VIRGO:2014yos}.
While most of the emitted radiation is expected to reach future null infinity,
where it can be detected by \ac{gw} observatories, a small fraction of the energy and angular 
momentum carried by \acp{gw} is absorbed by the \acp{bh}' horizons.
This effect is well known in the literature~\cite{Tagoshi:1997jy,Taylor:2008xy,Poisson:2014gka, Chatziioannou:2012gq, Chatziioannou:2016kem, Saketh:2022xjb}, 
and it is expected to be largely negligible for the dynamics of comparable-mass, 
circularized systems~\cite{Bernuzzi:2012ku}. 
Indeed, for spinning \acp{bh}, this effect enters 2.5 \ac{pn} orders beyond the leading radiation emitted towards infinity;
for nonspinning \acp{bh}, it is shifted even further to 4\ac{pn}.
However, for systems with large mass ratios~\cite{Hughes:2001jr,Isoyama:2017tbp,Maselli:2017cmm,Datta:2019epe,Datta:2024vll} or significant 
eccentricity~\cite{Datta:2023wsn},
the absorption of \acp{gw} by the black holes' horizons may have considerable impact on the system's dynamics
and emitted \ac{gw} signal. Furthermore, next generation detectors such as the Einstein Telescope~\cite{Punturo:2010zz}
may be able to distinguish between objects with different horizon properties~\cite{Mukherjee:2022wws}.
In addition, recent numerical experiments~\cite{Nelson:2019czq,Jaraba:2021ces} 
of \ac{bbh} scatterings have found that significant changes in the \ac{bh} spins 
can occur after such encounters.
While no detailed study of tidal torquing in this context has been performed~\cite{Scheel:2014ina}, it is reasonable to expect that
this effect is the one responsible for the observed changes in the \ac{bh} properties.

Analytical studies of tidal heating and torquing have a long-established history.
For a test-mass orbiting a much heavier \ac{bh} on a quasicircular orbit, the problem can be tackled within \ac{bhpt}~\cite{Poisson:1994yf,Tagoshi:1997jy,Mino:1997bx}.
The energy and angular momentum fluxes absorbed by the horizon have been computed up to relative 11\ac{pn} order~\cite{Fujita:2014eta} 
in the case of a particle orbiting a Kerr \ac{bh}, and up to 22.5\ac{pn} order for a Schwarzschild \ac{bh}.
In this context, an important feature of spinning \acp{bh} emerges from the perturbative computations:
an overall factor $(\Omega - \Omega_{\rm H})$, where $\Omega$ is the orbital frequency of the system
and $\Omega_{\rm H}$ the \ac{bh}'s horizon frequency, can be extracted from the fluxes. If the \ac{bh}'s spin
is aligned with the orbital angular momentum, this leads to an inversion in the flows of energy and angular momentum
as the orbital frequency exceeds the horizon's. When $\Omega < \Omega_{\rm H}$, momentum and rotational energy
are actually being extracted from the \ac{bh}, through a sort of Penrose process~\cite{Penrose:1969pc}; this regime is called 
superradiance.
Energy fluxes through the horizon of a Kerr \ac{bh} displaying this behavior have
been computed by numerically solving the Teukolsky equation~\cite{Teukolsky:1972my,Teukolsky:1973ha,Taracchini:2013wfa}.

Modeling horizon absorption for comparable-mass \acp{bbh} is significantly more challenging.
The first landmark result is due to Alvi~\cite{Alvi:2001mx}, who computed the leading order
effect on quasicircular inspirals. 
The calculations we present in this paper mainly build upon the line of work of Poisson and collaborators, who
in a series of works studied horizon absorption by tying it to the tidal interaction of a \ac{bh}
with its gravitational environment~\cite{Poisson:2004cw,Taylor:2008xy,Comeau:2009bz,Poisson:2009qj,Poisson:2014gka,Poisson:2018qqd}. 
The ``tidal heating/torquing'' monikers for the 
effect originate from the fact that this description is remarkably similar to the classical
gravitational interaction between celestial bodies~\cite{Hartle:1973zz,Hartle:1974gy}, where tidal deformations cause forces and torques
that exchange energy and momentum between the bodies themselves and the system.
In Ref.~\cite{Taylor:2008xy}, the authors consider the motion of a \ac{bh}
immersed in a wider, generic \ac{pn} field. Locally, the metric near the \ac{bh} is a deformation of
standard Schwarzschild (or Kerr) geometry that can be parametrized through a set of \emph{tidal moments} 
organized in a multipolar hierarchy~\cite{Poisson:2005pi,Yunes:2005ve}.
These are symmetric and trace-free tensorial functions of time that are a priori freely 
specifiable, and describe the local tidal environment of the \ac{bh}.
At leading order, they include quadrupolar electric-type ($\mce_{ab}$, even under parity)
and magnetic-type ($\mcb_{ab}$, odd under parity) moments. The former has a direct counterpart in Newtonian gravity,
corresponding to the quadrupole moment of the gravitational potential; the latter is instead a
purely relativistic effect. The tidal moments are the key pieces necessary to compute the
changes in the \ac{bh}'s mass and spin as a result of tidal interactions.
Assuming that the external field is generated by the companion in a binary system, Ref.~\cite{Taylor:2008xy} 
derived the equations of motion for the \ac{bh} and the (quadrupolar) tidal moments up to relative 
1PN order by matching the deformed \ac{bh} field with the \ac{pn} 
external field in an intermediate buffer zone where both descriptions are valid, 
after transforming them into the same coordinate system.
The moments were then used to calculate the tidal deformation of the \ac{bh} (see Sec. VIII of 
Ref.~\cite{Taylor:2008xy}), as well as the tidal heating 
and torquing it undergoes (see Sec. IX of Ref.~\cite{Taylor:2008xy}).

Ref.~\cite{Poisson:2014gka} expanded on the results of Ref.~\cite{Taylor:2008xy} by performing similar computations
for the case of a slowly rotating \ac{bh} of dimensionless spin $\left|\chi_1\right| = \left|S_1\right|/m_1^2 \ll 1$,
finding expressions for the quadrupolar tidal moments up to relative 1.5\ac{pn} order.
Importantly, while the assumption of slow spin is assumed to hold throughout Ref.~\cite{Poisson:2014gka}, 
the results are actually valid for all orders in $\chi_{1,2}$, as 1.5PN spin terms in the metric and equations 
of motion are known to be linear in the spins.

Building on these efforts, Refs.~\cite{Chatziioannou:2012gq, Chatziioannou:2016kem} derived expressions
relating the rates of change of a \ac{bh}'s mass and spin to the tidal moments up to 1.5\ac{pn} order
through a perturbative treatment of the Teukolsky equation;
however, they found discrepancies with previous test-mass results~\cite{Tagoshi:1997jy} when applying them to circularized
binary systems.
Later, Saketh et al., in Ref.~\cite{Saketh:2022xjb}, representing the \ac{bh} as a 
spinning particle with tidally induced quadrupole and octupole moments, calculated the energy and angular
momentum fluxes through the horizon as functions of the tidal tensors within effective world line theory~\cite{Goldberger:2020fot}, 
finding full agreement with earlier results when specializing to binaries on quasicircular orbits.
Both Refs.~\cite{Chatziioannou:2012gq,Chatziioannou:2016kem} and Ref.~\cite{Saketh:2022xjb} factorize in their
results a term involving the \ac{bh} horizon frequency, similar to that found in the extreme mass ratio limit, thereby predicting 
the possibility of superradiant orbits in the comparable-mass case too.

In this work, we expand on these results by computing analytical expressions for the horizon fluxes up to \ac{nnlo} 
on generic planar orbits. We then combine them with the state-of-the-art \ac{eob} model \TEOBResumS{}~\cite{Chiaramello:2020ehz,Nagar:2020xsk,Nagar:2021gss,Albanesi:2021rby,Nagar:2021xnh,Nagar:2024dzj} 
to perform an extensive exploration of their predictions on \ac{bbh} scattering dynamics,
and compare them with numerical simulations of such events.

The paper is organized as follows.
In Sec.~\ref{sec:analytics}, we present the analytical results for the horizon fluxes on generic orbits.
After summarizing the key steps of the computation, we evaluate
the \ac{bh} tidal moments and compute the horizon fluxes in harmonic and \ac{eob} coordinates.
In Sec.~\ref{sec:eob} we study their effect by evaluating them on trajectories
obtained with the \ac{eob} model \TEOBResumSDali{}. We show some examples of a few analytical representations
of the fluxes on unbound orbits, focusing in particular on the impact of tidal torquing during \ac{bbh} scatterings.
In Sec.~\ref{sec:nr} we compare the predictions of our model against results from \ac{nr} simulations
of equal mass, non spinning \ac{bbh} scatterings.
In Sec.~\ref{sec:astro}, we explore the possible astrophysical implications of our results, modeling the evolution of \acp{bh}
in globular clusters and providing order-of-magnitude estimates for the changes in their spins
resulting from repeated encounters.
Finally, in Sec.~\ref{sec:conclusions} we summarize our findings and discuss future developments.

Throughout the paper, we use geometric units $G=c=1$, occasionally keeping explicit powers of $c$ to 
organize \ac{pn} expansions. In addition, we denote the masses of the \acp{bh} as $m_1, m_2$, the
total mass as $M= m_1 + m_2$, the spins as $S_1 = \hat{S}_1/M^2, S_2=\hat{S}_2/M^2$ and their mass-rescaled 
counterparts as $\chi_1 = S_1/m_1^2, \chi_2 = S_2/m_2^2$. The mass ratio is defined as $q = m_1/m_2 \geq 1$, and the
symmetric mass ratio is $\nu = \mu/M = m_1 m_2 /M^2$.

\section{Tidal heating and torquing on generic planar orbits: analytical results} 
\label{sec:analytics}

In this section, we present the main analytical results of our work.
Our calculations below closely follow the one originally presented in Ref.~\cite{Taylor:2008xy}, 
and briefly sketched above. We combine results from this and other aforementioned works
to arrive at general formulae for the tidal heating and torquing. 
Schematically, we proceed as follows:
\begin{enumerate}
  \item[(i)] We evaluate the quadrupolar ($\mce_{ab} (t), \mcb_{ab} (t)$) 
  and octupolar ($\mce_{abc} (t),\mcb_{abc} (t)$) tidal moments of a \ac{bh} of mass $m_1$ due to its interaction 
  with the field of its companion in a binary system, up to 1.5\ac{pn} order~\cite{Poisson:2014gka,Poisson:2018qqd}.
  The expressions we start with are given in the coordinate system tied to the external \ac{pn} field and centered
  on its barycenter (hereafter known as the barycentric frame);
  \item[(ii)] We transform these moments to the \ac{bh}'s comoving rest frame (hereafter, the \ac{bh} frame).
  The transformation, derived in Ref.~\cite{Taylor:2008xy} (see Sec.s V-VI-VII there), consists of a shift 
  between barycentric time $t$ and the \ac{bh}'s proper time $\bar{t}$, 
  a coordinate displacement, and a time-dependent rotation that encodes the relative precession of the two systems' 
  coordinate axes. 
  Denoting quantities expressed in the \ac{bh} frame with an overbar, we thus find the tidal moments 
  $\bar{\mce}_{ab} (\bar{t}), \bar{\mcb}_{ab} (\bar{t}), \bar{\mce}_{abc} (\bar{t}), \bar{\mcb}_{abc} (\bar{t})$;
  \item[(iii)] From these, we compute the rates of change of the \ac{bh}'s mass and spin in the \ac{bh} frame
  using the results of Ref.~\cite{Saketh:2022xjb};
  \item[(iv)] We transform these rates back to the barycentric frame;
  \item[(v)] We map the results from harmonic to \ac{eob} coordinates using the coordinate transformations
  derived in~\cite{Damour:2007nc, Bini:2012ji}, and explore their possible factorization.
\end{enumerate}
Points (i) and (ii) are the subject of Sec.~\ref{sec:tidal_moments}, points (iii) and (iv) are discussed
in Sec.~\ref{sec:results}, while point (v) is the focus of Sec.~\ref{sec:eobcoords}.

\subsection{Transformation of the tidal moments}
\label{sec:tidal_moments}
Restricting the two-body problem to planar motion and working in center of mass coordinates, the positions and velocities
of the \acp{bh} are described by vectors $\bm{r}_{1,2} (t), \bm{v}_{1,2} (t)$. We define the relative position and
velocity vectors as $\bm{r} (t) = \bm{r}_1 (t) - \bm{r}_{2} (t)$ and $\bm{v} (t) = \bm{v}_1 (t) - \bm{v}_2(t)$.
Introducing the unit vector $\mbf{n} = (\cos\varphi (t), \sin\varphi(t), 0)$, where $\varphi (t)$ is the polar coordinate
of the relative position on the equatorial plane, as well as its perpendicular counterpart 
$\bm{\Phi} = (-\sin\varphi (t), \cos\varphi (t), 0)$, we have $\mbf{r} (t) = r(t) \mbf{n} (t)$, while the 
relative velocity becomes $\bm{v} (t) = \dot{r} (t) \bm{n}(t) + r(t) \dot{\varphi}(t) \bm{\varphi} (t)$.
The tidal moments in the barycentric frame $\mathcal{E}_{ab} (t)$, $\mathcal{B}_{ab} (t)$ are then given by
(see Eqs.~(7.14) and (7.15) of Ref.~\cite{Taylor:2008xy}, completed by Eqs.~(10.4-5) and~(12.10) of Ref.~\cite{Poisson:2014gka} 
for the 1.5\ac{pn} terms):
\begin{widetext}
\begin{eqnarray} 
  \label{eq:Eab}
  \mce_{ab} (t)&=& -\frac{3m_2}{r^3} \biggl\{ \biggl[
    1 + \dfrac{1}{c^2} \biggl(
      -\frac{3m_1^2}{2M^2} \dot{r}^2 + 2(r\dot{\varphi})^2 - \frac{5m_1+6m_2}{2r}    
    \biggr) - \dfrac{4}{c^3 r} \biggl(2m_1 \dot{r} + m_2 \chi_2 r \dphi\biggr)\biggr] n_\stf{ab} \nonumber \\
    & & -  \biggl[\dfrac{m_2}{M^2c^2} (2m_1+m_2) \dot{r} (r\dot{\varphi}) - \dfrac{2}{c^3 r} \biggl(\dfrac{8}{3} m_1 r \dphi + m_2 \chi_2 \dot{r}\biggr)\biggr] n_{(a} \Phi_{b)} \nonumber \biggr. \\
    & & \biggl. + \left[\dfrac{1}{c^2} (r\dot{\varphi})^2 - 2 \dfrac{m_2}{M c^3 r} \chi_2 (r \dot{\varphi})\right]\Phi_\stf{ab}
  \biggr\} +O(c^{-4}),\\
                        \mcb_{ab} (t) &=& -\frac{6 m_2}{r^3} \left[(r\dot{\varphi}) - \dfrac{m_2}{c r} \chi_2\right]\, l_{(a} n_{b)}  
  + O(c^{-2}),
  \label{eq:Bab}
\end{eqnarray} 
\end{widetext}
where $\bm{l} = (0,0,1)$ is a unit vector aligned with the $z$ direction and the orbital angular momentum $\bm{L}$, 
$\stf{}$ denotes the symmetric and trace-free part of a tensor, $n_{ab} = n_a n_b$ and $\Phi_{ab} = \Phi_a \Phi_b$. 
We consider here a non-precessing system, with both spins aligned with the orbital angular momentum: 
$\bm{S}_1 = S_1 \bm{l}, \bm{S}_2 = S_2 \bm{l}$. We allow the magnitudes $S_{1,2}$, as well as their dimensionless
counterparts $\chi_{1,2}$, to be negative; hereafter, we will thus refer to cases where $\bm{S}_{1,2}$ are parallel 
or anti-parallel to $\bm{L}$ as, respectively, positive and negative spins.
Note that the tidal moments are functions of time only; the coordinates $r, \varphi$ appearing in their expressions
are to be understood as functions of time themselves, describing the motion of the system.

The rates of change of the \ac{bh}'s mass and spin can be calculated from the \ac{bh} frame tidal moments,
$\bar{\mathcal{E}}_{ab} (\bar{t})$ and $\bar{\mathcal{B}}_{ab}(\bar{t})$.
So, to proceed we need to apply the aforementioned transformation between the two systems to Eqs.~\eqref{eq:Eab} and~\eqref{eq:Bab}.
The transformation from global time $t$ to local time $\bar{t}$ is parameterized by a function $A(t)$:
\begin{eqnarray}
t &=& \bar{t} + c^{-2} A(\bar{t}) + O(c^{-4}) \, ,\\
\dot{A} &=& \frac{m_2^2}{2M^2} \bigl[ \dot{r}^2 + (r\dot{\varphi})^2 \bigr] + \frac{m_2}{r} + O(c^{-2}) \, , \label{eq:ttbar}
\end{eqnarray}
while, for the case of spins aligned with the orbital angular momentum we are considering, the rotation 
is around the direction of $\bm{l}$ and is identified by the vector $\bm{R} (t)$, 
implicitly defined by the following differential equation:
\begin{align}
  \dot{R}^a= -\frac{m_2}{r^2}\left[  \frac{4m_1+3m_2}{2M} (r\dot{\varphi}) +  \dfrac{m_1 \chi_1 + m_2 \chi_2}{rc} \right. \nonumber \\
                               \left.+ O(c^{-2})\right] l^a \, .
\end{align}
Here and in the following we will strive to specify whether any function is of barycentric time $t$
or proper time $\bar{t}$ for clarity; the only exception is in \ac{nlo} and higher order terms, where
the distinction is formally negligible since $t = \bar{t}$ at \ac{lo}. Since both $A$ and $\bm{R}$ appear
only at \ac{nlo} in the transformations (see Eqs.~\eqref{eq:ttbar} above and~\eqref{eq:transf} below),
they can interchangeably be seen as functions of either, so their argument can be omitted. Time derivatives 
are denoted by overdots whenever it is clear from context whether they are taken with
respect to $t$ or $\bar{t}$, or when the distinction is inconsequential.

Overall, the transformation acting on the tidal moments to bring them from the barycentric to the \ac{bh} frame
is encoded in a rotation matrix $\mathcal{N}_{ab} (t)$, and is given by:
\begin{subequations}
  \label{eq:transfmoments}
  \begin{eqnarray}
  \bar{{\cal E}}_{ab}(\bar{t}) &=& {\cal N}_a^{\ c}(t) {\cal N}_b^{\ d}(t) 
    {\cal E}_{cd}(t), 
    \label{eq:transfmoments_E} \\ 
  \bar{{\cal B}}_{ab}(\bar{t}) &=& 
  {\cal N}_a^{\ c}(t) {\cal N}_b^{\ d}(t) 
    {\cal B}_{cd}(t),
\label{eq:transfmoments_B}
\end{eqnarray}
\end{subequations} 
with
\begin{equation} 
  \label{eq:transf}
  {\cal N}_{ab}(t) = \delta_{ab} - \frac{1}{c^2} \epsilon_{abc} R^c(t) + O(c^{-4}) \, .
\end{equation}
Notice that, to \ac{lo}, the transformation is an identity; i.e., any difference between tensors in the
two frames only appears starting at \ac{nlo}.

Taylor expanding Eqs.~\eqref{eq:transfmoments} up to $O(c^{-2})$ for the electric tidal moment $\mce_{ab}$
and to \ac{lo} for the magnetic tidal moment $\mcb_{ab}$ leads to:
\begin{eqnarray}
  \label{eq:bartobh}
\bar{\mce}_{ab}(\bar{t}) &=& \mce_{ab} \left. \right|_{t = \bar{t}} + c^{-2}\biggl[ A (\bar{t}) \partial_{t}
  \left. \mce_{ab} \right|_{t = \bar{t}} + \biggr. \nonumber \\
  &-& \biggl. 2 \epsilon_{cp(a} R^p (\bar{t}) \mce^c_{\ b)}\left.\right|_{t = \bar{t}}\biggr] + O(c^{-4}) \, , \\
\bar{\mcb}_{ab} (\bar{t})&=& \mcb_{ab} \left. \right|_{t = \bar{t}}+ O(c^{-2}) \, .
\end{eqnarray}
Finally, substituting Eqs.~\eqref{eq:Eab} and~\eqref{eq:Bab} into the expressions above, we obtain:
\begin{widetext}
\begin{eqnarray}
  \bar{\mce}_{ab} (\bar{t})&=& -\frac{3m_2}{r^3} \biggl\{
  \biggl[ 1 + \dfrac{1}{c^2} \biggl(
    -\frac{3m_1^2}{2M^2} \dot{r}^2 + 2(r\dot{\bar{\varphi}})^2 - \frac{5m_1+6m_2}{2r} -3A(\bar{t}) \dfrac{\dot{r}}{r}
    \biggr) - \dfrac{4}{c^3 r} \biggl(2m_1 \dot{r} + m_2 \chi_2 r \dbphi\biggr)\biggr] \bar{n}_\stf{ab} \biggr. \nonumber \\
  & & - \biggl.\biggl[\dfrac{1}{c^2} \biggl(\dfrac{(2m_1+m_2)m_2}{M^2} \dot{r} (r\dot{\bar{\varphi}}) -2\dfrac{A(\bar{t})}{r} (r \dot{\bar{\varphi}})\biggr) - \dfrac{2}{c^3 r} \biggl(\dfrac{8}{3} m_1 r \dbphi + m_2 \chi_2 \dot{r}\biggr) \biggr] \bar{n}_{(a} \bar{\Phi}_{b)} \biggr. \nonumber \\ 
  & & + \biggl. \left[\dfrac{1}{c^2}(r\dot{\bar{\varphi}})^2 - 2 \dfrac{m_2}{c^3 r} \chi_2 (r \dot{\bar{\varphi}})\right] \bar{\Phi}_\stf{ab} \biggr\} + O(c^{-4}) \\[1em]
                      \bar{\mcb}_{ab} (\bar{t}) &=& -\frac{6 m_2}{r^3} \left[(r\dot{\bar{\varphi}}) - \dfrac{m_2}{c r} \chi_2\right]\, \bar{l}_{(a} \bar{n}_{b)}  
  + O(c^{-2}), \, 
  \label{eq:momentsbh}
\end{eqnarray}
\end{widetext}
where all variables are here understood to be functions of proper time $\bar{t}$, and overdots indicate derivatives with respect
to the same.
The rotation shifts the azimuthal angle by $\delta \varphi (\bar{t}) = c^{-2} R^3 (\bar{t})$, and we 
collect this shift and that due to the different time coordinate in the definition of the \ac{bh} frame
phase variable $\bar{\varphi} (\bar{t}) = \varphi \left(t\right) + \delta \varphi (\bar{t}) = 
\varphi \left(t = \bar{t}\right) + c^{-2} \dot{\varphi} (\bar{t}) A (\bar{t}) + \delta \varphi (\bar{t})$. 
Correspondingly, the barred vectors $\bar{\bm{n}}, \bar{\bm{\Phi}}, \bar{\bm{l}}$ have analogous definitions to their barycentric 
counterparts, only with the substitution $\varphi \rightarrow \bar{\varphi}$ (plus they are functions
of $\bar{t}$).
The proper-time-derivative of $\bar{\varphi}$ is the angular velocity of the perturbing tidal field 
as seen in the \ac{bh} frame; it differs from the orbital frequency $\dot{\varphi} (t)$
in the barycentric frame because of both the time shift and the time-dependent rotation. 
The two are related by
\begin{align}
\dot{\bar{\varphi}} (\bar{t}) &= \dfrac{d \bar{\varphi}}{d \bar{t}} = \left. \dfrac{d \varphi}{dt} \right|_{t = \bar{t}} \dfrac{dt}{d\bar{t}} + c^{-2} \dfrac{dR^3}{d \bar{t}} = \nonumber \\
                           &= \left. \dfrac{d \varphi}{dt} \right|_{t = \bar{t}} \left(1 + c^{-2} \dot{A} (\bar{t})\right) + c^{-2} \dfrac{dR^3}{d \bar{t}}
\end{align}

The definition of the \ac{bh} frame phase variable conceals the dependence of the tidal moments on 
$\bm{R}$, but the function $A$ explicitly appears in $\bar{\mce}_{ab} (\bar{t})$ and 
$\bar{\mcb}_{ab} (\bar{t})$, as was found in Ref.~\cite{Taylor:2008xy}.

To compute the tidal heating and torquing up to fractional 1.5PN order with the formulas from Ref.~\cite{Saketh:2022xjb},
we need leading order octupolar tidal moments as well. We adapt the expressions valid for generic orbital motion
in a two-body system given in Eqs.~(9.4b) and~(9.9b) of Ref.~\cite{Poisson:2018qqd}:
\begin{subequations}
\begin{align}
\label{eq:octupoles}
\bar{\mce}_{abc} (\bar{t}) &= 15 \dfrac{m_2}{r^4} \bar{n}_{\langle a} \bar{n}_{b} \bar{n}_{c \rangle} \\[0.5em]
\bar{\mcb}_{abc} (\bar{t}) &= 30 \dfrac{m_2}{r^4} (r \dot{\bar{\varphi}}) \epsilon^{ij}{}_{(a} \left(\bar{n}_b \bar{n}_{c)} - \dfrac{1}{5} \delta_{bc)}\right) \bar{n}_i \bar{\Phi}_j
\end{align}
\end{subequations}
where $T_{(ab)}$ denotes the symmetric part of a tensor $T_{ab}$. Note that with respect to
the moments reported in Ref.~\cite{Poisson:2018qqd} we use a different normalization for the
magnetic tensor $\bar{\mcb}_{abc}$, to agree with the quasicircular expression given in 
Ref.~\cite{Saketh:2022xjb}.

\subsection{Computation of the horizon fluxes}
\label{sec:results}
Once the tidal moments in the \ac{bh} frame are known, the horizon fluxes can be computed by evaluating Eqs. (4.6-7) of Ref.~\cite{Saketh:2022xjb},
which we transcribe here in our notation for clarity:
\footnote{Note that we also correct a few typos in Eqs.~(4.6-7) of Ref.~\cite{Saketh:2022xjb}, stemming from one in their Eq.~(4.5) (the exchange of
the spin tensor's indices in the second term, which should read $3/(2J) O^{\mu \nu \lambda}_{\mathcal E} {\mathcal E}_{\mu \lambda}{}^{\rho} S_{\rho \nu}$).}
\begin{widetext}
  \begin{subequations}
    \label{eq:sb_fields}
    \begin{align}
    \dfrac{dm_1}{d \bar{t}} =& \dfrac{m_1^5}{2} \biggl\{
      f^1_{0} \left(\dot{\bar{\mce}}^{a b} \bar{\mce}_a {}^c \hat{S}_{b c} + \dfrac{1}{c^2}\dot{\bar{\mcb}}^{a b} \bar{\mcb}_a{}^c \hat{S}_{b c}\right) +
      f^3_{0} \left(\dot{\bar{\mce}}_{a}{}^c \bar{\mce}_b{}^d \hat{s}^a \hat{s}^b \hat{S}_{c d} + \dfrac{1}{c^2}\dot{\bar{\mcb}}_{a}{}^c \bar{\mcb}_b{}^d \hat{s}^a \hat{s}^b \hat{S}_{c d}\right)+ \nonumber \\
      &+\dfrac{m_1}{c^3} \biggl[
        f^0_{1} \left(\dot{\bar{\mce}}^{a b} \dot{\bar{\mce}}_{a b} + \dfrac{1}{c^2}\dot{\bar{\mcb}}^{a b} \dot{\bar{\mcb}}_{a b}\right) +
        f^2_{1} \left(\dot{\bar{\mce}}_{a}{}^c \dot{\bar{\mce}}_{b c} \hat{s}^a \hat{s}^b + \dfrac{1}{c^2}\dot{\bar{\mcb}}_{a}{}^c \dot{\bar{\mcb}}_{b c} \hat{s}^a \hat{s}^b\right) + \nonumber \\
        &+f^4_{1} \left(\dot{\bar{\mce}}_{a b} \dot{\bar{\mce}}_{c d} \hat{s}^a \hat{s}^b \hat{s}^c \hat{s}^d + \dfrac{1}{c^2}\dot{\bar{\mcb}}_{a b} \dot{\bar{\mcb}}_{c d} \hat{s}^a \hat{s}^b \hat{s}^c \hat{s}^d\right) + \nonumber \\
        &+ \dfrac{2}{3} \chi_1 f^{1}_{0} \left(\left(\bar{\mcb}_{a c d} \dot{\bar{\mce}}^{b c} - \dot{\bar{\mcb}}_{a c d}\bar{\mce}^{b c}\right) \hat{s}^a \hat{S}_b{}^d - \left(\bar{\mce}_{a c d} \dot{\bar{\mcb}}^{b c} - \dot{\bar{\mce}}_{a c d}\bar{\mcb}^{b c}\right) \hat{s}^a \hat{S}_b{}^d \right) + \nonumber \\
        &+ \dfrac{2}{3} \chi_1 f^{3}_{0} \left(\left(\bar{\mcb}_{b c k} \dot{\bar{\mce}}_{a}{}^d - \dot{\bar{\mcb}}_{b c k}\bar{\mce}_{a}{}^d\right) \hat{s}^a \hat{s}^b \hat{s}^c \hat{S}_d{}^k - \left(\bar{\mce}_{b c k} \dot{\bar{\mcb}}_{a}{}^d - \dot{\bar{\mce}}_{b c k}\bar{\mcb}^{a}{}^d\right) \hat{s}^a \hat{s}^b \hat{s}^c \hat{S}_d{}^k \right)
      \biggr]
    \biggr\} \\[1em]
    \dfrac{dS_1}{d\bar{t}} =& \dfrac{m_1^5}{2} \biggl\{
      -2f^1_0 \left(\bar{\mce}_{ab}\bar{\mce}^{ab} + \dfrac{1}{c^2}\bar{\mcb}_{ab}\bar{\mcb}^{ab}\right) + \left(3f^1_0 - f^3_0\right) \left(\bar{\mce}_{ac} \bar{\mce}_b{}^c \hat{s}^a \hat{s}^b + \dfrac{1}{c^2}\bar{\mcb}_{ac} \bar{\mcb}_b{}^c \hat{s}^a \hat{s}^b\right) + \nonumber \\
      &+ f^3_0 \left[\left(\bar{\mce}_{ab} \hat{s}^a \hat{s}^b\right)^2 + \dfrac{1}{c^2}\left(\bar{\mcb}_{ab} \hat{s}^a \hat{s}^b\right)^2\right] + \nonumber \\
      &-\dfrac{m_1}{c^3} \biggl[
        2f^0_1 \left(\dot{\bar{\mce}}^{a b}\bar{\mce}_a{}^c \hat{S}_{b c} + \dfrac{1}{c^2}\dot{\bar{\mcb}}^{a b}\bar{\mcb}_a{}^c \hat{S}_{b c}\right) + f^2_1 \left(\dot{\bar{\mce}}_a{}^c \bar{\mce}_b{}^d \hat{s}^a \hat{s}^b \hat{S}_{c d} + \dfrac{1}{c^2}\dot{\bar{\mcb}}_a{}^c \bar{\mcb}_b{}^d \hat{s}^a \hat{s}^b \hat{S}_{c d}\right) + \nonumber \\
        &- \dfrac{4}{3} \chi_1 f^1_0 \left(\bar{\mcb}_{a d k} \bar{\mce}^{b c} - \bar{\mce}_{a d k} \bar{\mcb}^{b c}\right)\hat{s}^a \hat{S}_b{}^d \hat{S}_c{}^k - \dfrac{4}{3} \chi_1 f^1_0 \left(\bar{\mcb}_{a c k} \bar{\mce}^{b c} - \bar{\mce}_{a c k} \bar{\mcb}^{b c}\right) \hat{s}^a \hat{S}_b{}^d \hat{S}_d{}^k + \nonumber \\
        &- \dfrac{4}{3} \chi_1 f^3_0 \left(\bar{\mcb}_{b c l} \bar{\mce}_a{}^d - \bar{\mce}_{b c l} \bar{\mcb}_a{}^d\right) \hat{s}^a \hat{s}^b \hat{s}^c \hat{S}_d{}^k \hat{S}_k{}^l
      \biggr]
    \biggr\}
  \end{align}
\end{subequations}
  \end{widetext}
The coefficients $f^k_{0,1}$ can be found in Eqs.~(3.35) of Ref.~\cite{Saketh:2022xjb}. For non-precessing
systems, the dimensionless spin vector is just $\hat{\bm{s}} = \bm{l} = (0,0,1)$, while the spin 
tensor is given by $\hat{S}^{ab} = \epsilon^{ab}{}_c \hat{s}^c$, where $\epsilon_{abc}$ is the 
three-dimensional Levi-Civita Tensor, with $\epsilon_{123} = +1$.

Plugging Eqs.~\eqref{eq:momentsbh} and~\eqref{eq:octupoles} into these, we find that the fluxes
in harmonic coordinates up to 1.5\ac{pn} are given by:
\begin{widetext}
\begin{subequations}
\begin{align}
  \dfrac{dm_1}{d \bar{t}} =& -\dfrac{8}{5} \dfrac{m_1^5 m_2^2}{r^6} \chi_1 \biggl\{
    \left(1+3\chi_1^2\right) \dbphi + \dfrac{1}{c^2} \biggl\{
      \left(1+3\chi_1^2\right) \biggl[
        \dfrac{7}{4} r^2 \dbphi^3 - (6M-m_1) \dfrac{\dbphi}{r} - 6 A(\tau) \dfrac{\dbphi \dot{r}}{r} - \dfrac{1}{2}\biggl(1 + \dfrac{5m_1}{M} - 5\nu\biggr) \dbphi \dot{r}^2 \nonumber \biggr. \biggr. \biggr. \\
        &- \biggl.\biggl.\biggl. \dfrac{1}{2} \biggl(1 - \dfrac{m_1}{M} + \nu\biggr) \left(r \dbphi \ddot{r} + r \dot{r} \ddot{\bar{\varphi}}\right)
        \biggr] + \dfrac{5}{4} r^2 \dbphi^3
    \biggr\} + \dfrac{1}{c^3} \biggl\{ 
      -\dfrac{5}{6} \left(2m_1\chi_1 + 3m_2\chi_2\right)\dbphi^2-m_1 \chi_1 \biggl(46 \dfrac{\dot{r}^2}{r^2} + 22 \dbphi^2\biggr) \left(1+\sigma_1\right) \nonumber \biggr. \biggr. \\
      &+ \biggl. \biggl. \left(1+3\chi_1^2\right) \biggl[ -16m_1 \dfrac{\dbphi \dot{r}}{r}
        -m_1\chi_1 \biggl(\dfrac{19}{2} \dfrac{\dot{r}^2}{r^2} + 4\dbphi^2\biggr)\left(1+\sigma_1\right) + \left(10 m_1\chi_1 - m_2\chi_2 - 18m_1 B_2\left(\chi_1\right)\right) \dfrac{\dot{r}^2}{r^2} \biggr. \biggr. \biggr. \nonumber \\
        &+ \biggl. \biggl.\biggl. \left(\dfrac{1}{3} m_1\chi_1 -\dfrac{7}{2} m_2\chi_2 - 8m_2B_2\left(\chi_1\right)\right)\dbphi^2 + m_2\chi_2 \dfrac{\ddot{r}}{r}
      \biggr] - \dfrac{2m_1 \left(1+\sigma_1\right)}{\chi_1} \biggl(\dbphi^2 + 3\dfrac{\dot{r}^2}{r^2}\biggr)
    \biggr\}
  \biggr\} \label{eq:dm1dtau} \\[1em]
                            \dfrac{dS_1}{d \bar{t}} =& -\dfrac{8}{5} \dfrac{m_1^5 m_2^2}{r^6} \chi_1 \biggl\{
    1+3\chi_1^2 + \dfrac{1}{c^2} \biggl\{ \dfrac{5}{4} r^2 \dbphi^2
      -\left(1+3\chi_1^2\right) \biggl[\dfrac{6M-m_1}{r} + 6 A(\tau) \dfrac{\dot{r}}{r} + 3\biggl(\dfrac{m_1}{M}-\nu\biggr) \dot{r}^2 - \dfrac{7}{4} r^2 \dbphi^2\biggr]
    \biggr\} \nonumber \biggr. \\
    &+ \biggl. \dfrac{1}{c^3} \biggl\{
      \dbphi \left(1+3\chi_1^2\right) \biggl[\dfrac{1}{3} m_1\chi_1 - \dfrac{7}{2} m_2\chi_2 - 8m_1 B_2 \left(\chi_1\right) - 4m_1\chi_1\left(1+\sigma_1\right)\biggr] 
      - 22m_1\chi_1\dbphi \left(1+\sigma_1\right) \nonumber \biggr.\biggr. \\
      &- \biggl.\biggl. \dfrac{5}{6} \dbphi\left(2m_1\chi_1 + 3m_2\chi_2\right) - 16 m_1 \left(1+3\chi_1^2\right) \dfrac{\dot{r}}{r}-\dfrac{2m_1\dbphi \left(1+\sigma_1\right)}{\chi_1}
    \biggr\}
  \biggr\} \label{eq:dS1dtau}
                    \end{align}
\label{eq:dmsdtau}
\end{subequations}
\end{widetext}
In the above, $\sigma_1 = \sqrt{1 - \chi_1^2}$, and the function $B_2$ is defined as
\begin{equation}
B_2 \left(\chi_1\right) = \Im \biggl[\psi^{(0)} \biggl(3 + 2 {\rm i} \dfrac{\chi_1}{\sigma_1}\biggr)\biggr] \, ,
\end{equation}
where $\psi^{(0)}$ is the digamma function.
These results are also tied to the \ac{bh} frame; again, all dynamical variables in the above expressions are
to be interpreted as functions of time $\bar{t}$. 
In order to implement them in the context of \ac{pn} or \ac{eob} dynamics, we need to transform them 
back to the barycentric frame, employing the inverse of the transformation used above. 
Inverting the matrix of Eq.~\eqref{eq:transf} and the relations of Eqs.~\eqref{eq:bartobh}
is straightforward, so finally we find:
\begin{widetext}
\begin{subequations}
  \label{eq:dmsdt}
\begin{align}
                                \dfrac{dm_1}{dt} =& -\dfrac{8}{5} \dfrac{m_1^3 m_2^2}{r^6}\chi_1 \biggl\{
    \dot{\varphi} \left(1+3\chi_1^2\right) + \dfrac{1}{c^2} \biggl\{
      \left(1+3\chi_1^2\right) \biggl[\dfrac{7}{4} r^2 \dot{\varphi}^3 - \biggl(1-\dfrac{m_1}{M}+\nu\biggr) \dfrac{r \ddot{r}\dot{\varphi} + r\dot{r}\ddot{\varphi}}{2} - 
      \biggl(15 - 5\dfrac{m_1}{M}+\nu\biggr) \dfrac{\dot{\varphi}}{2r} \nonumber \biggr. \biggr. \biggr. \\
      &- \biggl.\biggl.\biggl. \biggl(1+5\dfrac{m_1}{M}-5\nu\biggr) \dfrac{\dot{\varphi} \dot{r}^2}{2}  \biggr] + \dfrac{5}{4} r^2 \dot{\varphi}^3
    \biggr\} + \dfrac{1}{c^3} \biggl\{
      -\dfrac{5}{6} \left(2m_1 \chi_1 + 3m_2 \chi_2\right) \dot{\varphi}^2 - m_1 \chi_1 \biggl(46 \dfrac{\dot{r}^2}{r^2} + 22 \dot{\varphi}^2\biggr)\left(1+\sigma_1\right) \nonumber \biggr. \biggr. \\
      &+ \biggl. \biggl. \left(1+3\chi_1^2\right) \biggl[\left(m_1 \chi_1 + m_2 \chi_2\right)\dfrac{m_2}{r^3} + \left(10m_1 \chi_1 - m_2 \chi_2 - 18m_1 B_2 \left(\chi_1\right)\right)\dfrac{\dot{r}^2}{r^2} \nonumber \biggr. \biggr. \biggr. \\
      &+\biggl. \biggl. \biggl. \biggl(\dfrac{1}{3} m_1 \chi_1 - \dfrac{7}{2} m_2 \chi_2 - 8m_1 B_2 \left(\chi_1\right)\biggr) \dot{\varphi}^2 + m_2 \chi_2 \dfrac{\ddot{r}}{r}
        -16m_1 \dfrac{\dphi \dot{r}}{r}-m_1\chi_1 \biggl(\dfrac{19}{2} \dfrac{\dot{r}^2}{r^2} + 4 \dot{\varphi}^2\biggr)\left(1+\sigma_1\right) \biggr] \nonumber \biggr. \biggr. \\
        &-\biggl. \biggl. \dfrac{2m_1 \left(1+\sigma_1\right)}{\chi_1} \biggl(\dot{\varphi}^2 + 3 \dfrac{\dot{r}^2}{r^2}\biggr)
      \biggr\}
  \biggr\}
  \\[1em]
  \dfrac{d S_1}{d t} =& -\dfrac{8}{5} \dfrac{m_1^3 m_2^2}{r^6}\chi_1\biggl\{
     \left(1 + 3 \chi _1^2\right)-\frac{1}{c^2}  \biggl\{ \left(1+3\chi_1^2\right)
    \left[\dfrac{7M-2m_1}{r} + \biggl(1+5\dfrac{m_1}{M}-7\nu\biggr) \dfrac{\dot{r}^2}{2} - \biggl(5+2\dfrac{m_1}{M}+2\nu\biggr)\dfrac{r^2 \dot{\varphi}^2}{4}\right]  \biggr. \biggr. \nonumber \\
    &- \biggl. \biggl. \dfrac{5}{4} r^2 \dot{\varphi}^2 \biggr\} + \dfrac{1}{c^3} \biggl\{ \dphi \left(1+3\chi_1^2\right) \biggl[-4m_1 \chi_1 (1+\sigma_1)+ 
    \dfrac{1}{3} m_1\chi_1 - \dfrac{7}{2} m_2 \chi_2 - 8m_1 B_2 \left(\chi_1\right)\biggr] \biggr. \biggr. \nonumber \\
    &- \biggl. \biggl. 22m_1 \chi_1 \dphi \left(1+\sigma_1\right) -
    \dfrac{5}{6} \dphi \left(2m_1 \chi_1 + 3m_2 \chi_2\right) - 16m_1 \left(1+3\chi_1^2\right) \dfrac{\dot{r}}{r} - \dfrac{2m_1 \dphi (1+\sigma_1)}{\chi_1} \biggr\}
                    \biggr\} \, .
\end{align}
\end{subequations}
\end{widetext}
Notably, the results for the fluxes lose their explicit dependence on the function $A(t)$ that parametrizes
the time shift between the two systems (as well as the hidden one on $\bm{R} (t)$).
We have verified that, considering equatorial \emph{circular} orbits, these expressions reduce to those given
in Ref.~\cite{Saketh:2022xjb}.

\subsection{Results in EOB coordinates}
\label{sec:eobcoords}
Pure \ac{pn} dynamics is known to provide faulty results in the
high velocity regime that we will consider in the following sections.
Results obtained in Sec.~\ref{sec:results} therefore need to be implemented in an \ac{eob} formalism in order
for quantitative comparisons with numerical simulations to be possible.
This can be performed by simply applying the 1.5\ac{pn} canonical transformation from harmonic to \ac{eob}
coordinates to the equations describing the tidal heating and torquing in the barycentric frame.
We do this in two steps, first mapping from harmonic to \ac{adm} coordinates
using the results of Refs.~\cite{Damour:2007nc, Bini:2012ji}, and then applying the transformation from
\ac{adm} to \ac{eob} coordinates. For all calculations, we consider expressions only up to relative 1.5\ac{pn} order.
This means that, when moving from harmonic to \ac{adm} coordinates, the spin vectors are unchanged~\cite{Damour:2007nc};
the relative position vector $\mbf{x}_h$ receives corrections exclusively due to spin-orbit couplings,
as \ac{adm} and harmonic coordinates otherwise coincide at this order, and we switch from harmonic 
relative velocity $\mbf{v}_h$ to \ac{adm} momenta, $\mbf{p}_a$.
Conversely, the transformation from \ac{adm} to \ac{eob} coordinates is entirely independent of 
spin effects, which only enter at 2\ac{pn} order~\cite{Nagar:2011fx}.
We use rescaled, dimensionless variables (time, \ac{eob} momenta and radius), related to their physical counterparts 
(capitalized) by $p_\varphi = P_\varphi/(\mu M), p_r = P_r/\mu, r = R/M, t = T/M$. From this point on,
we will refer to the \ac{eob} rescaled coordinate distance by $r$, and use $r_{h}$ for the harmonic
dimensionful radius (and in general a subscript ${}_h$ for quantities expressed in harmonic coordinates).
For consistency, when using EOB coordinates we also switch to the dimensionless angular momentum variables
$\hS_{1,2} = S_{1,2}/M^2$. Throughout this work, we use the \emph{initial} total mass $M$ of the system 
as a simple scaling factor, never considering its evolution due to tidal heating; this is done for simplicity,
and to better focus on the angular momentum flux itself without the confounding effect of the variation of
both \acp{bh}' masses. Therefore, in particular, $\dot{\hS}_{1,2}=\dot{S}_{1,2}/M^2$.
As a final remark before moving on to the outcome of the transformation, in the previous section
we retained in the expressions for $dm_1/dt, dS_1/dt$ explicit second time-derivatives of the dynamical 
variables $r_h, \varphi_h$, without replacing them by means of the (harmonic) equations of motion.
However, when switching to \ac{eob} coordinates it is necessary to make use of them 
in order for the final results to only depend on the \ac{eob} canonical variables 
($r, p_{r}, \varphi, p_{\varphi}$). Since said second order derivatives only appear
starting at \ac{nlo}, Newtonian-level equations of motion are sufficient for this purpose.

The expressions we obtain after applying the transformation are:
\begin{widetext}
\begin{subequations}
  \begin{align}
  \dfrac{\dot{m}_1}{M} =& -\dfrac{8}{5} \nu^2 \biggl(\dfrac{m_1}{M}\biggr)^3 \dfrac{\chi_1}{r^6} \biggl\{
    \left(1+3 \chi _1^2\right) \frac{p_{\varphi }}{r^2} + \dfrac{1}{c^2}\dfrac{p_\varphi}{r^2} \biggl\{
    \left(1+3\chi_1^2\right) \biggl[\dfrac{p_\varphi^2}{r^2} \biggl(\dfrac{3}{4} + \dfrac{m_1}{2M} - 4\nu\biggr)
    -p_r^2 \biggl(\dfrac{7m_1}{2M} + 6\nu\biggr) \biggr. \biggr. \biggr. \nonumber \\
    &- \biggl. \biggl. \biggl.  \left(1-\dfrac{m_1}{M}-2\nu\right)\dfrac{2}{r}\biggr]+ \dfrac{5}{4} \dfrac{p_\varphi^2}{r^2} \biggr\}
    +\dfrac{1}{r^2c^3}\biggl\{\left(1+3\chi_1^2\right) \biggl[\chi_1 \biggl(10\dfrac{m_1}{M} p_r^2 + \dfrac{p_\varphi^2}{3r^2}\biggl(\dfrac{m_1}{M}+9\nu\biggr) + \dfrac{1}{2r} \biggl(4\dfrac{m_1}{M}+\nu\biggr)\biggr) \biggr. \biggr. \biggr. \nonumber \\
    &- \biggl. \biggl. \biggl. 2\dfrac{m_1}{M} \biggl(4\dfrac{p_\varphi^2}{r^2}+9p_r^2\biggr)B_2\left(\chi_1\right) +\chi_2 \biggl(\dfrac{p_\varphi^2}{2r^2} \biggl(6\nu - 5\dfrac{m_2}{M}\biggr) + \dfrac{1}{2r} \biggl(4\dfrac{m_2}{M}-3\nu\biggr) - \dfrac{m_2}{M} p_r^2\biggr) - \dfrac{64m_1}{3M} \dfrac{p_\varphi p_r}{r} \biggr. \biggr. \biggr. \nonumber \\
    &- \biggl. \biggl. \biggl.\dfrac{m_1\chi_1}{2M}\left(1+\sigma_1\right) \biggl(8\dfrac{p_\varphi^2}{r^2}+19p_r^2\biggr)\biggr] -46\dfrac{m_1}{M} \chi_1 p_r^2 (1+\sigma_1) - \dfrac{p_\varphi^2}{3r^2}\dfrac{m_1}{M} \chi_1 (71+66\sigma_1)-\dfrac{5}{2}\dfrac{m_2}{M} \chi_2 \dfrac{p_\varphi^2}{r^2} \biggr. \biggr. \nonumber \\
    &- \biggl. \biggl. \dfrac{2m_1 (1+\sigma_1)}{M\chi_1} \biggl(\dfrac{p_\varphi^2}{r^2}+3p_r^2\biggr)\biggr\}
    \biggr\} \label{eq:tidal-eob-m} \\[1em]
    \dfrac{\dot{S}_1}{M^2} =& -\dfrac{8}{5} \nu^2 \biggl(\dfrac{m_1}{M}\biggr)^3 \dfrac{\chi_1}{r^6} \biggl\{
      1+3 \chi _1^2+ \dfrac{1}{c^2}\biggl\{ \left(1+3\chi_1^2\right) \biggl[ \dfrac{p_\varphi^2}{2r^2} \biggl(\dfrac{5}{2} + \dfrac{m_1}{M} - 5\nu\biggr)-\dfrac{1}{2}p_r^2 \biggl(1+5\dfrac{m_1}{M}+11\nu\biggr) \biggr. \biggr. \biggr. \nonumber \\
      &- \biggl. \biggl. \biggl.  \biggl(1 - 2\dfrac{m_1}{M} - 3\nu\biggr)\dfrac{1}{r} \biggr] + \dfrac{5}{4} \dfrac{p_\varphi^2}{r^2} \biggr\}
                  +\dfrac{1}{r^2c^3} \biggl\{\left(1+3\chi_1^2\right) \biggl[\chi_1p_\varphi \left(-\dfrac{11m_1}{3M}+3\nu-4\dfrac{m_1}{M} \sigma_1\right)- 16\dfrac{m_1}{M} r p_r
      \biggr. \biggr. \biggr. \nonumber \\
      &+ \biggl. \biggl. \biggl.  \chi_2 p_\varphi\left(-\dfrac{7}{2}\dfrac{m_2}{M} + 3\nu\right) - 8\dfrac{m_1}{M} p_\varphi B_2(\chi_1) \biggr] 
      - \dfrac{m_1}{3M} p_\varphi (71+66\sigma_1)\chi_1 - \dfrac{5m_2}{2M}  \chi_2 p_\varphi-\dfrac{2m_1p_\varphi(1+\sigma_1)}{M\chi_1}\biggr\}
    \biggr\} \label{eq:tidal-eob-S}
  \end{align}
  \label{eq:tidal-eob}
\end{subequations}
\end{widetext}

The equation for the mass rate of change is up to \ac{nlo} proportional to the factor $p_\varphi/r^2$,
which to \ac{lo} corresponds to the orbital frequency $\dphi$; terms only depending on the radial
momentum appear at 1.5\ac{pn}. In the angular momentum flux, instead, we see that all terms at \ac{nnlo}
but one contain the same factor; the linear coupling between the angular momentum (or the orbital
frequency) and the \acp{bh}' spins can be understood as a form of spin-orbit interaction term.
In both fluxes, a single term at 1.5\ac{pn} is \emph{linear} in the radial momentum $p_r$.
These pieces come from contributions to the deformed \ac{bh} metric computed in Ref.~\cite{Poisson:2014gka}
(see Sec.~XII therein) that involve time-derivatives of the electric quadrupolar tidal moment.
While they only cause an uninteresting constant phase shift in $\mathcal{E}_{ab}$ on quasicircular
orbits, on generic orbits their impact on the horizon fluxes depends on the sign of the radial momentum
or velocity. For instance, on hyperbolic scattering trajectories such as the ones that are the focus
of Secs.~\ref{sec:eob} onwards of this work, they  break the symmetry about the time 
of closest approach.\footnote{Of course, this symmetry is only approximate for a \ac{bbh} system
on account of the loss of energy and angular momentum to \ac{gw} emission during the encounter, 
but this effect is almost negligible outside of very energetic, close scattering events.}

Note that all terms in Eqs.~\eqref{eq:tidal-eob} vanish when considering non-spinning \acp{bh}, save for one each in the energy and angular momentum fluxes
at 1.5\ac{pn}, that is proportional to $1/\chi_1$ (the spin in the denominator canceling that in the prefactor). 
These terms are also found in the quasicircular limit, where it is known that they can be factored out of the \ac{pn} expansion and into
a prefactor of the form $(\Omega_{\rm H}^1 - \Omega_T)$~\cite{Taracchini:2013wfa, Chatziioannou:2012gq, Chatziioannou:2016kem, Saketh:2022xjb}, 
where $\Omega_{\rm H}^1$ is the horizon frequency of the primary \ac{bh},
\begin{equation}
  \Omega_{\rm H}^1 = \dfrac{M\chi_1}{2m_1 \left(1 + \sigma_1\right)}.
\end{equation}
$\Omega_T$ is the orbital frequency of the perturbing tidal field, which coincides with the system's orbital
frequency at lowest order. This treatment is inspired by perturbative solutions of the Teukolsky equation \cite{Fujita:2014eta, Taracchini:2013wfa}.
It is meaningful for two reasons: first, it highlights the superradiance phenomenon, whereby a rotating \ac{bh} can actually inject 
angular momentum into the binary system, rather than absorb it. Taken only up to \ac{nlo}, both here and in the quasicircular case,
the expressions for $\dot{\hat{S}}_{1,2}$ would in fact always predict the \ac{bh} to spin down as a result of the tidal interaction, ceding 
momentum to the system, contrary to the absorption found already to \ac{lo} for Schwarzschild \acp{bh}~\cite{Taylor:2008xy}. 
This factorization at \ac{nnlo} shows that, for each \ac{bh}, the direction of the exchange of rotational energy actually depends on the 
relative sense of its intrinsic rotation and the system's motion, as seen in the \ac{bh}'s own rest frame.
Thus, as a binary evolves through a quasicircular inspiral, sign changes can occur in the angular momentum and energy fluxes.
This is relevant from a dynamical standpoint: horizon absorption, or exchange more generally, enters the right hand side of 
\ac{bbh} models; for \acp{bh} spinning very fast in the direction of the orbital angular momentum, this term could act as a ``stabilizing'', 
rather than dissipative, influence, late into the binary evolution.
Secondly, by setting the spins to zero in the factorized fluxes, their non-spinning counterparts can be recovered up to NLO, which corresponds to 2.5\ac{pn} 
above the leading order horizon flux in the spinning case (as a reminder, for non-spinning \acp{bh}, horizon absorption starts at 1.5\ac{pn} orders above
the spinning \ac{lo}). This means that higher-order non-spinning terms are hidden in the factorized spinning expressions.
For future clarity, we will refer to the factored out term involving the horizon frequency as either the superradiance prefactor, or the horizon
frequency prefactor.

The same factorization can be carried out in the generic-orbit expressions we have derived here as well.
Explicitly, in the equation for $dm_1/dt$ we identify the following 1.5\ac{pn} term:
\begin{equation}
  -\dfrac{2m_1 (1+\sigma_1)}{M\chi_1} \biggl(\dfrac{p_\varphi}{r^2} + 3 \dfrac{p_r^2}{p_\varphi}\biggr) = 
  -\dfrac{1}{\Omega_{\rm H}^1} \biggl(\dfrac{p_\varphi}{r^2} + 3 \dfrac{p_r^2}{p_\varphi}\biggr);
\end{equation}
in the expression for $dS_1/dt$ we find a similar one:
\begin{equation}
  -\dfrac{2m_1 (1+\sigma_1)}{M\chi_1} \dfrac{p_\varphi}{r^2} = -\dfrac{1}{\Omega_{\rm H}^1} \dfrac{p_\varphi}{r^2}.
\end{equation}
The leading factors in both fluxes can be reworked to feature the horizon frequency of the primary \ac{bh},
$\Omega_{\rm H}^1$; by separating the highlighted terms out of the \ac{pn} expansion in braces in Eqs.~\eqref{eq:tidal-eob}
and multiplying by the leading factor, we see explicitly that they indeed survive in the non-spinning limit.
Mimicking the quasicircular case, we can then proceed to factor out similar horizon frequency prefactors, finding:
\begin{widetext}
  \begin{subequations}
    \label{eq:eob_fact}
  \begin{align}
    \label{eq:mdot_eob_fact}
  \dfrac{\dot{m}_1}{M} = &-\dfrac{16}{5}\nu^2 \biggl(\dfrac{m_1}{M}\biggr)^4  \dfrac{1+\sigma_1}{r^6} \biggl[\Omega_{\rm H}^1 - \dfrac{1}{c^3}\biggl(\dfrac{p_\varphi}{r^2} + 3\dfrac{p_r^2}{p_\varphi}\biggr)\biggr] 
  \biggl\{\left(1+3\chi_1^2\right) \dfrac{p_\varphi}{r^2} + \dfrac{1}{c^2}\dfrac{p_\varphi}{r^2} \biggl[-\frac{2}{r} \left(1+3\chi_1^2\right) \biggl(\dfrac{m_2}{M} -2 \nu\biggr) \biggr. \biggr. \nonumber \\
  & \biggl. \biggl. -\dfrac{1}{2}p_r^2 \left(1+3\chi_1^2\right) \biggl(\dfrac{7m_1}{M}+12\nu\biggr) + \dfrac{p_{\varphi }^2}{r^2} \biggl(\left(1+3\chi_1^2\right) \biggl(\dfrac{3}{4} + \dfrac{m_1}{2M}-4\nu\biggr)+\dfrac{5}{4}\biggr)  \biggr] \biggr. \nonumber \\
  &+ \biggl. \dfrac{1}{r^2c^3}\biggl[\left(1+3\chi_1^2\right) \biggl[\chi_1 \biggl(\dfrac{10m_1}{M} p_r^2 + \dfrac{p_\varphi^2}{3r^2}\biggl(\dfrac{m_1}{M}+9\nu\biggr) + \dfrac{1}{2r} \biggl(\dfrac{4m_1}{M}+\nu\biggr)\biggr) - \dfrac{2m_1}{M} \biggl(4\dfrac{p_\varphi^2}{r^2}+9p_r^2\biggr)B_2\left(\chi_1\right)\biggr.\biggr.\biggr. \nonumber \\
  &+ \biggl.\biggl.\biggl.\chi_2 \biggl(\dfrac{p_\varphi^2}{2r^2} \biggl(6\nu - \dfrac{5m_2}{M}\biggr) + \dfrac{1}{2r} \biggl(\dfrac{4m_2}{M}-3\nu\biggr) - \dfrac{m_2}{M} p_r^2\biggr) -\dfrac{64m_1}{3M} \dfrac{p_\varphi p_r}{r}-\dfrac{m_1\chi_1}{2M}\left(1+\sigma_1\right) \biggl(8\dfrac{p_\varphi^2}{r^2}+19p_r^2\biggr)\biggr] \biggr. \biggr. \nonumber \\
  &- \biggl. \biggl. \dfrac{28m_1}{M} \chi_1 p_r^2 (1+\sigma_1) - \dfrac{p_\varphi^2}{3r^2}\dfrac{m_1}{M} \chi_1 (53+48\sigma_1)-\dfrac{5 m_2 \chi_2}{2M} \dfrac{p_\varphi^2}{r^2}\biggr]
  \biggr\} \\[1em]
\dfrac{\dot{S}_1}{M^2} =& -\dfrac{16}{5} \nu^2 \biggl(\dfrac{m_1}{M}\biggr)^4 \dfrac{1+\sigma_1}{r^6} \left(\Omega_{\rm H}^1 - \dfrac{p_\varphi}{r^2c^3}\right) \biggl\{
      1 + 3 \chi _1^2+ \dfrac{1}{c^2}\biggl[p_r^2 \left(1+3\chi_1^2\right) \biggl(-3+\dfrac{5m_2}{2M}  - \dfrac{11}{2} \nu \biggr) \biggr. \biggr. \nonumber \\
      &- \biggl. \biggl. \dfrac{p_{\varphi }^2}{2r^2} \biggl(\left(1+3\chi_1^2\right) \left(\dfrac{m_2}{M}+\dfrac{5}{2} \nu\right)+\dfrac{3}{2} \left(4+7 \chi _1^2\right)\biggr)- \dfrac{1+3\chi_1^2}{r} \left(\dfrac{2m_2}{M} - 1 - 3\nu\right)\biggr] \biggr. \nonumber \\
      &+ \biggl. \dfrac{1}{r^2c^3} \biggl[\left(1+3\chi_1^2\right) \biggl(\chi_1 p_\varphi\biggl(-\dfrac{11m_1}{3M}+3\nu-\dfrac{4m_1 \sigma_1}{M}\biggr)
      + \chi_2 p_\varphi \biggl(-\dfrac{7m_2}{2M} + 3\nu\biggr) - \dfrac{8m_1}{M} p_\varphi B_2(\chi_1) \biggr. \biggr. \biggr. \nonumber \\
      & \biggl. \biggl. \biggl.- \dfrac{16m_1}{M} rp_r\biggr) - \dfrac{m_1}{3M} p_\varphi (53+48\sigma_1)\chi_1 - \dfrac{5m_2}{2M} \chi_2 p_\varphi\biggr]
      \biggr\}
  \end{align}
\end{subequations}
\end{widetext}
Here we see a notable if predictable difference with respect to the quasicircular case: the factored out terms are not the same for the mass and angular 
momentum fluxes, with the former including an additional term proportional to $p_r^2$. This is due to the two fluxes no longer being tied by the \emph{rigid rotation
relation}, $\dot{m}_1 = M\Omega_{\rm T} \dot{\hS}_1$~\cite{Poisson:2004cw}, in the generic orbit case. 
For the angular momentum, the balance between the absorption through the horizon and the ``mechanical'', Penrose-like exchange is still determined only (at this order) 
by the azimuthal velocity; however, now, for the mass/energy, the radial motion enters
the equation as well, somewhat decoupling the two effects. One of our aims in this work is to explore tidal heating in scattering
encounters of \acp{bh}. Using the quasicircular limit as a physical guide, we would expect the switch into and out of superradiant 
dynamics to occur around periastron, where the orbital frequency of the system peaks; these results suggest that the sign
of the energy flux can already change with respect to the \ac{lo} before and after closest approach, when the \acp{bh}
have high radial velocity.

As in the quasicircular case, it is worth remarking here that the equations thus factorized formally include
terms beyond our working 1.5\ac{pn} order, found by multiplying the superradiance prefactors by the \ac{nlo}
and \ac{nnlo} terms in the following expansions.
By doing so, however, our expressions do not immediately reduce to the correct result for a non-spinning 
\ac{bh} that can be computed from Eqs.~(8.38) and~(8.39) of Ref.~\cite{Poisson:2004cw}.
This is expected: the \ac{nlo} and \ac{nnlo} parts of tidal frequency $\Omega_{\rm T}$ that appears in the 
superradiance prefactor in the quasicircular formulae enter the fluxes in this limit, and they are missing here.
Corrections to the generic prefactors connecting the spinning and non-spinning results could be derived by forcing 
agreement between them. We refer the interested reader to App.~\ref{app:nonspin} for
more details on this matter and a discussion of our efforts to this end.

\section{EOB dynamics}
\label{sec:eob}

\subsection{Phenomenology}
\label{sec:phenomenology}

\begin{figure*}
  \includegraphics[width=\textwidth]{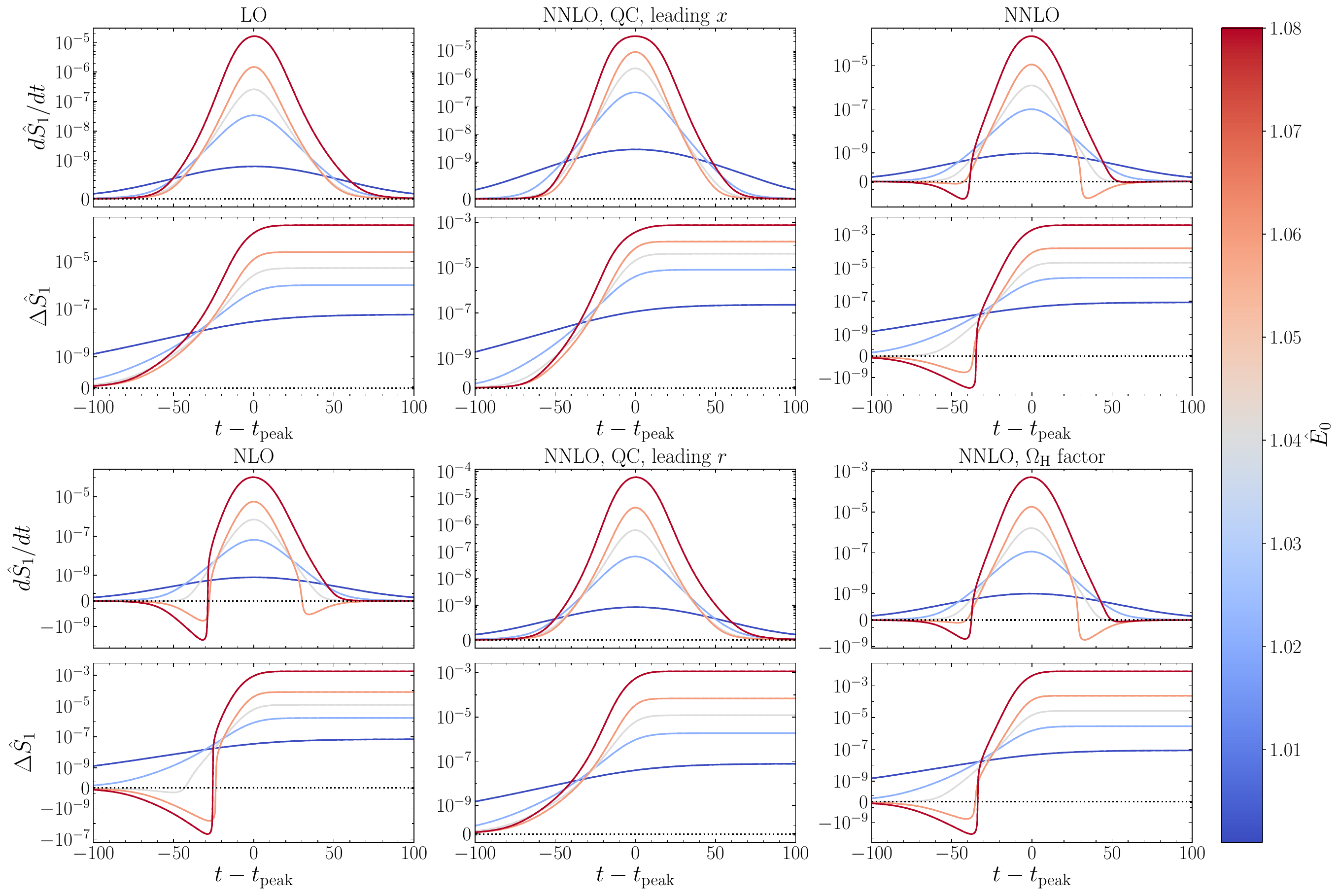}
  \caption{\label{fig:chi-06_spinflux} Instantaneous and cumulative (Eq.~\eqref{eq:def_delta}) variation of the spin $\hS_1$ 
  for an equal-mass system with $\chi_1 = \chi_2 = -0.6$
  and initial angular momentum $\lo = 6.5$, for several values of the initial energy $\eo$; 
  each plot zooms in on the time of closest approach in the scattering event, and refers to the analytical model specified at the top.
  The \ac{bh} spin overall increases in all cases (since it is negative, this means that its absolute magnitude
  is decreasing); although the flux is negative when $p_r$ is large starting at 1\ac{pn}. 
  The factorized \ac{nnlo} model predicts the largest effect (although by a rather slim margin) thanks
  to the superradiance prefactor being amplified when $\Omega_{\rm H} < 0$, with $\Delta \hS_1$ peaking above $10^{-2}$ 
  for the closest encounter in this example.}
\end{figure*}

\begin{figure*}
  \includegraphics[width=\textwidth]{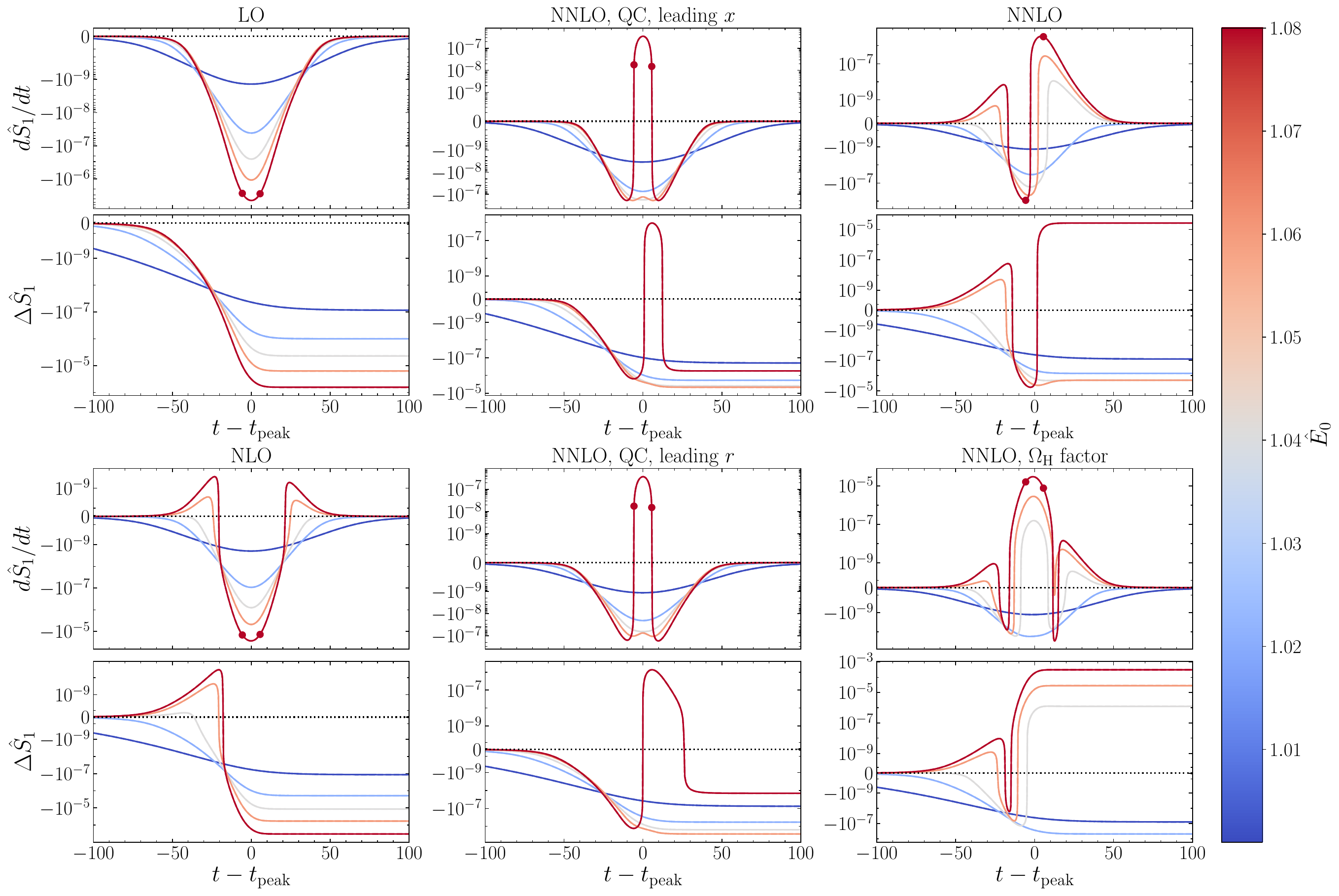}
  \caption{\label{fig:chi03_spinflux} Instantaneous and cumulative (Eq.~\eqref{eq:def_delta}) variation of the spin $\hS_1$ for an equal-mass system with $\chi_1 = \chi_2 = 0.3$
  and initial angular momentum $\lo = 5.5$, for several values of the initial energy $\eo$, according to the analytical model specified at the top 
  of each plot.
  Dots on the $\dot{\hS}_1$ lines mark times when the orbital frequency equals the \ac{bh}'s horizon frequency, $\Omega = \Omega_{\rm H}^1$.
  The quasicircular fluxes exhibit the predicted sign flip as the orbital frequency exceeds $\Omega_{\rm H}^1$ exactly thanks to a prefactor explicitly 
  enforcing it (the dots are not precisely on the 0 line because of the projection of small, discrete data points in logarithmic scale). The only noncircular
  expressions that should include this 1.5\ac{pn} effect are the \ac{nnlo} ones. However, the large \ac{nlo} term produces a positive flux when the \acp{bh} are
  far apart. The raw \ac{nnlo} model retains this sign all through the scattering in the most energetic orbit, producing a negative peak flux otherwise.
  The superradiance prefactor does instead cause a flip from negative to positive around periastron, but at much lower energy than the quasicircular models do.}
\end{figure*}

\begin{figure*}
  \includegraphics[width=\textwidth]{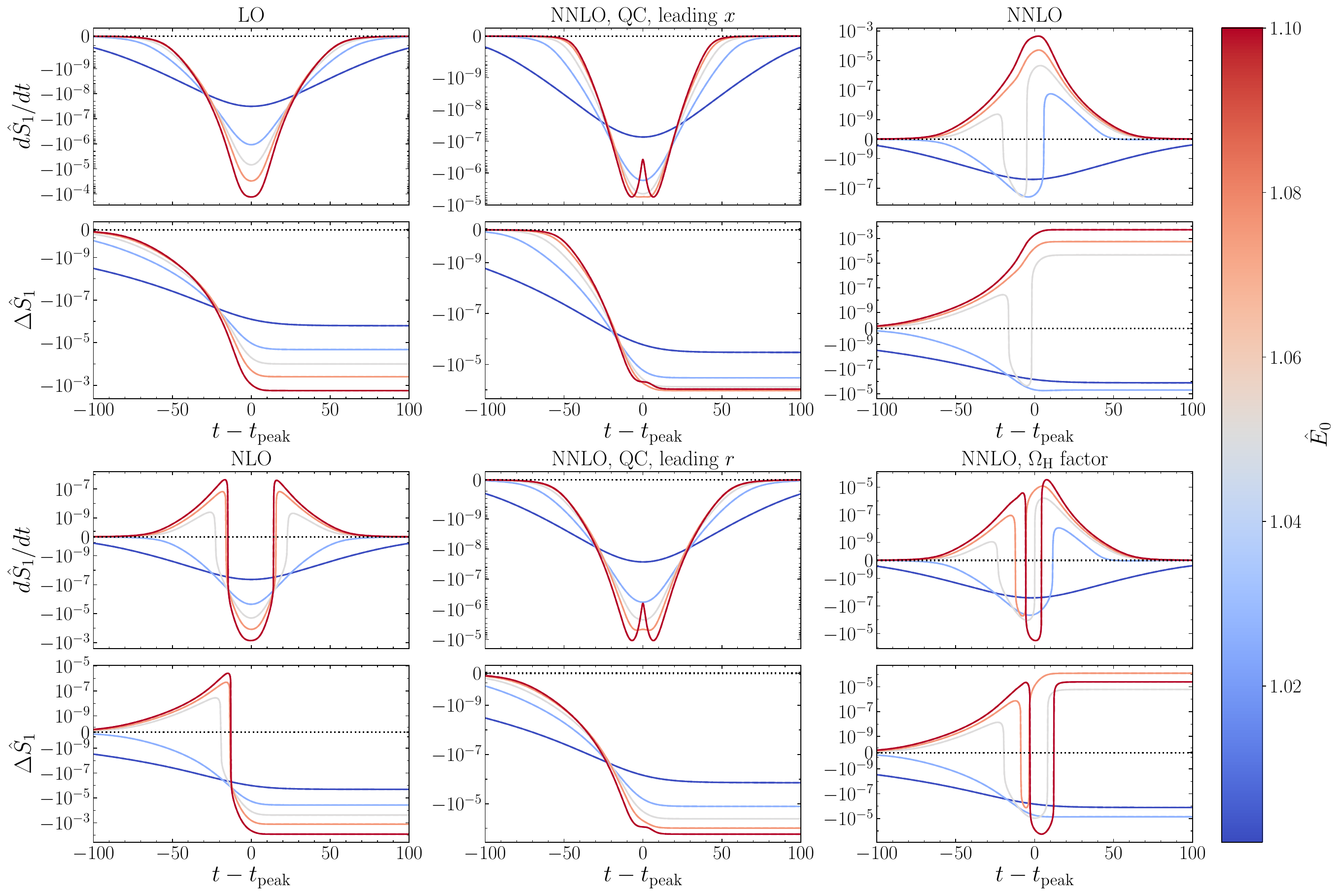}
  \caption{\label{fig:chi09_spinflux} Instantaneous and cumulative (Eq.~\eqref{eq:def_delta}) variation of the spin $\hS_1$ for an equal-mass system 
  with $\chi_1 = \chi_2 = 0.9$ and initial angular momentum $\lo = 5$, for several initial energies and analytical versions of the model, as specified at the top of each panel.
  In most cases the spin decreases as a result of the encounter, denoting superradiance, as would be expected for a fast-spinning \ac{bh} such as this. 
  Starting at \ac{nlo}, this is no longer true when the \acp{bh} are far apart, because of terms involving $p_r$ that cause a sign change for high energies.
  The raw \ac{nnlo} model, as in Fig.~\ref{fig:chi03_spinflux}, predicts a flow of angular momentum into the \ac{bh} all through the orbit for high enough $\eo$.
  The same happens using the factorized \ac{nnlo} version, but the superradiance prefactor forces a sign change in extremely close encounters, leading to negative peak and
  cumulative flux, counter to expectations stemming from the physics of quasicircular systems.}
  \end{figure*}

  \begin{figure*}
    \includegraphics[width=\textwidth]{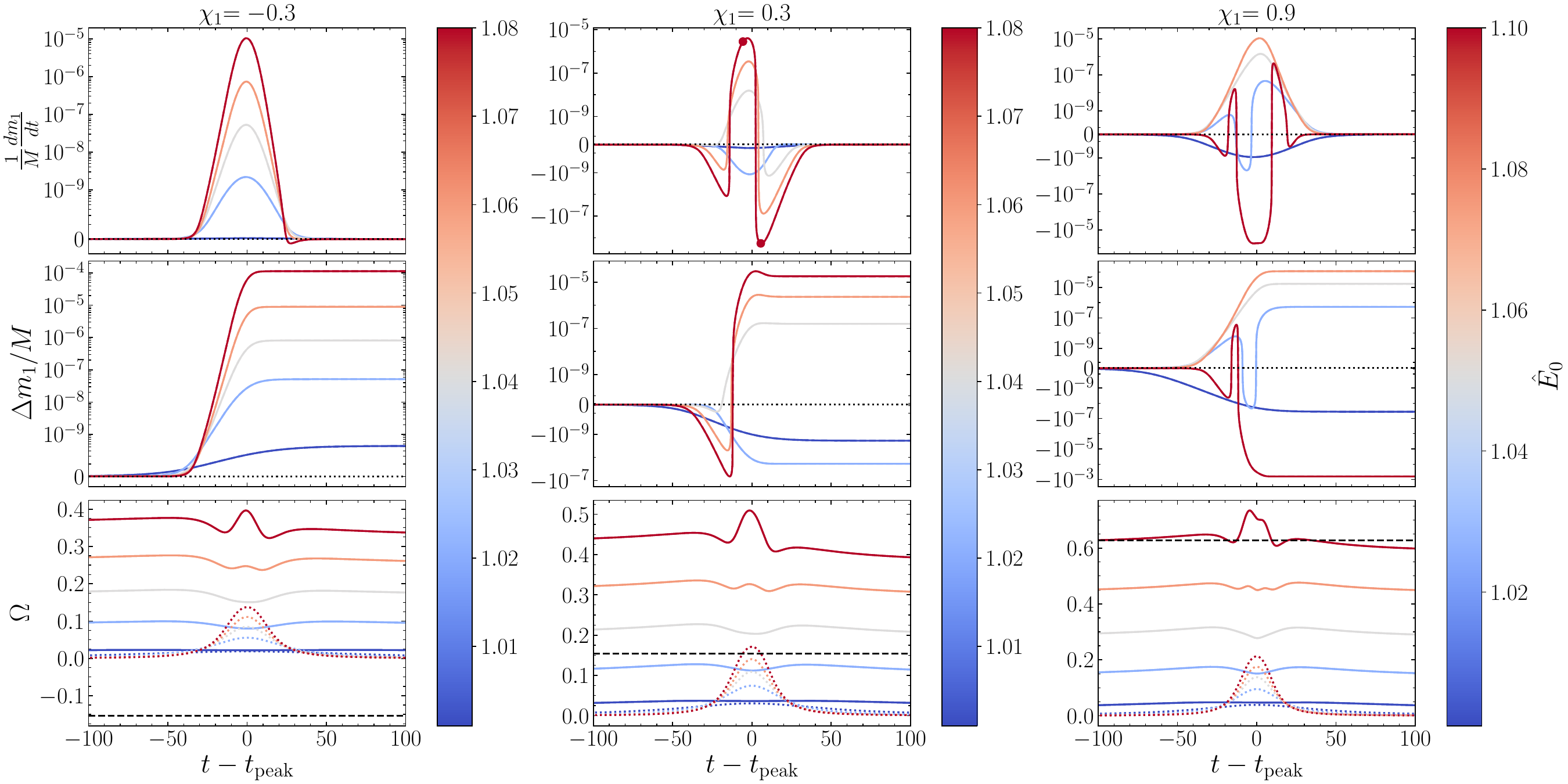}
    \caption{\label{fig:fact_mflux} Instantaneous and cumulative (calculated as in Eq.~\eqref{eq:def_delta}) variation of the \ac{bh} 
    mass $m_1$ for equal-mass systems with, left to right, 
    $\chi_1 = \chi_2 \in \left\{-0.3, 0.3, 0.9\right\}$ and initial angular momenta, respectively, $\lo = 6.5, 5.5,$ and $5$, for several 
    values of the initial energy $\eo$. The \ac{nnlo} factorized model is used in all cases.
    The bottom row of plots shows the horizon frequency (dashed black line), the orbital frequency (dotted lines), and the
    term of Eq.~\eqref{eq:mass_pref} that appears in the superradiance prefactor for $\dot{m}$ (full lines).
    The latter, differently from the pure orbital frequency, remains almost constant throughout the orbit in many cases;
    events close to the scattering-capture separatrix display a noticeable dip after periastron, and one or more peaks
    usually develop around closest approach. Notice how along certain orbits this factor remains close to the orbital frequency,
    causing the prefactor to suppress the energy flux in those cases.}
  \end{figure*}

This section is dedicated to an overview of the phenomenology of the mass and angular momentum fluxes we derived above, specifically
for the case of unbound binary orbits, i.e., hyperbolic scattering dynamics.
We evaluate and integrate on EOB dynamics several declinations of our generic orbit expressions: 
kept at \ac{lo}, up to \ac{nlo} (1\ac{pn}), and in their complete \ac{nnlo} (1.5\ac{pn}) version, both
with and without the superradiance prefactor. 
We also consider, to gauge the importance of non-circular terms, the factorized \ac{nnlo} quasicircular expressions
from Ref.~\cite{Saketh:2022xjb}, in two versions: as functions of only the \ac{pn} ordering parameter $x$, 
as in their Eqs.~(4.22-4.25), and using in their \ac{lo} prefactor the appropriate
power of $r$, consistently with our general expressions (meaning, $r^{-6}$ instead of $x^{12}$ in Eq.~(4.25) of~\cite{Saketh:2022xjb}).
Note that here we are simply evaluating the horizon flux expressions on solved, aligned-spin \ac{eob} dynamics, without considering 
the impact of the resulting time-varying spins (or masses) on the system during its evolution. As will be seen in the plots in
the coming pages, the cumulative changes in the \ac{bh} spin and mass rarely exceed
$10^{-3}$, so this simplification shouldn't significantly impact our results outside of the most extreme cases.

We use the \TEOBResumSDali{} model of Ref.~\cite{Nagar:2024dzj} for the orbital dynamics.
The commit hash used is tagged with the arXiv number of the paper (see also the Acknowledgements section).
The \ac{eob} Hamiltonian is:
\begin{equation}
  H_{\rm EOB} = M \sqrt{1 + 2\nu \left(\hat{H}_{\rm eff} - 1\right)},
\end{equation}
where $\hat{H}_{\rm eff} = H_{\rm eff}/\mu$ is the rescaled effective Hamiltonian, given
by:
\begin{align}
  \hat{H}_{\rm eff} &= \hat{H}^{\rm orb}_{\rm eff} + \hat{H}^{\rm SO}_{\rm eff} \\
  \hat{H}^{\rm orb}_{\rm eff} &= \sqrt{p_{r_*}^2 + A(r) \biggl(1 + \dfrac{p_\varphi^2}{r_c^2} + Q(r_c, p_{r_*})\biggr)} \\
  \hat{H}^{\rm SO}_{\rm eff} &= p_\varphi \left[\hat{S} G_S (r,p_{r_*}) + \hat{S}_* G_{S_*} (r, p_{r_*})\right] ,
\end{align}
where $\hat{S} = \hat{S}_1 + \hat{S}_2$ and $\hat{S}_* = 1/q \hat{S}_1 + q \hat{S}_2$.
In the equations above, $p_{r_*} = \sqrt{A/B}\, p_r$ is the canonical momentum associated with the
tortoise coordinate $r_*$. While they are formally equivalent, $p_{r_*}$ is used instead of $p_r$ 
because of the latter's divergence at the end of inspiralling dynamics. For consistency, we will use
$p_{r_*}$ as well in the analytic models for the horizon flux; since the two momenta differ only 
starting at \ac{nlo} (i.e., $p_{r_*} = p_r + O(c^{-2})$), but $p_r$ never appears in the \ac{lo} of Eqs.~\eqref{eq:tidal-eob},
there is formally no need to amend those equations to accommodate this substitution.
The definitions of the functions $r_c, A, B, Q, G_S, G_{S_*}$, as well as the equations of motion
and the radiation reaction model, can be found, e.g., in~\cite{Nagar:2024dzj}
and references therein. 
Initial conditions for hyperbolic-like orbits are specified by giving the 
initial separation, which we set at $r = r_0 = 3\ 000$, and the starting energy and orbital angular
momentum~\cite{Nagar:2020xsk}:
\begin{equation}
  \eo = \hat{H}_{\rm EOB}^0 = \dfrac{H_{\rm EOB}^0}{\mu}, \quad \lo = p_\varphi^0.
\end{equation}

\subsubsection{Spins}

Figs.~\ref{fig:chi-06_spinflux},~\ref{fig:chi03_spinflux} and~\ref{fig:chi09_spinflux} display the results 
for the angular momentum flux in equal-mass systems with $\chi_1 = \chi_2 \in \left\{-0.6, 0.3, 0.9\right\}$.
\footnote{In the following, we will always use the symbols $\chi_{1,2}$ to refer to each BH's \textit{initial}
mass-rescaled spin, which, as mentioned, remains constant throughout the evolution of the orbit as we are not
incorporating the horizon fluxes in a fully consistent way in the EOB model.}
In each block, the top plot is the instantaneous flux $\dot{\hS}$, while the bottom plot is the cumulative spin
variation found by integrating the flux:
\begin{equation}
  \label{eq:def_delta}
  \Delta \hS_{1,2} (t) \equiv \int_0^t dt \ \dot{\hS}_{1,2} \ = \dfrac{1}{M^2}\int_0^t dt \ \dot{S}_{1,2};
\end{equation}
we perform the integration using {\tt scipy}'s implementation of the trapezoidal rule.
The figures focus specifically on the primary \ac{bh}'s spin, but $\dot{\hS}_1 = \dot{\hS}_2$ if $q=1$ 
in an equal-spin binary. Below, we summarize the main properties of each analytical model.

\begin{enumerate}
  \item The \ac{lo} flux peaks at periastron, while remaining extremely small before and
  after. The sign of the flux is opposite to the spin's: positive spin \acp{bh} thus
  lose angular momentum to the system, while negative spin \acp{bh} absorb it, decreasing the magnitudes
  of $\hS_{1,2}$.
  Notably, for initially non-spinning \acp{bh}, the flux is identically zero at this order.

  \item Starting at \ac{nlo}, we start to see deviations from the expectations formed through the study of quasicircular systems,
  as for sufficiently high initial energies the angular momentum flux changes sign with respect to 
  its \ac{lo} when the \acp{bh} are far apart (compare the top and bottom left panels in each of Figs.~\ref{fig:chi-06_spinflux},
  ~\ref{fig:chi03_spinflux}, and~\ref{fig:chi09_spinflux}).
  This is due to the large, negative $O\left(c^{-2}\right)$ term proportional to the radial momentum in 
  Eq.~\eqref{eq:tidal-eob-S}, representing a novel finding of our generic noncircular expressions.
  The signs of the peak flux and the final cumulative change in $\hS_1$ are not affected by this;
  the inversion is confined to times before and after closest approach, when the \acp{bh} have
  high radial velocity. 
  Similar to the \ac{lo} fluxes, also at \ac{nlo} the expressions vanish for initially non-spinning systems. 

  \item Moving on to the \ac{nnlo} models, the quasicircular versions behave similarly to each other,
  with the one with $x$ in its leading term predicting a slightly larger effect. Each finds a flux 
  opposite the \ac{bh} spin's sign, except when the superradiance prefactor
  enforces a sign change around periastron if the orbital frequency exceeds the \ac{bh}'s horizon frequency,
  which, for any spin, only happens along sufficiently energetic orbits (in Fig.~\ref{fig:chi03_spinflux},
  only for the highest value of $\eo$; in Fig.~\ref{fig:chi09_spinflux}, not even then).
  
  \item The noncircular \ac{nnlo} expressions lead to more varied results.
  When the \ac{bh} spin is anti-aligned with the orbital angular momentum (see the rightmost plots of Fig.~\ref{fig:chi-06_spinflux}), 
  their behavior is simple: $\dot{\hS}_1 > 0$ at the peak, consistently across both analytic 
  versions, as is the case for the cumulative spin change. 
  This simplicity is related to the fact that, if both spins are negative, each \ac{bh} is in 
  retrograde motion with respect to the companion's intrinsic rotation; 
  thus, there is no inversion in the direction of the flux depending on the relative angular velocity, 
  and both the orbital and \ac{bh} angular momentum decrease in norm.
  The flux curves display increasing asymmetry around periastron as the orbit energy grows,
  both because of the \ac{nnlo} term linear in $p_r$ and the loss of energy due to \ac{gw}
  emission.
  Initially non-spinning \acp{bh} are predicted by the \ac{nnlo} flux expressions to acquire positive spin, 
  with similar phenomenology to the anti-aligned spin case as $\eo, \lo$ change.
  
  \item If the \ac{bh} spin is positive, the uniform negative sign of the spin flux on low-energy orbits,
  denoting superradiance, is replaced by a series of sign changes (see the top right panels in Figs.~\ref{fig:chi03_spinflux}
  and~\ref{fig:chi09_spinflux}): the \ac{nlo} radial terms, which are quadratic in $p_r$, cause an initially 
  positive flux; once the \ac{nnlo} terms instead begin to dominate the flux formula, including those linear 
  in $p_r$, a negative peak forms, that flips about the time of closest approach. The weight of the 
  \ac{nlo} and \ac{nnlo} contributions increases for fast spinning \acp{bh} (compare the plots for
  $\chi_1 = 0.3$ in Fig.~\ref{fig:chi03_spinflux} and $\chi_1 = 0.9$ in Fig.~\ref{fig:chi09_spinflux});
  on very energetic, close encounters, the flow of angular momentum for large $\chi$ eventually remains 
  positive throughout the orbit. This is in striking contrast with the prediction of the quasicircular
  models, that always find in these cases a flow of momentum out of the \ac{bh} and into the system,
  consistently with the horizon frequency being much larger than the orbital one.
  
    
  \item The factorization causes significant changes. In the bottom right plot of Fig.~\ref{fig:chi03_spinflux}, the angular momentum
  flux becomes negative earlier around periastron, only for the prefactor to again produce a positive
  peak at closest approach. Notice that the noncircular model predicts this to happen at much lower energies
  than the quasicircular ones, because of the fact that the former uses in its superradiance factor the 
  Newtonian approximation to the orbital frequency, $p_\varphi/r^2$, instead of the actual $\dot{\varphi}$.
  As the spin is increased, however, the prefactor and the \ac{pn} expansion combine into counterintuitive results,
  as seen for a $\chi_{1,2} = 0.9$ binary in Fig.~\ref{fig:chi09_spinflux}.
  Increasing $\eo$, it is the \ac{nnlo} terms in the expansion, including those linear in $p_r$,
  that eventually lead to an overall positive peak flux.
  On sufficiently energetic orbits, $\dot{\hS}_{1,2}$ would thus remain positive at all times,
  were it not for the superradiance prefactor forcing a sign change near closest approach.\footnote{
    The timing of the sign flips in $\dot{\hS}_1$ in the bottom right plot of Fig.~\ref{fig:chi09_spinflux} points to their origin:
    the grey and orange lines change back to positive \emph{before} or at periastron, indicating
    the terms linear in $p_r$ caused the negative sign.
    Instead, the red line temporarily dips below the $x$ axis approximately symmetrically 
    about the time of closest approach, when $p_\varphi/r^2$ in the superradiance prefactor
    exceeds the horizon frequency.
  }
  This is, again, at odds with the known physics of the quasicircular limit, where a \emph{negative} horizon
  flux becomes \emph{positive} at high orbital velocity.
\end{enumerate}

In summary, the key takeaways from this discussion should be that: (i) initially non-spinning \acp{bh} are predicted to acquire
angular momentum only if the \ac{nnlo} terms are included; (ii) the flux phenomenology for negative-spin \acp{bh} is
relatively simple at all orders, always leading to a cumulative increase of the \ac{bh} intrinsic angular momentum (i.e., a decrease in its magnitude);
(iii) up to \ac{nlo}, the cumulative change in $\hS_{1,2}$ is always negative for positive spins;
(iv) the inclusion of \ac{nnlo} terms significantly complicates the picture, with the factorized
model predicting at times counterintuitive results, especially in the most energetic orbits for large spins.
This is likely due to the fact that we are employing and assessing these expressions outside their expected range of reliability, 
and in a regime where our physical intuition, informed only by results tied to a very special edge case 
(that of test mass, quasi-circular evolutions), might not be applicable.
We do anticipate here, however, that the factorization of the flux expressions appears to help
in more accurately reproducing the spin-up observed in \ac{nr} simulations of scattering Schwarzschild \acp{bh}; 
we discuss this in more detail in Sec.~\ref{sec:nr}.

\subsubsection{Masses}

Several of these points still apply when moving on to discussing the rate of change of the \acp{bh}' masses:
\begin{enumerate}
\item At \ac{lo}, $\dot{m}_1$ has opposite sign to $\chi_1$ throughout each orbit.
\item Starting at \ac{nlo}, terms involving the radial momentum invert the sign of $\dot{m}_1$ 
  before and after closest approach with respect to the leading term, everywhere except for the lowest
  initial energies.
\item For \acp{bh} with negative spins, the phenomenology again remains straightforward at \ac{nnlo}: 
      save for the (small, temporary) effect of the $p_r$-dependent terms, the \acp{bh}' masses increase 
      as a result of the interaction.
\item For positive spins, the \ac{nnlo} non-factorized flux becomes dominated by the large 1.5\ac{pn} terms,
      leading in energetic, close encounters to uniformly positive energy flux and an increase in $m_1$.
      Low-$\eo$, high-$\lo$ orbits instead retain negative peak and cumulative change in mass.
\item The asymmetry caused by the \ac{nnlo} contributions linear in the radial momentum is particularly
      evident when inspecting the factorized model (see the central and rightmost plots in the top 
      row of Fig.~\ref{fig:fact_mflux}).
\end{enumerate}

It is interesting here to explore the effect of the superradiance prefactor, as it also now depends on the 
radial momentum, by Eq.~\eqref{eq:mdot_eob_fact}. The additional correcting term is the square of
the ratio of radial velocity to angular velocity, in their Newtonian approximations:
\begin{equation}
  \label{eq:mass_pref}
\dfrac{p_\varphi}{r^2} + 3 \dfrac{p_r^2}{p_\varphi} \simeq \dfrac{p_\varphi}{r^2} \left(1 + 3 \dfrac{\dot{r}^2}{r^2 \dot{\varphi}^2}\right).
\end{equation}
In the spin flux prefactor, the horizon frequency is compared with a term that approximates
the orbital frequency, being negligible at large distances and exhibiting a peak around closest
approach. Here, the combination of the radial and angular terms produces instead a total that remains
quite large throughout the orbit, especially at higher energies; see the bottom row of Fig.~\ref{fig:fact_mflux}.
For negative-spin \acp{bh}, $\Omega_{\rm H} < 0$; the difference in the superradiance prefactor thus
is always positive, and often enhanced with respect to the case of positive spins, leading the factorized
model to predict a somewhat larger effect than its expanded counterpart.
As $\eo$ increases, so does the value of Eq.~\ref{eq:mass_pref}; it eventually reaches and passes
the horizon frequency if the \ac{bh} spin is aligned with the orbital angular momentum.
When the two are very close, the prefactor essentially suppresses the rate of change of the mass 
during most of the orbit (see, e.g., some of the $\chi_1 = 0.3$ and $\chi_1 = 0.9$ plots in the bottom row 
of Fig.~\ref{fig:fact_mflux}). 
Differently from the spin flux, the superradiance prefactor here, above the required energy threshold,
changes sign throughout the entire evolution of the system, rather than just temporarily around periastron.
The combination of angular and radial momentum in Eq.~\eqref{eq:mass_pref} can in very close encounters
develop several peaks around periastron, resulting in $\dot{m}_{1,2}, \Delta m_{1,2}$ curves that appear quite erratic
in such edge cases (see the bottom right plot of Fig.~\ref{fig:fact_mflux}).

Assessing the reliability of these expressions for $\dot{m}$ is tricky, as the role of the 
radial momentum in determining its overall sign through the superradiance prefactor is a 
novel prediction that has as of yet no test-mass counterpart that might act as a guide
of sorts, as perturbative calculations did in the case of quasi-circular binaries.
Furthermore, as already stressed, the \ac{nlo} and \ac{nnlo} terms beyond the prefactor
also cause sign flips themselves in some cases, and the analytic models to find very 
different predictions for the cumulative change in mass following an encounter;
this is likely telling us that we might be beyond the scope of these \ac{pn} expressions.

\subsection{Parameter space investigation}
\label{sec:param_space}

\begin{figure*}
  \centering
  \includegraphics[width=0.49\textwidth]{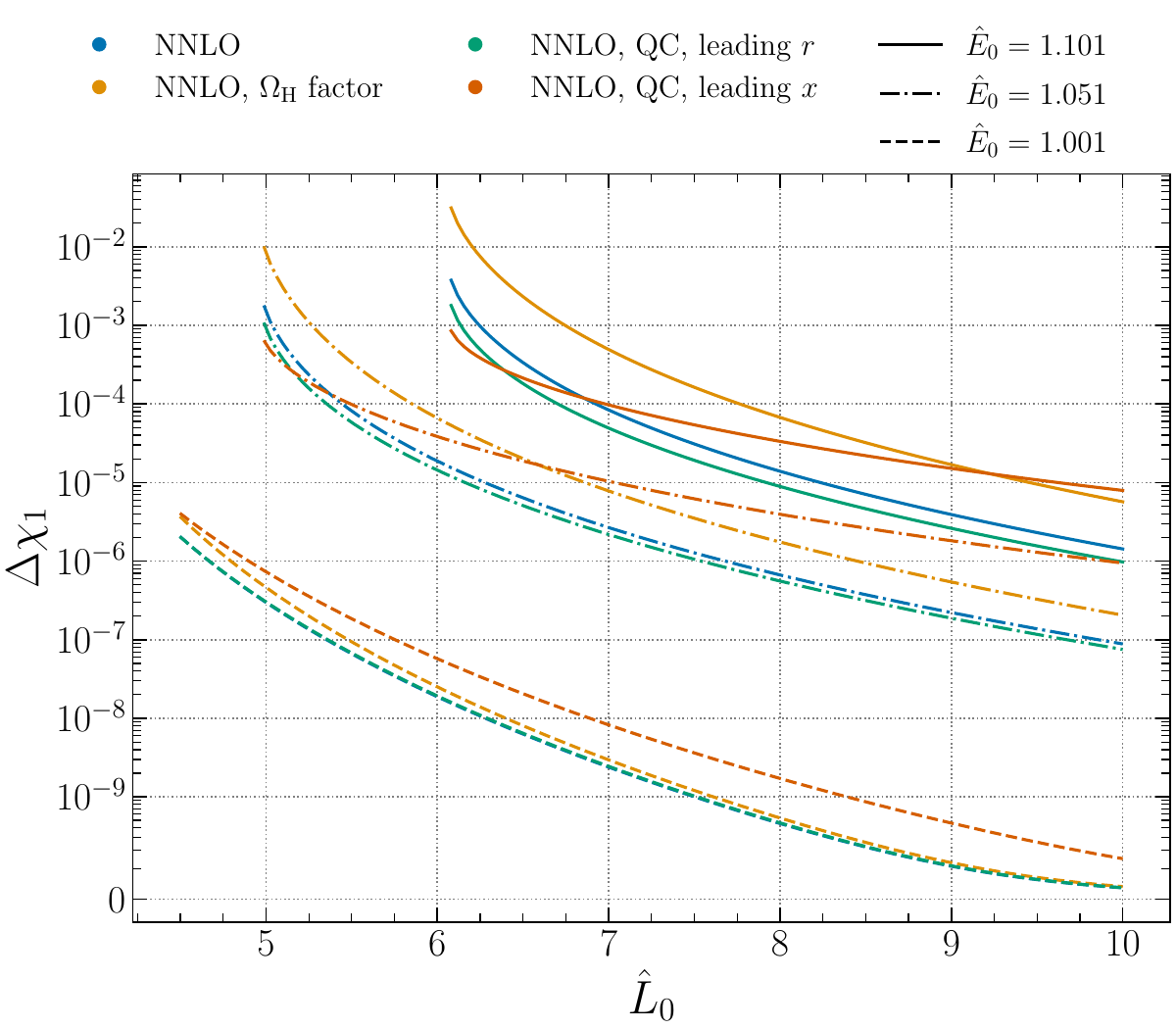}
  \includegraphics[width=0.49\textwidth]{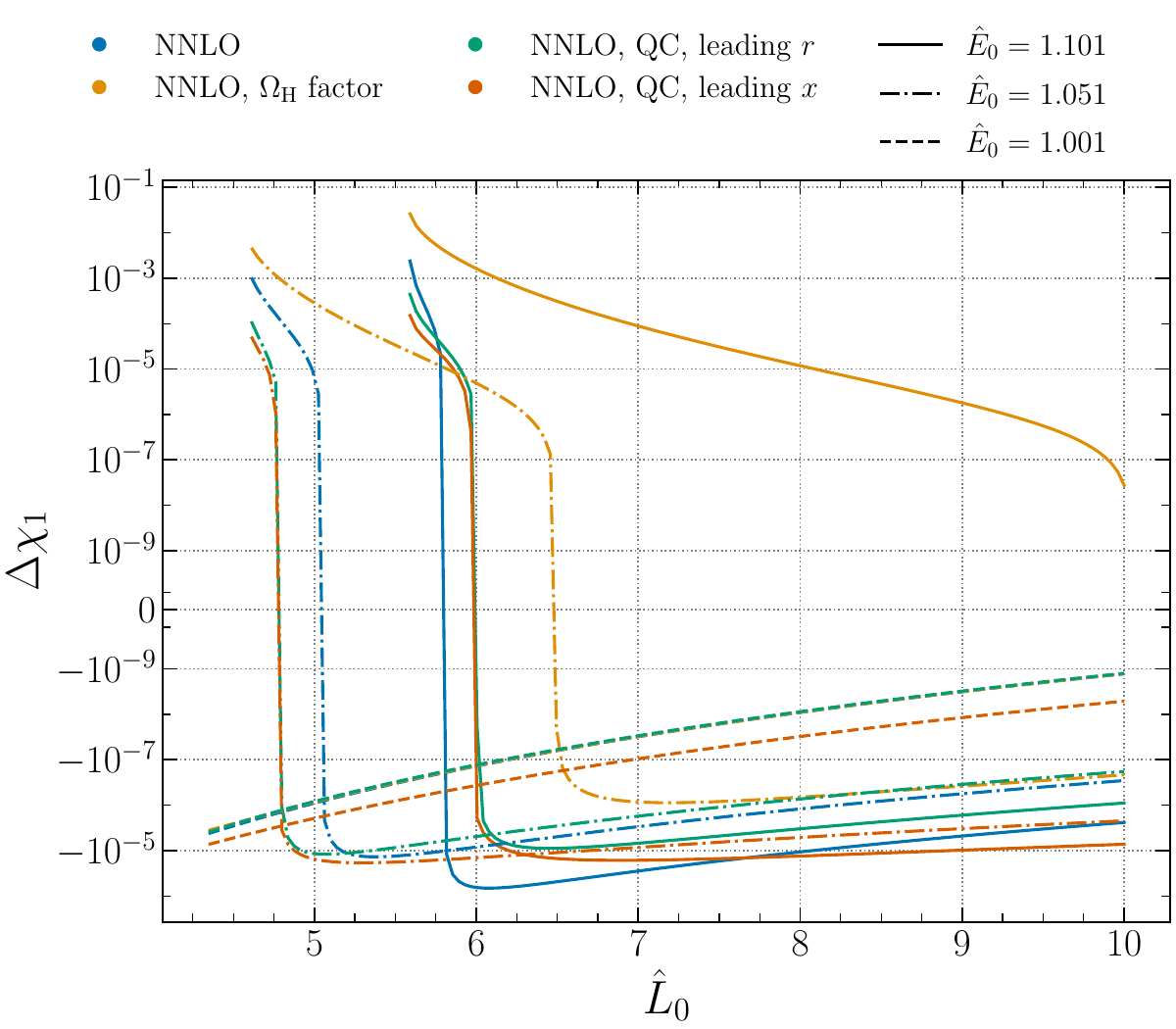}
  \caption{\label{fig:orders_q1_chi003} Total variation of the dimensionless spin $\chi_1$ due to horizon absorption,
  according to all the \ac{nnlo} analytical models of $\dot{m}_{1,2}, \dot{\hS}_{1,2}$, for equal-mass binaries.
  $\Delta \chi_1 = \chi_1^{\rm final} - \chi_1$ is plotted against the initial orbital angular momentum $\lo$ for three values of the starting
  energy $\eo$; the \acp{bh} are initially non-spinning in the left panel, and they have $\chi_1 = \chi_2 = 0.3$ 
  on the right.
  The differences between the models grow with the initial energy. The \ac{nnlo} model with a superradiance prefactor
  predicts the largest spin-up for initially non-spinning \acp{bh} among the noncircular models; only the quasicircular
  version using $x = \left(\Omega\right)^{2/3}$ in its leading term is larger for low-energy, high-$\lo$ orbits.
  Spinning \acp{bh} lose angular momentum due to superradiance, except for close encounters at high energy, where
  \ac{nnlo} models predict a sign change; notice, however, that each model does so for different initial parameters.}
  \end{figure*}

After inspecting the angular momentum and energy fluxes predicted by our expressions in a 
few examples in the last section, we now move on to discussing the results of a more systematic exploration. 
We use the \TEOBResumSDali~model to evolve the dynamics of a large number of systems 
spanning the available parameter space for unbound, non-precessing binaries, with dimensionless spins 
$\chi_{1,2} \in \left\{0, \pm 0.3, \pm 0.6, \pm 0.9\right\}$, mass ratios $q \in \left\{1, 2, 4, 8\right\}$, 
and initial orbital angular momenta $\lo \in \left[2.5, 10\right]$. 
We also vary the initial energy from $\hat{E}_{\rm min} = 1.0001$ to an upper
limit of $\hat{E}_{\rm max}^{q = 1} = 1.2$ for equal-mass systems. When $q > 1$, we instead calculate an upper 
threshold that corresponds to the same $\nu$-rescaled effective energy as the equal-mass case:
\begin{align}
\hat{E}_{\rm max} (q) &= \sqrt{1 + 2 \nu (\hat{H}_{\rm eff}^{q = 1} - 1)}, \\
\text{with  } \hat{H}_{\rm eff}^{q = 1} &= \dfrac{\left(\hat{E}_{\rm max}^{q = 1}\right)^2 - 1}{2\nu} - 1.
\end{align}
This yields progressively lower maximum initial energies as $q$ grows, down to 
$\eo \simeq 1.083$ for $q = 8$. 
Initial data resulting in a merger are discarded.\footnote{This includes both immediate capture,
as well as systems that become bound after the first encounter but still complete a few orbits
before merger; these are recognized by their energy falling below 1.}
For each system, we compute the maximum angular momentum and mass fluxes, as well as the 
final cumulative variations in $\hS_{1,2}, m_{1,2}$ and $\chi_{1,2}$, for each analytic version 
considered in the previous section: up to \ac{lo}, \ac{nlo}, \ac{nnlo}
with and without superradiance prefactor, and using the quasi-circular expressions 
of~\cite{Saketh:2022xjb} with either leading $x$ or $r$.
In particular, we calculate the final value of the mass-rescaled spin $\chi_{1,2}^{\rm final}$ by:
\begin{equation}
  \chi_{1,2}^{\rm final} = M^2\dfrac{\hS_{1,2} + \Delta \hS_{1,2}}{(m_{1,2} + \Delta m_{1,2})^2} ,
\end{equation}
where $\Delta m_{1,2}, \Delta \hS_{1,2}$ here are the integrated mass and spin variations at the end of the orbit
evolution:
\begin{equation}
  \Delta m_{1,2} \equiv \int_0^{t_{\rm end}} dt\  \dot{m}_{1,2}, \qquad \Delta \hS_{1,2} = \int_0^{t_{\rm end}} dt \dot{\hS}_{1,2} .
\end{equation}
As a reminder, we neglect in the EOB equations of motion the effect of the changing masses and spins;
as we will see, outside of a few exceptional cases the total effect of horizon absorption is too small
to have a visible impact.

\begin{figure}
   \includegraphics[width=0.49\textwidth]{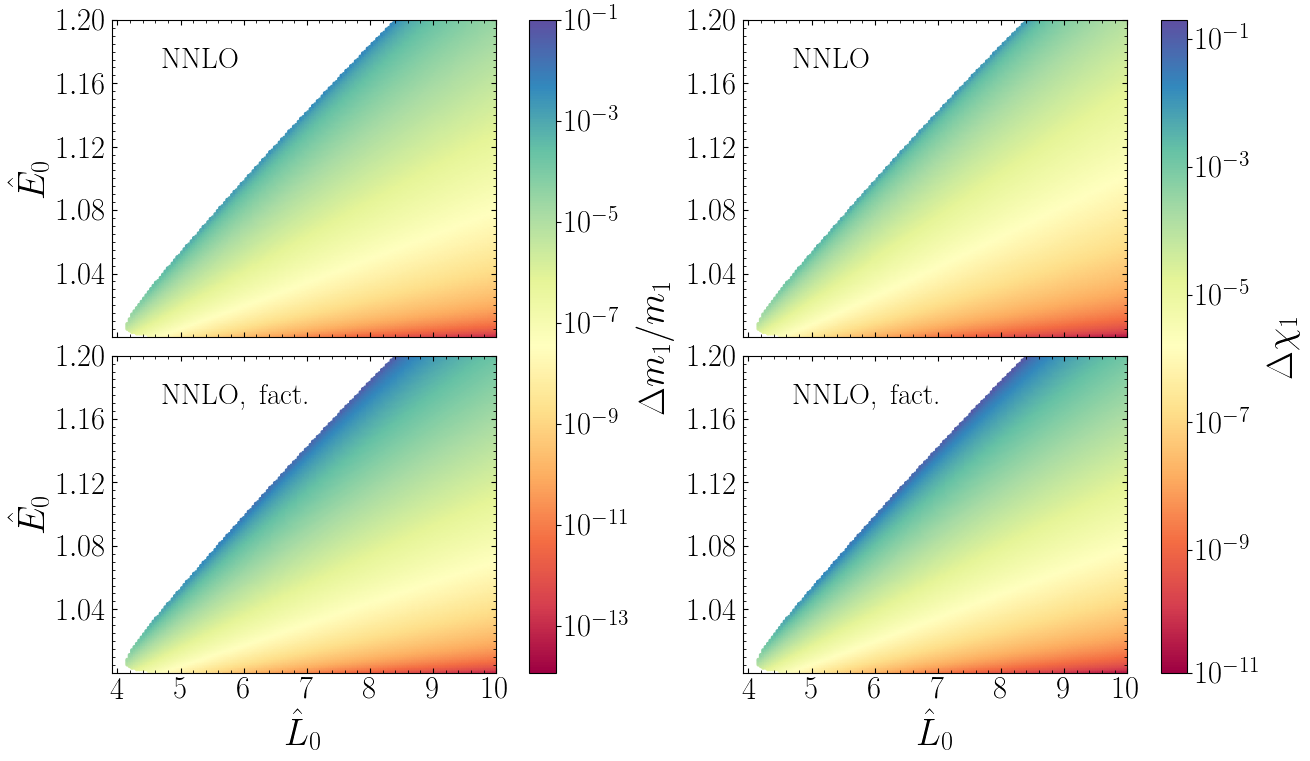}
   \caption{\label{fig:ns_mchispace}Relative cumulative mass variation (left) and spin-up (right)
   for initially non-spinning equal-mass scattering \acp{bbh} across the $\eo, \lo$ parameter space; 
   raw \ac{nnlo} model in the top row, factorized version in the bottom plots.
   In the vast majority of cases the analytical models predicts a very small cumulative increase in the \ac{bh} mass
   and spin. Significant tidal heating only happens
   in very energetic encounters, where however our \ac{pn} models are likely unrealiable.
   The factorized model typically predicts values greater by a factor of up to $\sim 10$ than 
   the raw one, thanks to the additional terms created by the superradiance prefactor for initially non-spinning
   \acp{bh}.}
\end{figure}

\begin{figure}
  \includegraphics[width=0.49\textwidth]{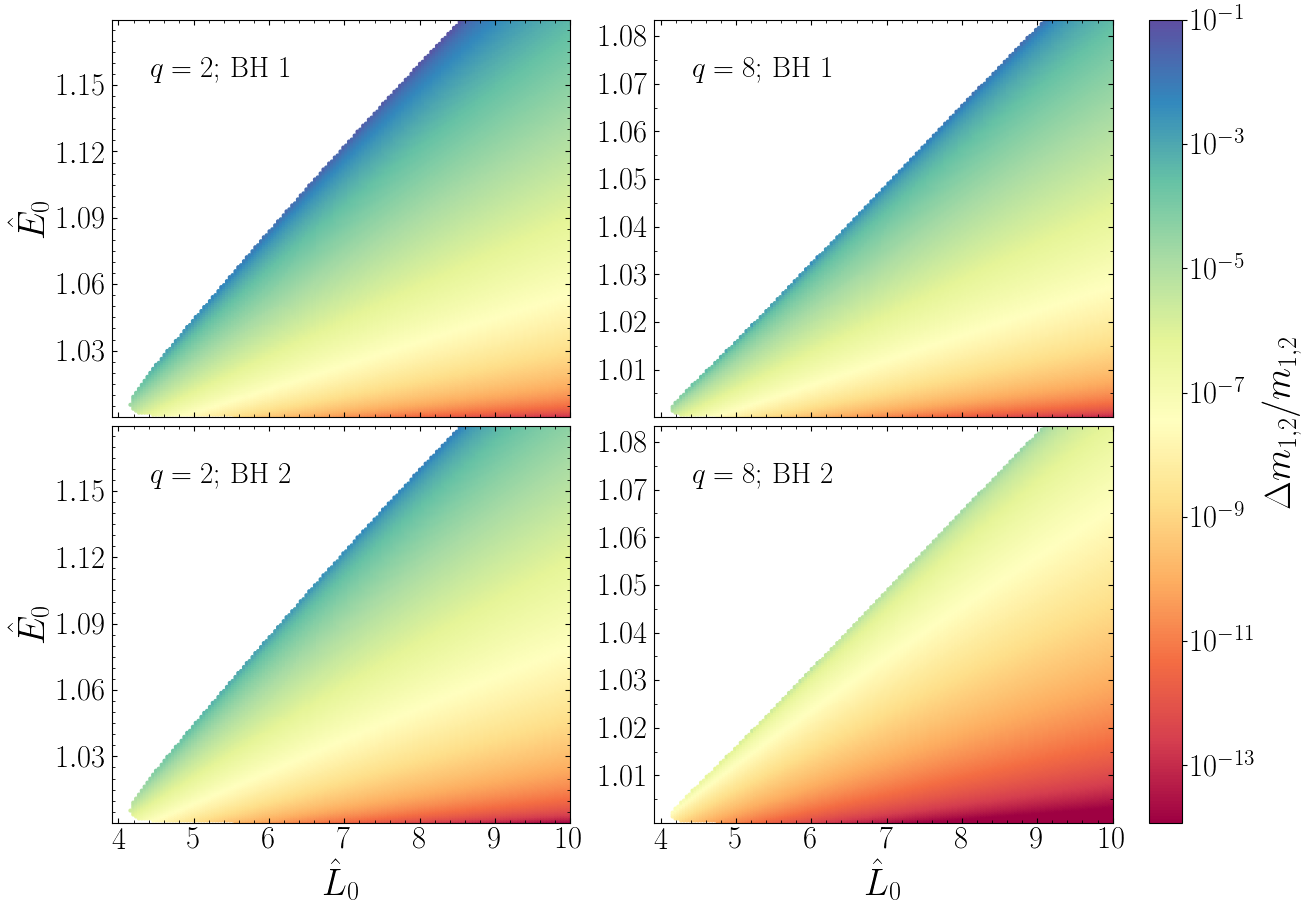}
  \caption{\label{fig:ns_m_q28_space} Relative cumulative mass variation (color) for initially non-spinning \acp{bh}, 
  according to the NNLO factorized model, across the $\eo, \lo$ parameter space. Top: primary \ac{bh}; bottom:
  secondary. Mass ratios $q = 2$ (left) and 8 (right). The color scale is everywhere the same as the left panels of
  Fig.~\ref{fig:ns_mchispace}. The overall behavior is the same as the equal-mass case.
  The effect is suppressed for the secondary \ac{bh} due to the leading mass factor (Eq.~\eqref{eq:eob_fact});
  this is increasingly evident as $q$ grows.}
\end{figure}

\begin{figure*}
  \includegraphics[width=\textwidth]{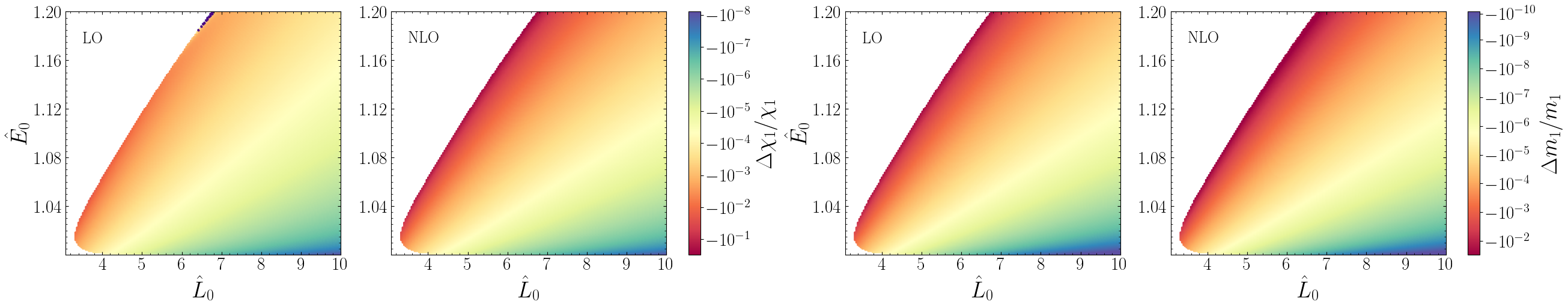}
  \caption{\label{fig:chi09_lonlospace}
  Tidal torquing (leftmost two plots) and heating (rightmost plots) across the $\eo, \lo$ parameter space
  for equal-mass binaries with initial spins $\chi_1 = \chi_2 = 0.9$, as found using the \ac{lo}
  and \ac{nlo} noncircular models.
  The results are similar in their dependence on the orbit parameters, the main difference being the 
  larger final cumulative variations in $\chi_1, m_1$ predicted by the \ac{nlo} model.
  For positive spins both models give negative peak and cumulative energy and angular momentum fluxes in 
  all cases. The dark purple patch at the top of the leftmost plot is a region where the dimensionless
  $\chi_1 = S_1/m_1^2$ actually decreases as a result of the encounter; this is due to the different 
  rates of change for the mass and spin at \ac{lo} (respectively, $\sim p_\varphi r^{-8}$ and $\sim r^{-6}$).}
\end{figure*}

\begin{figure*}
  \includegraphics[width=\textwidth]{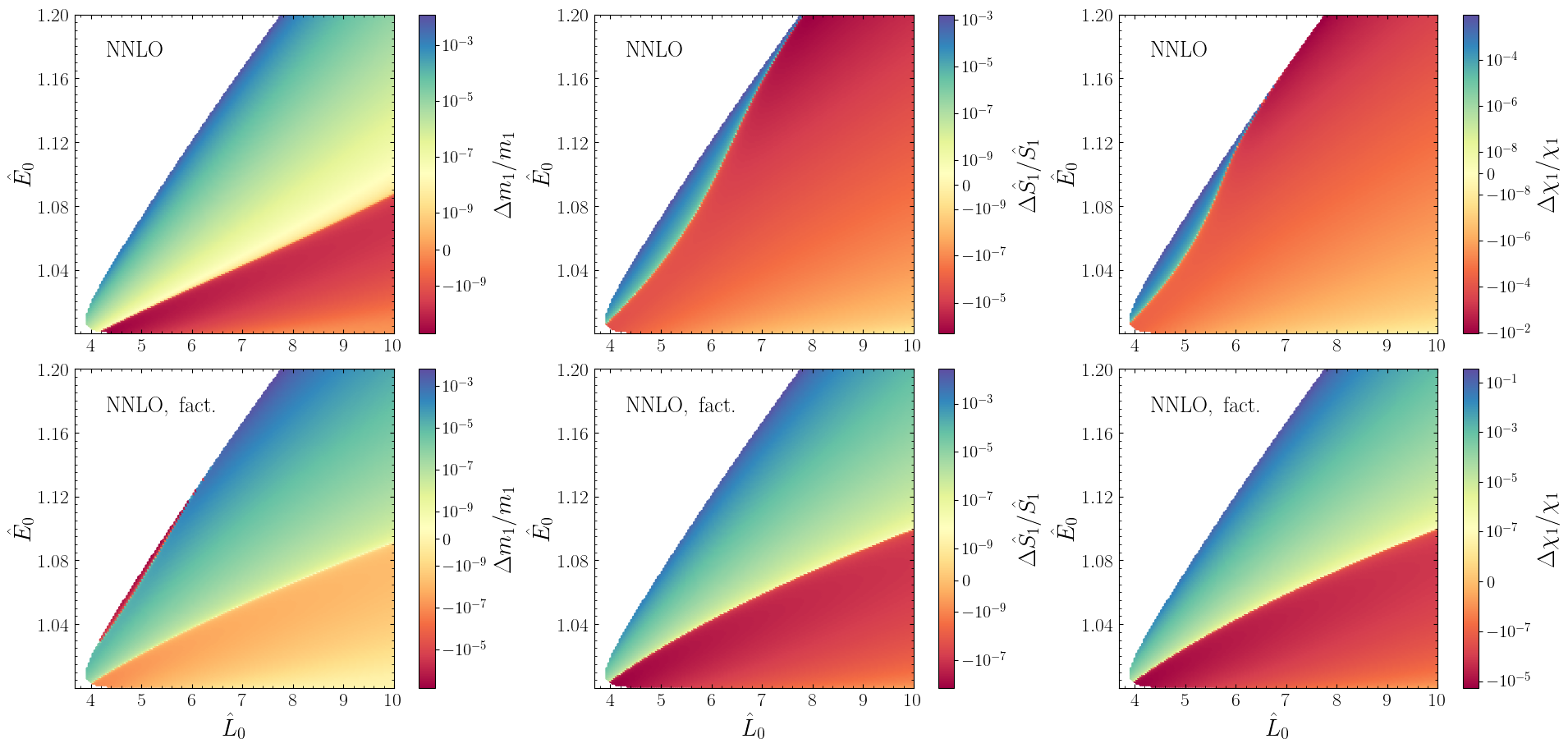}
  \caption{\label{fig:chi03_nnlospace}
  Tidal heating and torquing in scatterings of equal-mass, equal-spin binaries with $\chi_1 =
  \chi_2 = 0.3$ as calculated using the noncircular \ac{nnlo} models (raw in the top row,
  factorized in the bottom). The initial energy $\eo$ and orbital angular momentum $\lo$ are
  on the coordinate axes; the color represents, moving left to right, the relative cumulative variation
  in the \ac{bh} mass $\Delta m_1/M$, spin $\Delta \hS_1$ and mass-rescaled spin $\Delta \chi_1$.
  The models mostly agree on the sign of the mass change, except in a small region near 
  the scattering-capture separatrix where the factorized version finds a second sign change;
  they completely disagree at higher energies when looking at the angular momentum flux,
  with the factorized model also finding a larger maximum value ($\sim 10^{-2}$ against $\sim 10^{-3}$).
  Due to the overall small change in the mass, the normalized spin $\chi_1$'s behavior
  mimics that of $\hS_1$.}
\end{figure*}

\begin{figure*}
  \includegraphics[width=\textwidth]{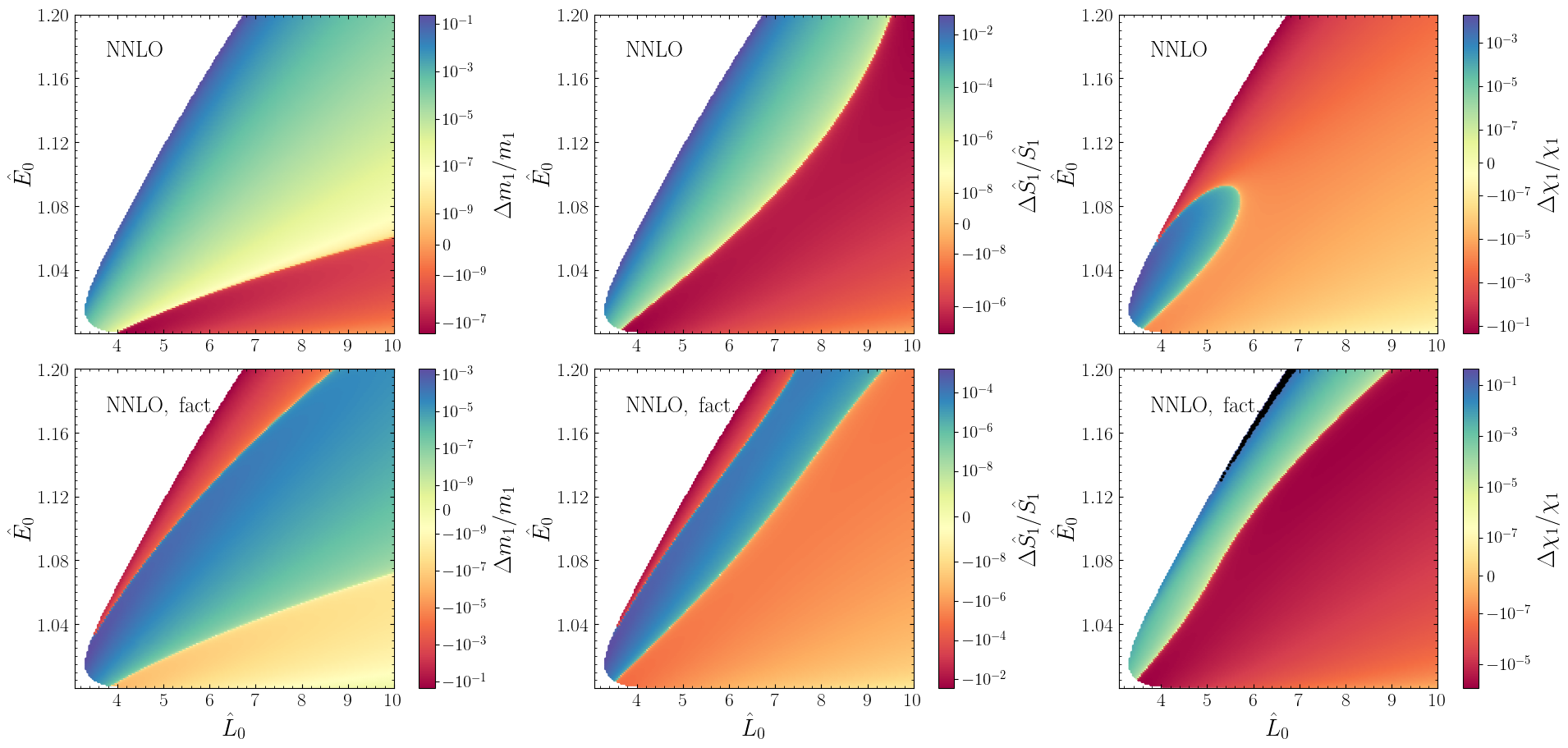}
  \caption{\label{fig:chi09_nnlospace} 
  Tidal heating and torquing in scatterings of equal-mass, equal-spin binaries with 
  $\chi_1 = \chi_2 = 0.9$ across the $\lo,\eo$ parameter space. The color represents,
  left to right, the relative cumulative variation in the \ac{bh} mass $\Delta m_1/M$, spin $\Delta \hS_1$, and
  mass-rescaled spin $\Delta \chi_1$ as a result of horizon absorption; we use the raw \ac{nnlo}
  model in the top row, and the factorized version in the bottom.
  The models differ wildly in their outcomes, outside of the bottom right of each plot
  (low energy, high angular momentum). The factorized model finds two sign changes in both
  $\Delta m_1$ and $\Delta \hS_1$ when moving to high $\eo$, with decreasing mass and spin
  along the scattering-capture separatrix; the maximum values of $\Delta m_1, \Delta \hS_1$
  are similar to the non-factorized model in order of magnitude, but with opposite sign.
  These discrepancies result in very different results concerning the sign of $\Delta \chi_1$.
  Of course, \ac{pn} models should not be trusted in the high velocity regime in the absence
  of solid data for comparisons. In fact, the factorized model even finds unphysical final values
  of $\chi_1 > 1$ in the black patch in the bottom right plot.}
\end{figure*}

Beginning with the simplest case, consistently with the discussion in the previous section,
\acp{bh} with initially negative spins are predicted across the parameter space to decrease
the magnitude of their angular momentum as a consequence of the tidal interaction with the 
companion, and gain mass from the absorption of energy. 
The cumulative variation in $\chi$ after the scattering event spans several orders of magnitude,
depending on the analytical model used for the flux and the parameters of the system itself.
The spin-down is particularly correlated with the distance of closest approach
in the case of negative spins (see Sec.~\ref{sec:universality} and Fig.~\ref{fig:rmindeltas}). 
Downstream of this, we see that $|\Delta \chi_{1,2}| = |\chi_{1,2}^{\rm final} - \chi_{1,2}|$ is typically smaller for low $\eo$ and large $\lo$: 
for example, in equal-mass systems with $\chi_1 = \chi_2 = -0.6$, $|\Delta \chi_{1,2}|$ increases by 2-3 orders of 
magnitude as $\lo$ decreases from $10$ to the threshold of direct capture; meanwhile, the peak value 
of $|\Delta \chi_{1,2}|$ grows from $\simeq 10^{-5}-10^{-4}$ at the lowest initial energy to $\simeq 10^{-2}-10^{-1}$ 
when $\eo \simeq 1.1$. These figures refer to the \ac{nnlo} factorized model, but this behavior
is common to all of them, and it is inherited from analogous phenomenology displayed by the changes 
in the masses $\Delta m_{1,2}$ and spins $\Delta \hS_{1,2}$.
Note that such large values for $\Delta \chi_{1,2}$ are only reached at very high energies, and even then only
by systems that graze the separatrix between scattering and dynamical capture.
Spin-ups above $\sim 10^{-2}$ (and comparable changes in the mass) are the 
exception rather than the norm, which is consistent with our approximations.

The \ac{nnlo} spin flux expression with a superradiance prefactor is consistently the one
predicting the largest effect among the noncircular models, which is due to both the ``hidden"
inclusion of higher order terms, and the size of the prefactor itself (here $\Omega_{\rm H} < 0$;
see the left panel of Fig.~\ref{fig:orders_q1_chi003} for a direct comparison between models
in the phenomenologically similar case of initially non-spinning \acp{bh}). 
At the lowest energies, the various analytical models don't differ much in their predictions; as $\eo$ 
increases, a clearer hierarchy develops, with cumulative variation larger by as many as
two orders of magnitude at higher \ac{pn} order than at \ac{lo} when using the 
generic-orbit expressions. 
The quasicircular ones cut through this, especially the version using $x$ 
in its leading factor, which often leads to completely different behavior as the parameters 
of the system are changed. For example, the quasicircular $x$ flux is the largest in the
low-energy portion of the parameter space, at times by a sizable margin (though all models agree
on an extremely small effect when this happens).

As for initially non-spinning \acp{bh}, similar considerations apply, except of course
for the \ac{lo} and \ac{nlo} analytical models that because of their low \ac{pn} order 
predict no absorption of energy or angular momentum in the initial absence of spin; 
see Figs.~\ref{fig:ns_mchispace} and~\ref{fig:ns_m_q28_space}. Again, because
of its extra terms stemming from the factorization of the 1.5\ac{pn} non-spinning factors,
the \ac{nnlo} model utilizing the horizon frequency prefactor here typically predicts
a cumulative $|\Delta \chi_{1,2}|$ exceeding the closest competitor by a factor of 10 or more (see the 
left panel of Fig.~\ref{fig:orders_q1_chi003}).
The highest final dimensionless spin among our investigations for an initially 
non-spinning \ac{bh} is $\chi_1^{\rm final} \simeq 0.175$, reached by the primary in a $q = 2$ system
with $\eo \simeq 1.16, \lo \simeq 8.1$. Restricting to equal-mass binaries, the largest
spin-up is found with similar parameters: $\chi^{\rm final}_{1,2} = 0.162$, with $\eo = 1.19, \lo = 8.2$.

The complex phenomenology seen in Sec.~\ref{sec:phenomenology} for positive spins emerges in the 
results of the systematic investigation of the parameter space. The \ac{lo} and \ac{nlo} models 
lead to a simple picture: the \acp{bh} overall lose angular momentum and mass (superradiance), the magnitude of 
the effect determined by the closeness of the encounter (high energy and low angular momentum thus 
lead to larger absolute cumulative changes in $m_{1,2}, \hS_{1,2}, \chi_{1,2}$). 
Results with the \ac{nnlo} models are instead somewhat reminiscent of the expected phenomenology in
the quasicircular case, with the spin-down effect reversed if the \acp{bh} reach close enough 
separation and high enough velocity; this can be seen 
in Figs.~\ref{fig:chi03_nnlospace}-\ref{fig:chi09_nnlospace}. 
Each analytical version of the model however places the threshold of superradiance differently,
visibly so in the case of the angular momentum flux, where the horizon frequency prefactor 
pushes it down to lower energies compared to the \ac{pn} expanded model.
The factorized model actually even finds a second sign change in $\Delta m_{1,2}$ and $\Delta \hS_{1,2}$ for high 
$\chi$ (Fig.~\ref{fig:chi09_nnlospace}), due to the large \ac{nnlo} term in the expansion; 
this mostly happens in the more extreme regions of the parameter space, where our 
\ac{pn} expressions, and possibly even the \ac{eob} dynamics~\cite{Albanesi:2024xus}, are likely unreliable, so the 
physical reality of this feature is uncertain in the absence of numerical data.
It is interesting to observe that these additional sign flips do not carry over to the dimensionless spin 
$\chi$, which continues to increase above the first threshold as seen in the bottom right panel of 
Fig.~\ref{fig:chi09_nnlospace}; however, with initially high spin, we do find cases where the final
state has $\chi_1 > 1$. Of course, our description of such a system is clearly inadeguate, as
this large a change in $\chi_1$ (and $m_1$, which in this regime decreases by $\sim 0.1M$) would
certainly impact the dynamics; seeing however as horizon absorption intensifies with increasing
\ac{bh} spin, a model incorporating the changing $m, \hS$ into the orbital dynamics would likely
find similar results, if not more extreme.

Before moving on, let us remark briefly on the relationship between the energy and angular momentum fluxes
and the \ac{bh} masses and mass ratio. Considering Eqs.~\eqref{eq:eob_fact}, the leading mass-dependent factor
differs slightly for spinning and non-spinning \acp{bh}. In the former case, both $\dot{m}_1$ and $\dot{\hat{S}}_1$
are proportional to $\eta_{\rm S} (q) = \nu^2 (m_1/M)^3 = q^5/(1 + q)^7$ (taking into account the fact that $\Omega_{\rm H}^1 \propto m_1^{-1}$);
in the latter, we have $\eta_{\rm NS} (q) = \nu^2 (m_1/M)^4 = q^6/(1 + q)^8$ instead. The fluxes for the secondary
\ac{bh} scale with $\eta (1/q)$; the ratio between the two bodies is thus $\eta_{\rm S} (1/q)/\eta_{\rm S}(q) = q^{-3}$,
or $\eta_{\rm NS} (1/q)/\eta_{\rm NS}(q) = q^{-4}$ in the non-spinning case.
Focusing, for definiteness, on the spinning case, $\eta_{\rm S}$ peaks for $q = 2.5$,
with $\eta_{\rm S}(2.5)/\eta_{\rm S}(1) \simeq 1.94$, and $\eta_{\rm S}(1/2.5)/\eta_{\rm S}(1) \simeq 0.124$.
So, tidal heating and torquing of the primary \ac{bh} are enhanced for moderate mass ratios ($q \lesssim 7$) with respect
to the equal-mass case. Their effect instead progressively weakens for the secondary as the mass ratio grows; 
due to the steep dependence on $q$, this is true also when looking at relative mass and spin variations (see Fig.~\ref{fig:ns_m_q28_space}).

While this simple analysis suggests that we should expect tidal heating and torquing to be strongest among the $q = 2$ or $q = 4$
data in our parameter space exploration, as we will further emphasize in the next section, the underlying binary dynamics
has a deciding role in determining the strength of the effect. In particular, to \ac{lo}, $\dot{m}_1 \sim r^{-8}$ and
$\dot{\hat{S}}_1 \sim r^{-6}$. For hyperbolic encounters, the lower limit on the distance of closest approach is the 
\ac{lr} radius; in \TEOBResumSDali{}, its location is pushed inward
as the mass ratio approaches 1~\cite{Nagar:2018zoe}. So, for $q > 1$, the mass factor and the dynamics affect the fluxes in opposite ways,
partially compensating each other; on the closest encounters, in particular, even small reductions in $r_{\rm LR}$ are
magnified, so the peak cumulative fluxes we find for $q = 2$ are not enhanced by as much as the mass factor suggests. 
Further from the scattering-capture separatrix, we do observe shifts, with cumulative fluxes growing more steeply
with $\eo$ and sign changes, when they occur, pushed to lower energies.

\subsection{Universality}
\label{sec:universality}

\begin{figure*}
  \includegraphics[width=\textwidth]{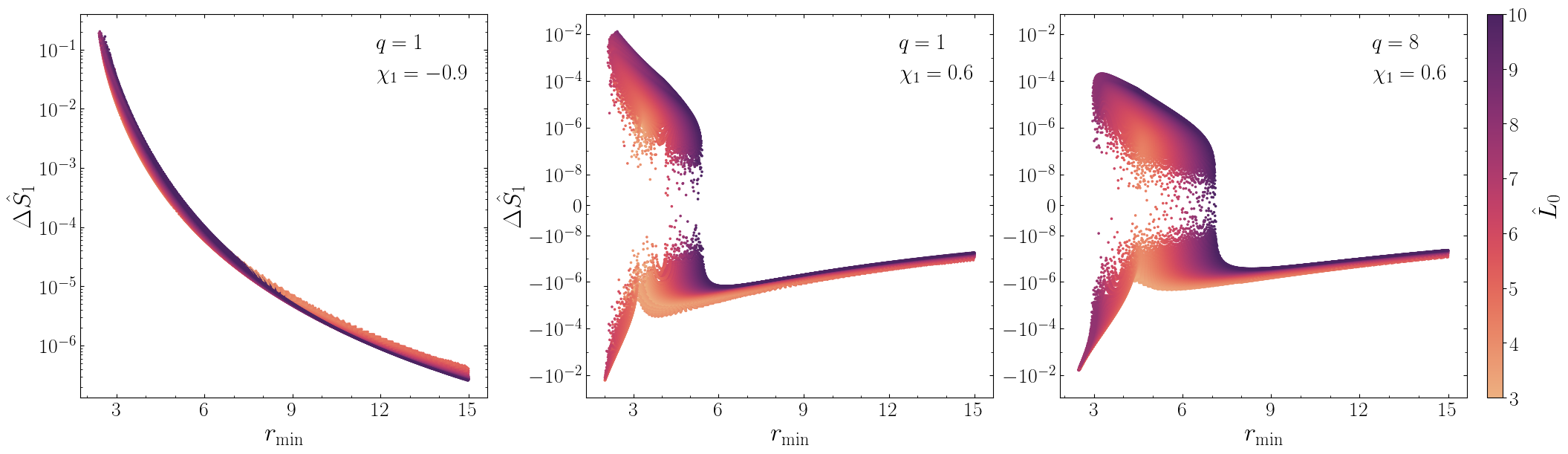}
  \caption{\label{fig:rmindeltas} 
  Cumulative change in the primary \ac{bh} spin $\hS_1$ as a function of the distance of closest approach $r_{\rm min}$
  for binary systems with varying mass ratio and initial primary spin $\chi_1$, including all available values of 
  $\eo, \lo$ (represented by the color scale) and $\chi_2$. The \ac{nnlo} factorized model is used here.
  Because of its leading $\sim r^{-6}$ term, the total effect of the absorption of angular momentum
  by the black hole is a simple function of $r_{\rm min}$, with limited deviations as the
  other parameters of the system vary, if $\chi_1 < 0$ (leftmost plot). The same is true when $\chi_1 > 0$ (middle and rightmost plot),
  but only for orbits with large periastron distance; $\Delta \hS_1$ otherwise changes sign
  twice as the encounters become closer, breaking the degeneracy. In particular, the color scale highlights the fact that the first sign flip occurs 
  earlier for higher $\lo$ and/or larger mass ratio, consistently with the form of the superradiance prefactor, 
  which compares the horizon frequency $\Omega_{\rm H}^1 \propto m_1^{-1}$ with $p_\varphi/r^2$.}
\end{figure*}

\begin{figure*}
  \includegraphics[width=\textwidth]{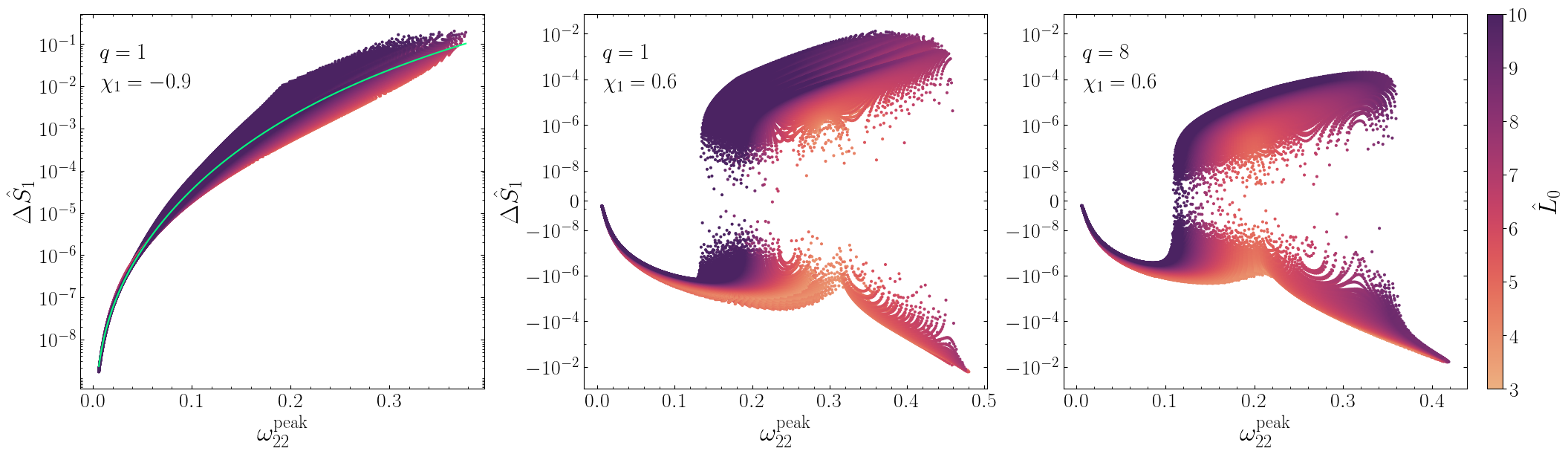}
  \caption{\label{fig:omgdeltas} 
  Cumulative change in the primary \ac{bh} spin $\hS_1$ as a function of the peak frequency of the 
  $(2,2)$ mode of the gravitational waveform. As in Fig.~\ref{fig:rmindeltas}, three illustrative 
  example plots are shown; for each, the intrinsic parameters are specified on the plot area, 
  and all available values of $\eo,\lo$ and $\chi_2$ are included. The green line is the result of
  the fit of the data to the function of Eq.~\eqref{eq:fitfunction}.
  Results here are similar to those of Fig.~\ref{fig:rmindeltas}, although all around less clear
  cut. 
  Some of the increased dispersion of the data points is due to the frequency scale naturally
  relegating orbits with large periastron to the left of the plot and giving more space to 
  the closest encounters; but mostly the reason is that there isn't an equally direct 
  analytical bond between the horizon fluxes and $\omega_{22}^{\rm peak}$ as there is
  with $r_{\rm min}$.}
\end{figure*}

\begin{figure}
  \includegraphics[width=0.49\textwidth]{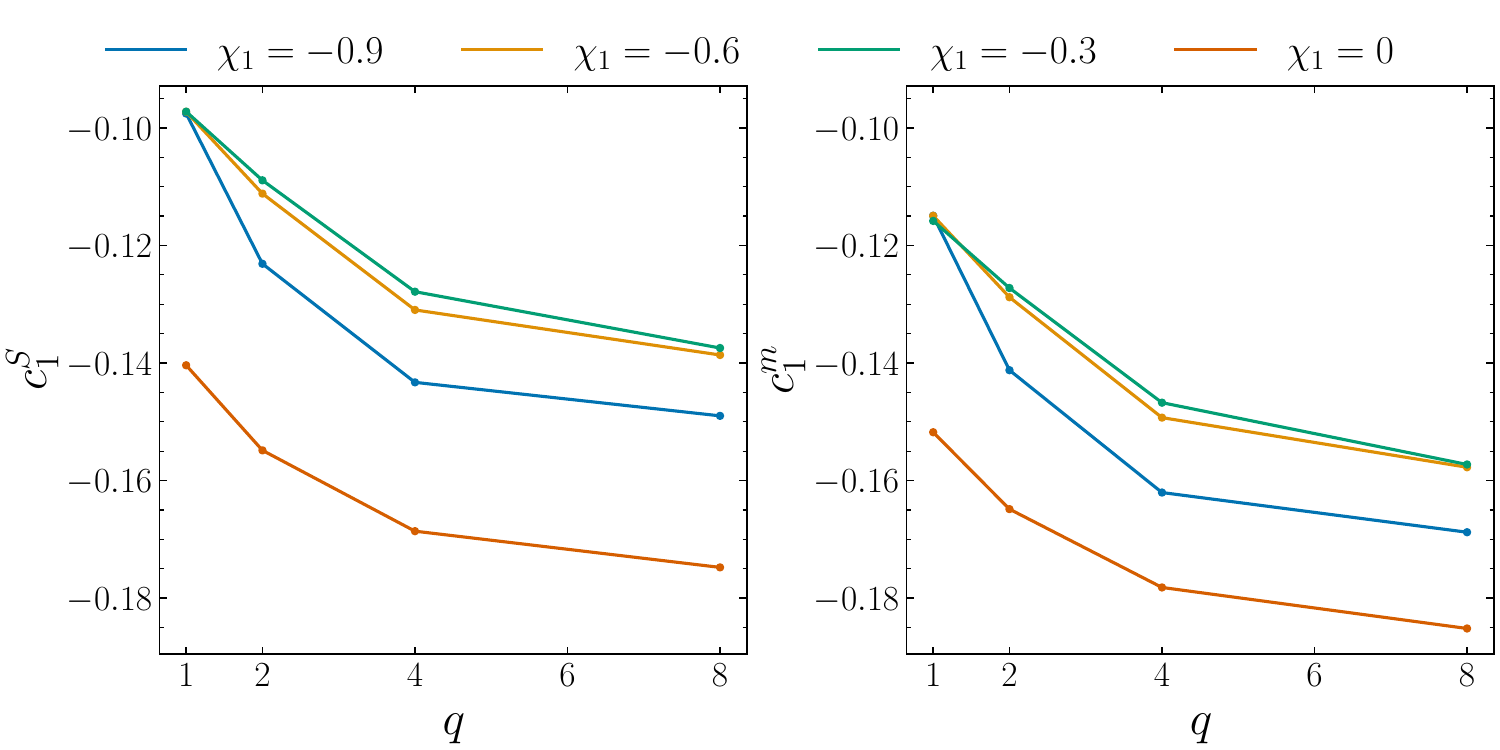}
  \caption{\label{fig:fit_par} Results for the fit parameter $c_1$ of Eq.~\eqref{eq:fitfunction}
  as a function of the mass ratio $q$ and the primary spin $\chi_1$; for the spin variation on the left,
  the mass on the right.
  In both cases, the leading exponent $c_1$ decreases as $q$ grows, with the values for $\Delta m_1$
  systematically lower than those for $\Delta \hS_1$.
  We also find a milder, opposite trend with $|\chi_1|$, with the exception 
  of the case of initially non-spinning \acp{bh}, which consistently have slightly lower values.
  The parameter $c_0$ is always found to be around $\sim -10$, so these negative values of $c_1$
  mean that $\Delta m_1$ and $\Delta \hS_1$ are rapidly approaching 0 as $\omega_{22}^{\rm peak}$ does.}
\end{figure}

The \ac{lo} contributions to both the angular momentum and energy fluxes through the \ac{bh}
horizons have simple mathematical forms, even for generic orbits:
\begin{equation}
  \dot{\hS}_{1} \sim \dfrac{m_{1}^5 m_2^2}{M^7} \chi_1 r^{-6}; \qquad \dfrac{\dot{m}_{1}}{M} \sim \dfrac{m_{1}^5 m_2^2}{M^7} \chi_{1} p_\varphi r^{-8}.
\end{equation}
Because of this, and despite the complexity of the complete \ac{nnlo} analytical expressions,
the cumulative changes in a \ac{bh}'s mass and spin as a result of horizon absorption during
a scattering event display a rather strikingly simple correlation with the distance of closest
approach $r_{\rm min}$. As shown for the angular momentum flux in Fig.~\ref{fig:rmindeltas}, 
this is especially true when the \ac{bh}'s spin is negative, or initially vanishes. 
Positive spins and their more involved phenomenology muddle this picture in close encounters, 
where sign flips in the fluxes occur; however, at least in the case of the \ac{nnlo} factorized model, 
recognizable patterns in the data can be easily traced back to the analytical expression. 
Similar conclusions are reached when investigating this using other versions of the analytical 
model, or turning to the cumulative change in the \ac{bh} masses.

It would be preferable to translate this correlation into one involving only gauge-invariant
quantities, instead of the coordinate-dependent $r_{\rm min}$. Dynamical variables that have
this property include the system's total energy and angular momentum, as well as the impact 
parameter $b = \hat{E}_0/\hat{L}_0$, and the scattering angle; none of these display such a clean relationship 
with the horizon fluxes though, with the latter two merely acting as proxies of the former,
without cutting through the complexity seen in the parameter space plots included in these pages.
So we turn instead to properties of the waveform generated by the system, 
as calculated by the \TEOBResumSDali{} model: the peak wave
amplitude and frequency can be taken as indicators of the closeness of a hyperbolic encounter.
The former, specifically the $(2,2)$ mode amplitude, appears to be an imperfect 
descriptor; probable cause for this is the appearance of multiple peaks around closest 
approach as the energy of the orbit increases. 
The frequency $\omega_{22}^{\rm peak}$
\footnote{If the $(2,2)$ multipole of the \ac{gw} strain as defined in Eq.~(1) of~\cite{Nagar:2024dzj}
is $h_{22} = A_{22} {\rm e}^{-{\rm i} \phi_{22}}$, the (dimensionless) frequency we consider here is
$\omega_{22} \equiv M \omega_{22}^{\rm phys} \equiv \dot{\phi}_{22}$.}
instead leads to comparable results to $r_{\rm min}$, 
as seen in Fig.~\ref{fig:omgdeltas}. We do find wider curves, indicating that, especially
for high peak frequencies, the two quantities are not as tightly bound. 
This can be traced to the additional step required to analytically link them:
the cumulative flux is greatly influenced by its peak value, which, at first
approximation and in the case of the angular momentum, is given by the leading order term 
$\sim r^{-6}$, evaluated at closest approach;
meanwhile, the peak frequency at lowest order, neglecting non-circular contributions,
is double the orbital frequency, which by Kepler's Law is $\dot{\varphi} \sim r^{-3/2}$
at periastron.
Particularly at higher mass ratios, the $r_{\rm min}$ - $\omega_{22}^{\rm peak}$ data
can indeed be fitted to a power law with exponent close to this rough estimate.

Seeing this, we attempt to fit the energy and angular momentum cumulative flux data to
the peak frequency. For each pair $(q, \chi_1)$ with nonpositive $\chi_1$ we fit
both $\Delta m_1$ and $\Delta \hS_1$ to a function of the form:
\begin{align}
  \label{eq:fitfunction}
  \log X &= c_0 (\omega_{22}^{\rm peak})^{c_1} \dfrac{1 + c_2 \omega_{22}^{\rm peak}}
  {1 + c_3 \omega_{22}^{\rm peak}} \\
  \quad X &= \Delta \hS_1 \text{ or } \dfrac{\Delta m_1}{M} \nonumber
\end{align}
with $c_1$ parametrizing the leading power law, and the rational factor catching any 
deviation from it at larger values of the frequency. Fig.~\ref{fig:omgdeltas} displays
in its leftmost panel an example of the resulting curve for the spin variation;
Fig.~\ref{fig:fit_par} the fitted values for $c_1$.
Notable, and repeated in every other instance, are the accurate reproduction of the 
low-frequency data by the fit, but also their dispersion around the line at higher 
$\omega_{22}^{\rm peak}$; this is somewhat concealed visually by the logarithmic scale, 
but deviations from the fitted function span more than an order of magnitude in $\Delta \hS_1$.

\section{Numerical relativity comparison}
\label{sec:nr}

\begin{table*}
  \begin{tabular}{cc|cc|cc|cc|ccc}
     $|p|/M$ & $\theta_{\rm p}$ [rad] & $\hat{E}^{\rm NR}_{0}$ & $\hat{L}^{\rm NR}_{0}$ & $\hat{E}_{0}^{\rm EOB, NNLO}$ & $\hat{L}^{\rm EOB, NNLO}_{0}$ &  $\hat{E}_{0}^{\rm EOB, NNLO~fact}$ & $\hat{L}^{\rm EOB, NNLO~fact}_{0}$ & $\chi_f^{\rm NR}$  & $\chi_f^{\rm EOB, NNLO}$  & $\chi_f^{\rm EOB, NNLO~fact}$ \\
\hline
  $0.245$ & $0.080$ & $1.1091$ & $7.8254$ & $1.1570$ & $8.1677$ & $1.1297$ & $8.3484$ & $0.00026$  & $0.00026$  & $0.00026$\\
  $0.245$ & $0.073$ & $1.1091$ & $7.1944$ & $1.1344$ & $7.1178$ & $1.0996$ & $6.7455$ & $0.00089$  & $0.00089$  & $0.00089$\\
  $0.245$ & $0.067$ & $1.1091$ & $6.5631$ & $1.1253$ & $6.6569$ & $1.1048$ & $6.5133$ & $0.0035$  & $0.0035$  & $0.0035$\\
  $0.245$ & $0.061$ & $1.1091$ & $5.9315$ & $1.0975$ & $5.9759$ & $1.0896$ & $5.8675$ & $0.016$  & $0.0069$  & $0.016$\\
\hline
  $0.3675$ & $0.080$ & $1.2378$ & $11.7381$ & $1.2729$ & $10.9906$ & $1.2343$ & $11.5333$ & $0.00082$  & $0.00082$  & $0.00082$\\
  $0.3675$ & $0.072$ & $1.2378$ & $10.5016$ & $1.2619$ & $10.1897$ & $1.2574$ & $11.4485$ & $0.0028$  & $0.0028$  & $0.0028$\\
  $0.3675$ & $0.063$ & $1.2378$ & $9.2643$ & $1.2461$ & $9.5503$ & $1.2302$ & $9.7840$ & $0.011$  & $0.011$  & $0.011$\\
  $0.3675$ & $0.055$ & $1.2378$ & $8.0264$ & $1.1964$ & $8.2894$ & $1.1922$ & $8.2833$ & $0.054$  & $0.017$  & $0.054$\\
\hline
  $0.49$ & $0.100$ & $1.4019$ & $19.5688$ & $1.3705$ & $18.0349$ & $1.3650$ & $19.2148$ & $0.00043$  & $0.00028$  & $0.00043$\\
  $0.49$ & $0.085$ & $1.4019$ & $16.6054$ & $1.3494$ & $15.1652$ & $1.3306$ & $15.0202$ & $0.0019$  & $0.0019$  & $0.0019$\\
  $0.49$ & $0.070$ & $1.4019$ & $13.6381$ & $1.3261$ & $12.5187$ & $1.3364$ & $14.0956$ & $0.011$  & $0.011$  & $0.011$\\
  $0.49$ & $0.054$ & $1.4019$ & $10.6678$ & $1.2924$ & $10.8465$ & $1.2782$ & $10.5281$ & $0.1$  & $0.026$  & $0.1$\\
  \hline
  \end{tabular}
  \caption{Parameters of the \ac{nr} simulations of binary black hole scatterings considered in this work, 
  estimated from Ref.~\cite{Jaraba:2021ces} using the initial data generated via {\tt TwoPunctures}~\cite{Ansorg:2004ds,Daszuta:2021ecf},
  and the optimal initial conditions and final spin of the \acp{bh} as calculated by our \ac{eob} model
  using the \ac{nnlo} expressions for the tidal torquing.}
  \label{tab:nr}
\end{table*}

\begin{figure*}
  \includegraphics[width=0.9\textwidth]{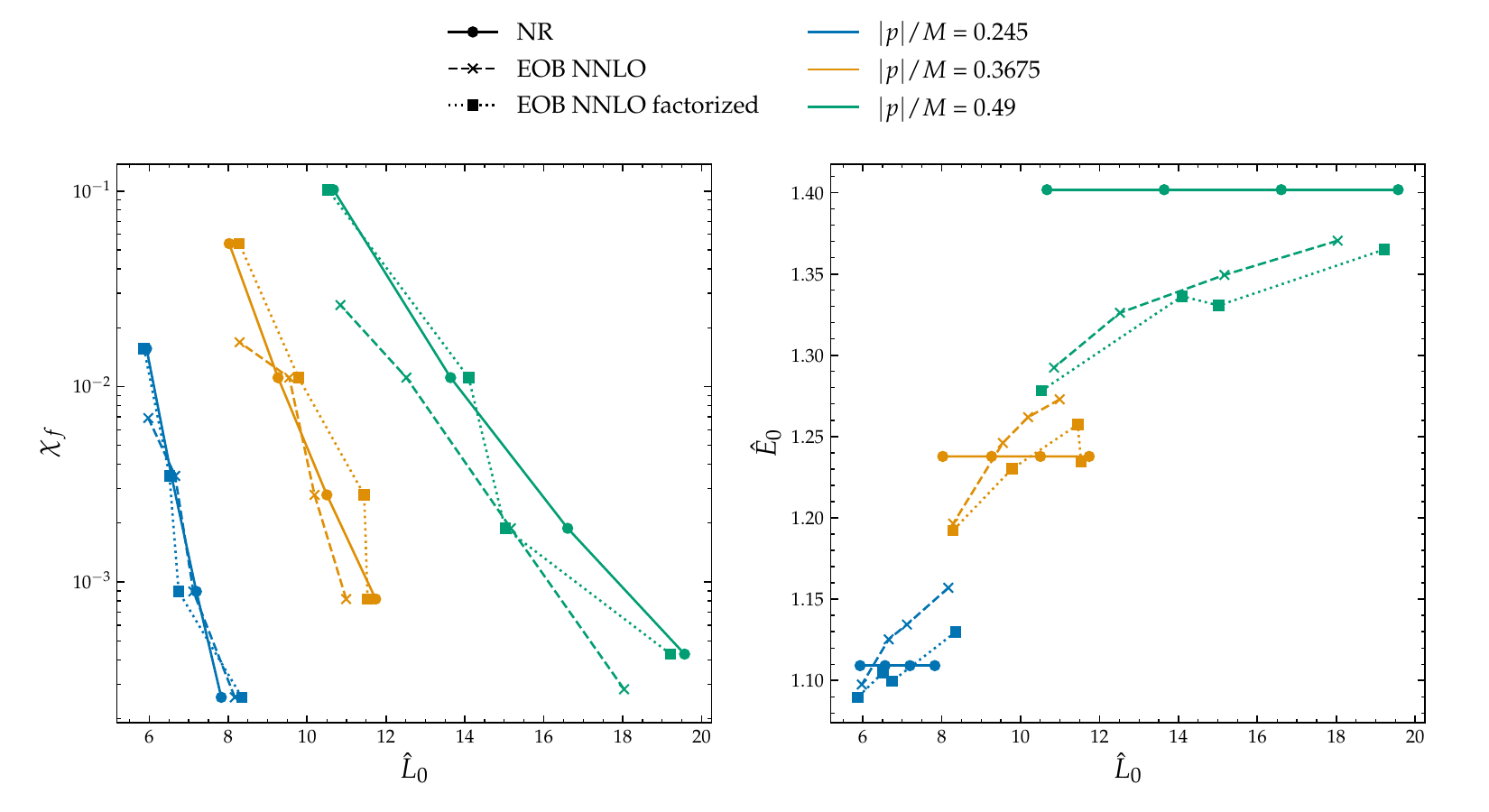}
  \caption{Left: comparison of the final spin of the \acp{bh} in the NR simulations of Ref.~\cite{Jaraba:2021ces}, as estimated from their
  Table II, and the EOB model with NNLO tidal torquing as a function of the
  initial orbital angular momentum. Right: comparison between the \ac{nr} initial data (energy and angular momentum) and the values that
  minimize the difference between the final spin of the \acp{bh} in the \ac{nr} simulations and the EOB model.
  The overall behavior of the \ac{nr} data is reproduced: the factorized \ac{nnlo} expressions are able to recover the correct final spin for all the
  scattering points considered, whereas the non-factorized version fails to capture the more extreme spins found 
  at high energy and low angular momenta.
  The largest variations in the initial data are of the order of $9\%$ for both energy and angular momentum, 
  predictably corresponding to the points with the largest spin-up. For these configurations, we consistently find that for $|p|/M = 0.49$ the initial energy
  should be decreased with respect to the nominal \ac{nr} value to find agreement; this is because of the \ac{eob} model predicting 
  a lower threshold between scattering and capture.
  \label{fig:eobnr_scatterings}}
\end{figure*}

We now turn to the comparison of our semi-analytical model with \ac{nr} simulations 
of binary black hole scatterings, and in particular to the simulations of Ref.~\cite{Jaraba:2021ces}.
In this work, the authors performed, among others, four sets of simulations of equal-mass, initially nonspinning \ac{bbh} scatterings.
For each set, they fixed the magnitude of the initial linear momenta of the \acp{bh} $p_{1,2} = |\mbf{p}_{1,2}|$ 
and varied the angle between the initial linear momenta and the separation vector of the \acp{bh}, $\theta_{\rm p}$.
Initial data were generated using {\tt TwoPunctures}~\cite{Ansorg:2004ds,Daszuta:2021ecf}, and the evolutions performed using the {\tt Einstein Toolkit}~\cite{EinsteinToolkit:2024_05} library.
We restrict our comparisons to the sets with $p/M = 0.245, 0.3675, 0.49$, and neglect those with $p/M=0.75$. This is because
our \ac{eob} code does not reliably evolve the dynamics of the system at such high energies.
Table \ref{tab:nr} summarizes the parameters of the simulations that we consider\footnote{The initial values of \ac{adm} mass and angular momentum were
estimated by choosing four uniformly spaced points in the $\theta_p$ range specified in Tab~I of Ref.~\cite{Jaraba:2021ces} for each $p_i$, 
and recomputing the initial data with a stand alone implementation of {\tt TwoPunctures}~\cite{Daszuta:2021ecf}} and the predicted final values of
the spin of the \acp{bh} (see Tab.~II of Ref.~\cite{Jaraba:2021ces}).

As was already observed in Ref.~\cite{Jaraba:2021ces}, and Ref.~\cite{Nelson:2019czq} before that, considerable spin-up of the \acp{bh} can be measured in the
simulations, with the final spins reaching $\chi \sim 0.1$ after the encounter.
Following from Sec.~\ref{sec:phenomenology}, this increase in spin can only be caused by terms beyond $1$\ac{pn}, 
as up to that point the analytical models identically vanish if $\chi_{1,2} = 0$. 
This is a clear indication that the \ac{nnlo} terms in Eqs.~\eqref{eq:mdot_eob_fact} and \eqref{eq:tidal-eob} 
are crucial to accurately describe the dynamics of the system in the strong-field regime.
To perform our comparisons, then, we integrate the \ac{eob} dynamics and compute the tidal torquing
on the \acp{bh} using the \ac{nnlo} expressions (both factorized and not). 
Notably, rather than directly employing the initial \ac{nr} energy and angular momentum to initialize the \ac{eob} dynamics, 
we allow for small variations around these values and minimize the loss function:
\begin{equation}
  \mathcal{L} = |\chi^{\rm NR}_{f} - \chi^{\rm EOB}_f (\hat{E}^{\rm NR}_0(1 + \epsilon_E), \hat{L}^{\rm NR}_0(1 + \epsilon_L) )| \, ,
\end{equation}
with $\epsilon_{E,L} \in [-0.1, 0.1]$\footnote{Notably, $\mathcal{L}$ does not account for the uncertainty in the \ac{nr} data,
aiming to minimize the difference between the \ac{nr} and \ac{eob} predictions to numerical precision. A better,
more complete comparison would involve accounting for such an uncertainty, but this is beyond the scope of this work.}.
This approach allows us to account for numerical uncertainties in the computation of initial data, 
initial junk radiation during evolutions, gauge differences (particularly in the initial separation where $\hat{E}_0$ and $\hat{L}_0$ are defined), 
as well as potential inaccuracies in extracting the final spin from simulations. 
Additionally, this optimization procedure absorbs some inconsistencies inherent to the \ac{eob} dynamics, 
such as our neglecting the variation of the spins and masses during the evolution.
Therefore, this step should be seen as an attempt to estimate the system's dynamics that best aligns with 
the \ac{nr} results for a given choice of analytical flux model.
A more rigorous comparison would involve optimizing the initial data of the \ac{eob} evolution to match 
another gauge-invariant observable from \ac{nr}, such as the scattering angle. 
With the underlying dynamics then fixed, the final spin could be compared to the \ac{nr} prediction 
using different versions of the tidal torquing expressions. 
Unfortunately, due to the lack of scattering angle data from Ref.~\cite{Jaraba:2021ces}, 
we are unable to carry out this comparison here.

Our results are presented in Fig.~\ref{fig:eobnr_scatterings}. The \ac{eob} predictions closely follow 
the behavior of the \ac{nr} data across all initial momenta considered. 
Notably, the non-factorized \ac{nnlo} expressions predict a lower spin-up for the scattering points 
with lower initial angular momentum. In contrast, the factorized expressions successfully recover the correct 
final spin for all scattering points, suggesting that in this regime, the inclusion of the superradiance prefactor 
is essential for accurately modeling the observed spin-up. 
This finding also indicates that \ac{nr} data from scatterings of initially non-spinning black holes 
could be used to discriminate between different analytical representations of tidal torquing.

The largest variations in the initial data are of the order of $9\%$ for both energy and angular momentum,
predictably corresponding to the points with the largest energy and lowest angular momentum.
Interestingly, for these configurations, while the ``optimal'' values of angular momentum are typically not very far from the nominal ones,
the energy is often underestimated by a few percent. This is because, at such high energies, the \ac{eob} dynamics
would often predict a merger, which is not the outcome of the simulations investigated.

Overall, these findings are rather remarkable: \TEOBResumS{} employs no \ac{nr} information beyond the quasi-circular limit
and we are currently relying on pure \ac{pn} expressions to model the tidal torquing of the \acp{bh}.

\section{Astrophysical implications}
\label{sec:astro}

\begin{figure*}
  \includegraphics[width=\textwidth]{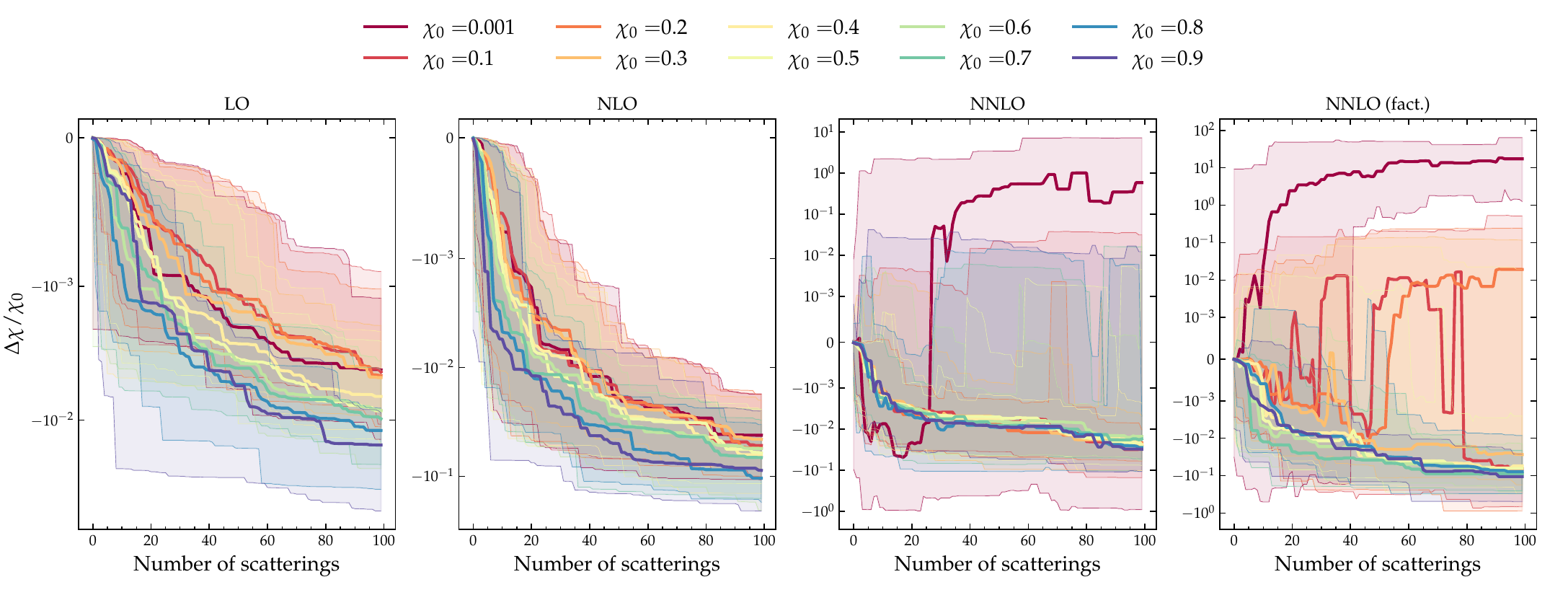}
  \caption{Evolution of the spin of a single \ac{bh} of initial mass $50 M_{\odot}$ in a dense cluster 
  after $N_s$ scatterings. The initial spin magnitude of the \ac{bh} is varied between $0.001$ and $0.9$, and 
  the results are averaged over $100$ repetitions. The shaded regions represent the $90\%$ intervals for the spin values
  reached by the \ac{bh}; the solid lines represent the median values. The four different panels correspond to the
  different analytical models for the tidal torquing and heating employed, as described in the main text.
  The effect of multiple scatterings is -- on average -- to decrease the spin magnitude of the \ac{bh}, with the
  largest variations observed for the largest initial spins if the \ac{lo} and \ac{nlo} expressions are employed. 
  The \ac{nnlo} models predict the more complex behavior, with almost non-spinning \acp{bh} rapidly acquiring significant spins,
  slowly spinning \acp{bh} ($\chi^0_{\rm bh} \leq 0.2$) increasing their spin magnitude up to relative $10\%$, 
  and highly spinning \acp{bh} sometimes spinning up after a scattering event.
  \label{fig:cluster}}
\end{figure*}

\acp{bh} in dense environments, such as globular clusters, can scatter off each other multiple times
before forming a bound system and merging~\cite{1993Natur.364..423S, RevModPhys.50.437, Rodriguez:2019huv}.
Tidal heating and torquing, studied in the main body of this work, can have implications for the properties
of the \acp{bh} formed in these environments.
In particular, in Sec.~\ref{sec:param_space} we observed how the spins of the \acp{bh} can change significantly after highly
energetic scattering events, depending on the initial configuration of the system (mass ratio, spins, energy and angular momentum).
In this section, we follow the evolution of a single \ac{bh} embedded in a (simulated) dense cluster.
Varying the initial spin magnitude of the \ac{bh}, we estimate the effect of multiple scatterings
on its evolutionary path. 

\subsection{Cluster properties}
The properties of \acp{bh} in dense globular clusters are not well known.
Two-body relaxation, dynamical friction, three body interactions and other processes determine the
evolution of the cluster and the intrinsic properties and number of the \acp{bh} within it~\cite{Rodriguez:2019huv}.
Investigating the formation and evolution of \acp{bh} in globular clusters is a complex problem that
requires detailed numerical simulations of N body evolutions~\cite{Joshi_2000, Pattabiraman_2013, Rodriguez:2019huv}.

Here, we aim to provide a simple, order-of-magnitude estimate of the effect of tidal torquing and heating
on the spins of \acp{bh} in a dense cluster.
For simplicity, then, we assume that the \acp{bh} in the cluster have masses uniformly
distributed in the range $[11, 80] M_{\odot}$~\cite{Zevin:2020gbd}, and spins uniformly distributed in the range $[-0.9, 0.9]$.
We make no assumptions regarding the shape, density or size of the cluster, as these quantities will not be
relevant for the estimates we are interested in.
We focus on a single \ac{bh} of mass $m_{\rm bh}$ and initial spin magnitude $\chi^0_{\rm bh} = a^0_{\rm bh}/m^0_{\rm bh}$ 
and follow its evolution in the cluster,
assuming that it undergoes $N_s$ separate scatterings with other \acp{bh}.
The intrinsic properties of the companion \ac{bh} are randomly drawn from the distributions described above. The orbital
configurations are instead chosen by sampling the initial energy and angular momentum of the system
from the ranges $[1.001, 1.1]$ and $[4, 20]$ respectively, fixing the initial separation to $r_0 = 10\ 000 M$.
Notably, for each scattering, we randomly draw the sign of the initial spin of the target black hole, in order
to account for all possible configurations.
Given these initial conditions, we integrate the \ac{eob} equations of motion for the system, and compute the
mass and spin variations due to tidal heating and torquing, $\Delta m_{\rm bh}$ and $\Delta \chi_{\rm bh}$, as done in the
rest of the paper.
We then update the mass and spin magnitude of the target \ac{bh}, and repeat the process for $N_s=100$ scatterings, discarding
the results whenever the system merges or becomes bound to another \ac{bh} instead of scattering.
For each initial value of spin, we repeat the process $100$ times in order to explore different realizations 
of the cluster and obtain a statistically meaningful estimate.

\subsection{Results}

Figure \ref{fig:cluster} shows the results of this investigation, employing the different expressions
for the tidal torquing and heating derived in the main body of this work at different \ac{pn} orders.
Unsurprisingly, up to \ac{nlo}, tidal torquing causes the magnitude of the \ac{bh} (dimensionless) spin to \textit{decrease}.
The effect is more pronounced for \acp{bh} with larger initial spin magnitudes, which also show a wider spread in the final
value after $N$ scatterings. While not very efficient in reducing $|\chi_{\rm bh}|$, the effect
of torquing can nevertheless be significant after $\sim 10$ scatterings, with median relative changes varying 
from a few fractions of percent up to $\sim 5-8\%$ (for $\chi^0_{\rm bh}=0.9$ and \ac{nlo} fluxes).
As also discussed in the previous sections, the picture is significantly more complicated when \ac{nnlo} terms are included in the model.
The clear hierarchy in the effect of the initial spin magnitude is lost, and while \textit{on average} multiple scatterings
lead to a decrease of the spin magnitude for $|\chi^0_{\rm bh}| \geq 0.2$, there exist paths through the cluster where the \ac{bh}
can actually spin up after a scattering event. The number of such paths, however, decreases with increasing $|\chi^0_{\rm bh}|$.
Additionally, initially (almost) non-spinning \acp{bh} can rapidly acquire
significant spins, with variations up to $1000\%$ when the factorized version of the flux is employed.

Our results are clearly dependent on the specific assumptions made in the construction of the cluster and the
initial conditions of the scattering events. While, for simplicity, we assumed the initial energy and momentum distributions
to be uniform, in reality these quantities should be determined by the dynamics of the cluster. As a consequence, the configurations
leading to very large spin-ups (or spin-downs) might be even rarer, given that they typically require the encounters to be
highly energetic and close to the transition between scattering and capture~\cite{Kankani:2024may,Albanesi:2024xus,Long:2024ltn}.

\section{Conclusions}
\label{sec:conclusions}

In this work we computed the mass and angular momentum fluxes due to tidal heating and torquing
for \ac{bbh} systems moving on generic planar orbits.
Building upon the work of Refs.~\cite{Taylor:2008xy,Poisson:2014gka,Saketh:2022xjb}, we derived 
\ac{pn} expressions valid up to \ac{nnlo}, or 1.5\ac{pn} orders above the leading fluxes for Kerr \acp{bh},
in modified harmonic and \ac{eob} coordinates.
We found that it is possible to factor out in these expressions a term involving the \ac{bh}'s horizon frequency.
This operation might prove physically significant as it emulates the results in the quasicircular limit, 
where a similar extracted prefactor parametrizes the transition into and out of superradiant dynamics, i.e.,
the direction of the flow of energy and angular momentum between each \ac{bh} and the binary system.
Interestingly, the noncircular factors we obtained in the mass and angular momentum fluxes are different.

We carefully investigated the properties of our expressions by evaluating them on a few selected 
scattering trajectories computed with the \TEOBResumSDali{} model.
Starting at \ac{nlo}, the noncircular models introduce novel effects in the 
phenomenology of both $\dot{S}_{1,2}$ and $\dot{m}_{1,2}$, most notably regarding the sign of the fluxes.
For \acp{bh} co-rotating with the binary, a complex pattern of sign changes emerges around periastron as the system parameters vary, 
due to the interplay of the \ac{nnlo} terms in the expansion and, when used, the superradiance prefactor.
The resulting physics appears counterintuitive in some cases at high energy, especially for fast-spinning \acp{bh}.

An in-depth investigation of the parameter space of scattering binaries revealed that the cumulative effect 
of the tidal interaction on the scattering \acp{bh} is, as expected, usually quite small; significant 
spin-up/down occurs only in highly energetic, close encounters.
The analytic models considered, however, predict in this regime very different qualitative features if the
\ac{bh} spins are aligned with the orbital angular momentum.
Most relevant among these are the boundaries in the parameter space of the regions where superradiance occurs.
This variance, and the possibility of even unphysical results ($\chi > 1$), points to our \ac{pn} results 
being unreliable on close, energetic encounters, highlighting the need for a more careful analytic treatment, 
beyond the simple use of the superradiance prefactor.

We compared our model with \ac{nr} simulations of scatterings of initially non-spinning \acp{bh}, finding that the \ac{nnlo} terms
are crucial to accurately describe the spin-up observed in the simulations. Allowing for small variations around the
nominal values of initial \ac{nr} energy and angular momentum, the factorized version of the fluxes
was able to correctly reproduce the final spin of the \acp{bh} in all the scattering points considered.
However, for the most challenging configurations (high energy, low angular momentum), the optimal
energy values we recover typically underestimate the \ac{nr} ones. This is a symptom of the \ac{eob}
dynamics incorrectly locating the threshold between scattering and dynamical capture~\cite{Albanesi:2024xus}.
Still, this is a noteworthy result: using a \ac{bbh} model calibrated on quasicircular inspirals and purely \ac{pn}
flux expressions, we were able to find orbital configurations that lead to the correct spin-up. In particular,
the better performance of the factorized model suggests that this treatment of the raw \ac{pn} expansions
could be a key step in the construction of an effective prescription for the horizon exchange for use in \ac{bbh}
models. Clearly, this is contingent on this pattern being repeated in further testing, especially in the case
of \emph{initially spinning} \acp{bh}. However, \ac{nr} data suitable for such an investigation is not available at the moment, 
as most of the current \ac{nr} simulations of \ac{bbh} scatterings do not reach the high energies required to 
clearly observe these effects~\cite{Damour:2014afa,Hopper:2022rwo,Rettegno:2023ghr}.

Finally, we investigated the effect of multiple scatterings in a dense cluster on the spin of a single \ac{bh}.
We found that, while on average tidal torquing leads to a decrease in the spin magnitude of the \ac{bh}, the \ac{nnlo} terms
can lead to significant spin-ups for some configurations.

Our work provides a solid foundation for future investigations of the impact of horizon absorption
on the dynamics of both bound and unbound \ac{bbh} systems.
In particular, given that tidal torquing is the only mechanism that can lead to a change in the
spin magnitude of the \acp{bh}, it appears natural to attempt to inform the analytical structure of these fluxes -- exploring
resummations and factorizations~\cite{Nagar:2011aa} -- by comparing the spin and mass variations with numerical simulations of scattering and 
merging \acp{bbh}. This will allow us to improve the accuracy of our models during the plunge phase, and to study the impact of
horizon absorption on the phase evolution of the \ac{gw} signal.
Additionally, we plan to perform careful comparisons with numerical test mass results, to
investigate the novel effect in $\dot{m}$ that is predicted by \ac{pn} theory, and to perform \ac{nr} simulations
of \textit{spinning} \ac{bbh} scatterings, to probe the complex phenomenology that we observed in Sec.~\ref{sec:param_space}.
Finally, we will implement these flux models in \TEOBResumS{} in a self-consistent way, allowing for the evolution of the spins
and masses of the \acp{bh} to be informed by the tidal torquing and heating. Then, leveraging the model developed in 
Ref.~\cite{Albertini:2024rrs}, we will assess the importance of these variations on the dynamics of large mass ratio 
inspirals in the context of LISA (Laser Interferometer Space Antenna).

\acknowledgments
The authors would like to thank S.~Bernuzzi, A.~Nagar, D.~Radice, S.~Albanesi, M.~Panzeri and G.~Pratten for 
useful discussions. DC and RG are grateful to M.V.S.~Saketh for kindly helping to clarify some passages in the calculations
of Ref.~\cite{Saketh:2022xjb}.
DC and RG wish to acknowledge S.~Grossard, E.~Vedder, S.~Carpenter, K.~R.~Amstutz, C.~E.~Aitchison for inspiring
them thougout the development of this work.
The version of \TEOBResumS{} employed in this work is available at 
\url{https://bitbucket.org/teobresums/teobresums/src/GIOTTO/}
and is tagged with the arXiv number of this work.
DC acknowledges support from the Italian Ministry of University and Research (MUR) via the PRIN 2022ZHYFA2, {\it GRavitational wavEform models for coalescing compAct binaries
with eccenTricity} (GREAT).
RG acknowledges support from NSF Grant PHY-2020275
(Network for Neutrinos, Nuclear Astrophysics, and Symmetries (N3AS)).

\appendix

\section{Factorized fluxes in harmonic coordinates}
\label{app:hfact}

In this Appendix we write out for completeness the \ac{nnlo} results for the rates of change of mass
and spin in harmonic coordinates using a superradiance prefactor.
Following Sec.~\ref{sec:eobcoords}, we find in each of the Eqs.~\eqref{eq:dmsdt} a 1.5\ac{pn} term proportional
to $\chi_1^{-1}$, and we shift it into a leading factor containing the horizon frequency
$\Omega_{\rm H}^1$. We obtain:

\begin{widetext}
  \begin{subequations}
    \label{eq:dmsdt_hfact}
  \begin{align}
    \dfrac{dm_1}{dt} =& -\dfrac{16}{5} \dfrac{m_1^6 m_2^2}{r^6} (1+\sigma_1) \biggl[\Omega_{\rm H}^1 - \dfrac{1}{c^3} \biggl(\dphi + 3 \dfrac{\dot{r}^2}{r^2 \dphi}\biggr)\biggr]
     \biggl\{
      \dot{\varphi} \left(1+3\chi_1^2\right) + \dfrac{1}{c^2} \biggl\{
        \left(1+3\chi_1^2\right) \biggl[\dfrac{7}{4} r^2 \dot{\varphi}^3 \nonumber \biggr. \biggr. \biggr. \\
        &- \biggl. \biggl. \biggl. \biggl(1-\dfrac{m_1}{M}+\nu\biggr) \dfrac{r \ddot{r}\dot{\varphi} + r\dot{r}\ddot{\varphi}}{2}
        -  \biggl(15 - \dfrac{5m_1}{M}+\nu\biggr) \dfrac{\dot{\varphi}}{2r} - \biggl(1+\dfrac{5m_1}{M}-5\nu\biggr) \dfrac{\dot{\varphi} \dot{r}^2}{2}  \biggr] + \dfrac{5}{4} r^2 \dot{\varphi}^3
      \biggr\} \nonumber \biggr. \\
      &+ \biggl. \dfrac{1}{c^3} \biggl\{
        -\dfrac{5}{6} \left(2m_1 \chi_1 + 3m_2 \chi_2\right) \dot{\varphi}^2 - 4m_1 \chi_1 \biggl(7 \dfrac{\dot{r}^2}{r^2} + 4 \dot{\varphi}^2\biggr)\left(1+\sigma_1\right) \nonumber \biggr. \biggr. \\
        &+ \biggl. \biggl. \left(1+3\chi_1^2\right) \biggl[\left(m_1 \chi_1 + m_2 \chi_2\right)\dfrac{m_2}{r^3} + \left(10m_1 \chi_1 - m_2 \chi_2 - 18m_1 B_2 \left(\chi_1\right)\right)\dfrac{\dot{r}^2}{r^2} \nonumber \biggr. \biggr. \biggr. \\
        &+\biggl. \biggl. \biggl. \biggl(\dfrac{1}{3} m_1 \chi_1 - \dfrac{7}{2} m_2 \chi_2 - 8m_1 B_2 \left(\chi_1\right)\biggr) \dot{\varphi}^2 + m_2 \chi_2 \dfrac{\ddot{r}}{r}
          -16m_1 \dfrac{\dphi \dot{r}}{r}-m_1\chi_1 \biggl(\dfrac{19}{2} \dfrac{\dot{r}^2}{r^2} + 4 \dot{\varphi}^2\biggr)\left(1+\sigma_1\right) \biggr] 
        \biggr\}
    \biggr\}
    \\[1em]
    \dfrac{d S_1}{d t} =& -\dfrac{16}{5} \dfrac{m_1^6 m_2^2}{r^6}(1+\sigma_1) \biggl(\Omega_{\rm H}^1 - \dfrac{1}{c^3} \dphi\biggr)
    \biggl\{
       \left(1 + 3 \chi _1^2\right)-\frac{1}{c^2}  \biggl\{ \left(1+3\chi_1^2\right)
      \biggl[\dfrac{7M-2m_1}{r} + \biggl(1+\dfrac{5m_1}{M}-7\nu\biggr) \dfrac{\dot{r}^2}{2} \nonumber \biggr. \biggr. \biggr. \\
      &- \biggl. \biggl. \biggl. \biggl(5+\dfrac{2m_1}{M}+2\nu\biggr) \dfrac{r^2 \dot{\varphi}^2}{4}\biggr] - \dfrac{5}{4} r^2 \dot{\varphi}^2 \biggr\} + \dfrac{1}{c^3} \biggl\{ \dphi \left(1+3\chi_1^2\right) \biggl[-4m_1 \chi_1 (1+\sigma_1)+ 
      \dfrac{1}{3} m_1\chi_1 - \dfrac{7}{2} m_2 \chi_2 \nonumber \biggr. \biggr. \biggr. \\
      &- \biggl. \biggl. \biggl. 8m_1 B_2 \left(\chi_1\right)\biggr] -  16m_1 \chi_1 \dphi \left(1+\sigma_1\right) -
      \dfrac{5}{6} \dphi \left(2m_1 \chi_1 + 3m_2 \chi_2\right) - 16m_1 \left(1+3\chi_1^2\right) \dfrac{\dot{r}}{r}\biggr\}
    \biggr\} \, .
  \end{align}
  \end{subequations}
  \end{widetext}

\section{Non-spinning limit}
\label{app:nonspin}

Picking up from the end of Sec.~\ref{sec:analytics}, this Appendix is dedicated to an exploration
of the relationship between \ac{nnlo}, generic-orbit expressions for the mass and spin evolutions
for spinning \acp{bh} and the \ac{nlo} results for nonrotating \acp{bh}.

As mentioned in the main body of this work, the apparently simple task 
of obtaining the non-spinning limit from the spinning expressions is in fact not trivial. 
It was noticed in Ref.~\cite{Poisson:2004cw} that the different scaling properties of the energy and angular 
momentum fluxes of spinning and nonspinning \acp{bh} 
prevent a straightforward $\chi \rightarrow 0$ limit of the Kerr case from correctly recovering the results 
for Schwarzschild. The two cases can be reconciled, in the quasicircular limit, by using the superradiance prefactor
and the rigid rotation relation, $\dot{m} = \Omega \dot{S}$~\cite{Poisson:2004cw, Chatziioannou:2012gq}.
Indeed, it is easy to verify that the factorization leads to the correct
non-spinning results up to \ac{nlo}, despite that being formally 2.5\ac{pn} orders above
the spinning \ac{lo}, beyond our working \ac{pn} order. By taking Eq.~(4.16) and~(4.19) of 
Ref.~\cite{Saketh:2022xjb}, setting $\chi_1 = \chi_2 = 0$ and expanding the tidal frequency
(their $\Omega$ corresponding to what we called $\Omega_{\rm T}$ above)
up to \ac{nlo} according to their Eq.~(4.12), one arrives precisely at Eq.~(9.4) of Ref.~\cite{Taylor:2008xy}.
Here, we attempt to perform a similar exercise for horizon flux expressions
valid on generic orbits.

Working in modified harmonic coordinates, we calculate noncircular expressions for the 
rates of change of the mass and spin of nonrotating \acp{bh} by substituting the tidal moments of Eqs.~\eqref{eq:momentsbh} into
Eqs.~(8.38) and~(8.39) of Ref.~\cite{Poisson:2004cw}:\footnote{These are just
the terms of Eqs.~\eqref{eq:sb_fields} that depend on the coefficient $f^0_1$,
the only one of the $f_k^l$'s that does not vanish when $\chi_{1,2} \rightarrow 0$
(see Eqs.~(3.35) of Ref.~\cite{Saketh:2022xjb}).}
\begin{widetext}
  \begin{subequations}
  \begin{align}
    \dfrac{dm_1}{d\bar{t}} =& \dfrac{32}{5} \dfrac{m_1^6 m_2^2}{r^6} \biggl\{ 
      \dbphi^2 + 3 \dfrac{\dot{r}^2}{r^2} + \dfrac{1}{c^2} \biggl[
        \left(1-\dfrac{4m_1}{M}+2\nu \right) \dfrac{3\dot{r}^4}{r^2} + 3 r^2 \dbphi^4 - \left(6M-m_1\right) \dfrac{\dbphi^2}{r} - \biggl(\dfrac{m_2}{M}+\nu\biggr) \dbphi^2 r \ddot{r} \biggr. \biggr. \nonumber \\
        &- \biggl. \biggl. 2 \left(9M+m_1\right) \dfrac{\dot{r}^2}{r^3} + \left(9-\dfrac{5m_1}{M}-\nu\right) \dbphi^2 \dot{r}^2 + \left(\dfrac{m_1}{M}-\nu\right) \dfrac{6\dot{r}^2 \ddot{r}}{r} - \biggl(11-\dfrac{m_1}{M}+\nu\biggr) r \dot{r} \dbphi \ddot{\bar{\varphi}} + r^2 \ddot{\bar{\varphi}}^2 \biggr. \biggr. \nonumber \\
        &- \biggl. \biggl. 6 A (\bar{t}) \biggl(4\dfrac{\dot{r}^3}{r^3} + \dfrac{\dot{r}\dbphi^2}{r}-\dfrac{\dot{r} \ddot{r}}{r^2}\biggr)
      \biggr] + \dfrac{16m_1}{3c^3} \biggl[-12 \dfrac{\dot{r}^3}{r^3} - 3\dfrac{\dot{r} \dbphi^2}{r} + \dfrac{\dot{r} \ddot{r}}{r^2} + \dbphi \ddot{\bar{\varphi}}\biggr]
    \biggr\}\label{eq:mdot_ns_bh} \\[1em]
    \dfrac{dS_1}{d\bar{t}} =& \dfrac{32}{5} \dfrac{m_1^6 m_2^2}{r^6} \biggl\{
      \dbphi + \dfrac{1}{c^2} \biggl[
        3r^2 \dbphi^3 - \biggl(\dfrac{m_2}{M}+\nu\biggr) \dfrac{r}{2} \left(\dbphi \ddot{r} + \ddot{\bar{\varphi}}\dot{r}\right) - \biggl(1+\dfrac{5m_1}{M}-5\nu\biggr) \dfrac{\dot{r}^2 \dbphi}{2} - \left(6M-m_1\right) \dfrac{\dbphi}{r} \biggr. \biggr. \nonumber \\
        &- \biggl. \biggl. 6 A(\bar{t}) \dfrac{\dot{r} \dbphi}{r} 
      \biggr] + \dfrac{8m_1}{3c^3} \biggl[-6\dfrac{\dot{r} \dbphi}{r} + \ddot{\bar{\varphi}}\biggr]
    \biggr\}\label{eq:sdot_ns_bh}
  \end{align}
    \label{eq:msdot_ns_bh}
  \end{subequations}
\end{widetext}
We can then map them into the barycentric frame by means of the transformation used in Sec.~\ref{sec:results}:
\begin{widetext}
  \begin{subequations}
  \begin{align}
    \dfrac{dm_1}{dt} =& \dfrac{32}{5} \dfrac{m_1^6 m_2^2}{r^6} \biggl\{
      \dphi^2 + 3 \dfrac{\dot{r}^2}{r^2} + \dfrac{1}{c^2} \biggl[
        \biggl(1-\dfrac{7m_1}{M}+5\nu\biggr) \dfrac{3\dot{r}^4}{2r^2} - \biggl(-7+\dfrac{m_1}{M}+\nu\biggr)\dfrac{r^2 \dphi^4}{2} - \biggl(8M-3m_1+M\nu\biggr) \dfrac{\dphi^2}{r} \biggr. \biggr. \nonumber \\
        &- \biggl. \biggl. \biggl(\dfrac{m_2}{M}-\nu\biggr) \dphi^2 r \ddot{r} - \left(21M-m_1\right) \dfrac{\dot{r}^2}{r^3} + 4\biggl(2-\dfrac{m_1}{M}\biggr) \dphi^2 \dot{r}^2 + 6 \biggl(\dfrac{m_1}{M}-\nu\biggr) \dfrac{\dot{r}^2 \ddot{r}}{r} - \biggl(11-\dfrac{m_1}{M}+\nu\biggr) r \dot{r} \dphi \ddot{\varphi} \biggr. \biggr. \nonumber \\
        &+ \biggl. \biggl. r^2 \ddot{\varphi}^2
      \biggr] + \dfrac{16m_1}{3c^3} \biggl[-12 \dfrac{\dot{r}^3}{r^3} - 3\dfrac{\dot{r} \dphi^2}{r} + \dfrac{\dot{r} \ddot{r}}{r^2} + \dphi \ddot{\varphi}\biggr]
    \biggr\}\label{eq:mdot_ns} \\[1em]
    \dfrac{dS_1}{dt} =& \dfrac{32}{5} \dfrac{m_1^6 m_2^2}{r^6} \biggl\{\dot{\varphi} + \dfrac{1}{c^2} \biggl[
      3r^2 \dot{\varphi}^3 - \biggl(\dfrac{m_2}{M}+\nu\biggr) \dfrac{r}{2} \left(\dot{\varphi} \ddot{r} + \ddot{\varphi}\dot{r}\right)  +  \biggl(\dfrac{15M}{2} - \dfrac{5}{2}m_1 + M\nu\biggr)\dfrac{\dot{\varphi}}{r} \biggr. \biggr. \nonumber \\
      &- \biggl. \biggl. \biggl(1 + \dfrac{5 m_1}{M} -5 \nu\biggr) \dfrac{\dot{\varphi} \dot{r}^2}{2} 
    \biggr] + \dfrac{8m_1}{3c^3} \biggl[-6\dfrac{\dot{r} \dphi}{r} + \ddot{\varphi}\biggr]
    \biggr\} \label{eq:sdot_ns}
  \end{align}
    \label{eq:msdot_ns}
  \end{subequations}
\end{widetext}
Here we find in each of $dm_1/dt$ and $dS_1/dt$ a \ac{nnlo} term that vanishes in the
quasicircular limit (where $\dot{r} = \ddot{r} = \ddot{\varphi} = 0$). Using the Newtonian 
equations of motion, we can also see that these all contain odd powers of the radial velocity $\dot{r}$, since
$\ddot{\varphi} = -2\dot{r}\dphi/r$ at \ac{lo}.

Now, as mentioned, setting $\chi_{1,2}$ to $0$ in Eqs.~\eqref{eq:dmsdt_hfact}
does not lead to the expressions~\eqref{eq:msdot_ns}. 
An obvious structural difference between the quasicircular flux expressions and our generic ones
is the inclusion in the former of the tidal frequency $\Omega_{\rm T}$ in the superradiance prefactor
up to \ac{nnlo} (although the 1.5\ac{pn} terms vanish in the absence of spin), which the latter lack. 

We attempt to reconcile the non-spinning and spinning expressions by writing the prefactors generically as 
$\left(\Omega_{\rm H}^{1,2} - \Omega_{\rm T}^{X_{1,2}}\right)$, with either $X = m$ or $X = S$, and 
making the following ans\"{a}tze for the noncircular, non-spinning tidal frequencies
in the barycentric frame:
\begin{subequations}
\begin{align}
\Omega_{\rm T}^{S_{1}} (t) &= \dot{\varphi}\biggl[1 + \dfrac{1}{c^2} \delta \omega_{1} (t) \biggr]\\
\Omega_{\rm T}^{m_{1}} (t) &= \Omega_{\rm T}^{S_{1}} (t) + 3 \dfrac{\dot{r}^2}{r^2 \dot{\varphi}}\biggl[1 + \dfrac{1}{c^2} \delta \rho_{1} (t)\biggr]
\end{align}
\end{subequations}
By requiring that Eqs.~\eqref{eq:dmsdt_hfact} reduce to Eqs.~\eqref{eq:msdot_ns} when setting $\chi_{1,2} = 0$ and
re-expanding up to $O(c^{-3})$ with respect to the leading contribution, we find:
\begin{subequations}
\begin{align}
  \delta \omega_1 (t) =& -\dfrac{m_2}{M^2} \left(m_1 r^2 \dot{\varphi}^2 - M\dot{r}^2\right)-\dfrac{16m_1}{3c} \dfrac{\dot{r}}{r}\\
  \delta \rho_1 (t)   =& -\dfrac{m_1}{M} \dfrac{5M+6m_1}{r} + \biggl(25+\dfrac{3m_1}{M}-6\nu\biggr) r^2 \dot{\varphi}^2 \nonumber \\
                      &- 3\nu \dot{r}^2 + \dfrac{16m_1}{c} \dfrac{r^3 \dphi^2-M}{r^2 \dot{r}} 
\end{align}
\end{subequations}
Notably, we made use of the harmonic gauge equations of motion to \ac{lo} here to simplify the otherwise
much more involved $\delta \omega$ and $\delta \rho$.
Notice the appearance in the corrections to the tidal frequencies as well of terms linear in the radial
velocity at \ac{nnlo}.
Reconstructing the $\Omega_{\rm T}$ with these results, we find that they reduce
to the correct quasicircular limit, i.e., Eq.~(4.12) of Ref.~\cite{Saketh:2022xjb},
up to 1\ac{pn}.

Performing the same calculations in the \ac{bh} frame, one might expect
the tidal frequency $\Omega_{\rm T}^{S_{1,2}} (\bar{t})$ to coincide with $\dbphi$; however,
this is not the case:
\begin{align}
  \delta \omega_1 (\bar{t}) =& -\dfrac{m_2}{2M^2} \left(M+m_1\right) \left(r^2 \dbphi^2 - \dot{r}^2 - \dfrac{M}{r}\right) \nonumber \\
  &-\dfrac{16m_1}{3c} \dfrac{\dot{r}}{r} \vphantom{A^{A^{A^{A^{A^{A}}}}}}
\end{align}
Restricting to quasicircular orbits, we do find though that 
the correction $\delta \omega_1 (\bar{t})$ vanishes, so that indeed $\Omega_{\rm T}^{S_{1,2}} 
= \dbphi$ in that limit.

In conclusion, a comparison of the spinning and non-spinning limits might shed some light
on currently unknown 2.5\ac{pn} order terms and beyond entering the horizon fluxes in their 
superradiance prefactors. However, in the absence of more solid general results at those orders,
we choose not to test them along with the other flux models considered in the main text.

\end{document}